\documentclass[seceq]{ptptex}
\usepackage{overcite}
\usepackage{bm}
\usepackage{graphicx}
\usepackage{wrapft}
\usepackage{here}
\usepackage{longtable}
\usepackage{supertabular}

\def\mbf#1{\hbox{\boldmath $#1$}}
\def\eq#1{Eq.\ (\ref{#1})}

\def\bR{{\mbf R}}

\def\bk{{\mbf k}}

\def\bq{{\mbf q}}
\def\br{{\mbf r}}

\def\bx{{\mbf x}}

\def\CF{{\cal F}}
\def\CG{{\cal G}}

\def\CW{{\cal W}}

\def\H{{\scriptstyle \frac{1}{2}}}
\def\3H{{\scriptstyle \frac{3}{2}}}
\def\5H{{\scriptstyle \frac{5}{2}}}
\def\7H{{\scriptstyle \frac{7}{2}}}
\def\9H{{\scriptstyle \frac{9}{2}}}

\def\omeg{{\omega_\rho}}
\def\zet{{\zeta_\rho}}
\def\alph{{\alpha_\rho(r)}}

\notypesetlogo                       

\title{
A Practical Method to Solve Cut-off Coulomb Problems
in the Momentum Space}
\subtitle{
Application to the Lippmann-Schwinger Resonating-Group Method
and the $pd$ Elastic Scattering}

\author{Yoshikazu \textsc{Fujiwara} and Kenji \textsc{Fukukawa}
}

\inst{
Department of Physics, Kyoto University, Kyoto 606-8502, Japan
}

\abst{
A practical method to solve cut-off Coulomb problems of two-cluster systems
in the momentum space is given. When a sharply cut-off Coulomb force 
with a cut-off radius $\rho$ is introduced at the level of constituent 
particles, 
two-cluster direct potential of the Coulomb force becomes in general 
a local screened Coulomb potential. The asymptotic Hamiltonian
yields two types of asymptotic waves; one is an approximate Coulomb wave
with $\rho$  in the middle-range region, and the other 
a free (no-Coulomb) wave in the longest-range region.
The constant Wronskians of this Hamiltonian can be calculated 
in either region.
We can evaluate the Coulomb-modified nuclear phase shifts for the
screened Coulomb problem, using the matching condition
proposed by Vincent and Phatak for the sharply cut-off Coulomb problem.
We apply this method first to an exactly solvable model 
of the $\alpha \alpha$ scattering with the Ali-Bodmer potential 
and confirm that a complete solution is obtained with a finite $\rho$.
The stability of nuclear phase shifts with respect to 
the change of $\rho$ in some appropriate range
is demonstrated in the $\alpha \alpha$ resonating-group method (RGM) using 
the Minnesota three-range force. An application to the $pd$ elastic 
scattering is also discussed.
}

\PTPindex{200, 205}  

\begin{document}

\maketitle

\section{Introduction}

In the momentum representation, incorporation
of the long-range Coulomb force always poses problems.
In particular, three-body scattering problems involving
the Coulomb force are still under intensive investigations.
\cite{Gl09, Gl10, Sk09, Wi09e, Wi09b}
Here we mainly consider a much simpler problem of
solving the Lippmann-Schwinger (LS) equations in the momentum
representation, in which the Coulomb force is included 
in the two-cluster resonating-group method (RGM).
In this particular case, the longest range direct
potential consists of a nuclear direct potential and the  
long-range Coulomb potential in the error function form
when simple harmonic-oscillator shell-model wave functions 
are employed for clusters.
We introduce a sharp cut-off radius $\rho$ for the Coulomb
force acting between constituent particles. 
We can solve the LS equations and obtain the $T$-matrix
in the standard procedure. A problem is how to extract the
correct nuclear phase shifts from this $T$-matrix or the
phase shifts, including the effect of the screened Coulomb force.
Here we propose a simple method,
taking examples of the $\alpha \alpha$ RGM and 
the proton-deuteron ($pd$) elastic scattering
using the quark-model baryon baryon interaction.

The standard procedure to solve the Coulomb problem,
including the short-range nuclear potential and
the long-range Coulomb force is well established
as long as two-body problems are concerned.
In the treatment in terms of the distorted waves,
the relative-wave function $\psi_\ell(r)$ between two clusters are
solved numerically in the configuration space,
including the complete Coulomb force.
The nuclear phase shift $\delta^N_\ell$ is then obtained
from the asymptotic form of the relative wave function
through the so-called matching condition
\begin{eqnarray}
{\rm tan}~\delta^N_\ell=-\frac{W[F_\ell, \psi_\ell]_\rho}
{W[G_\ell, \psi_\ell]_\rho}\ ,
\label{eq1-1}
\end{eqnarray}
where $F_\ell=F_\ell(k, r)$ and $G_\ell=G_\ell(k, r)$ stand 
for the regular and irregular Coulomb wave functions, respectively,
and $W[f, g]=f(k,r)(\partial/\partial r)g(k, r)-g(k,r)(\partial/\partial r)f(k, r)$
is the Wronskian for functions $f(k, r)$ and $g(k, r)$.
The Wronskian values in \eq{eq1-1} are evaluated at the relative distance
$r=\rho$, which should be taken large enough to avoid the effect
of the nuclear force in the short-range region.  
Quite naturally, this standard procedure should be modified
in the momentum representation, 
if we try to solve three-body problems like the $pd$ scattering, 
and also the Lippmann-Schwinger RGM (LS-RGM)
equations with the Coulomb interaction. 
In these applications, the basic
ingredient is the $T$-matrix, which is usually formulated in the
momentum space. The Born term of the $T$-matrix is already singular
for the diagonal part of the initial and final momenta, $q_f=q_i$.

A practical method to deal with the Coulomb force 
in the momentum representation is to use the cut-off or screened 
Coulomb force.
In the early work by Vincent and Phatak \cite{Vi74}, 
the Coulomb force in the $\pi^\pm+\hbox{}^{16}\hbox{O}$ scattering 
is assumed to be a sharply cut-off Coulomb force
\begin{eqnarray}
\omega_\rho(r)=\frac{2 \eta k}{r}\theta(\rho-r)\ ,
\label{eq1-2}
\end{eqnarray}
where $\eta=\alpha/\hbar v$ is the Sommerfeld parameter
and $\theta$ is the step function.
Since the relative wave function has a Coulomb-free asymptotic behavior,
the asymptotic wave is composed of the nuclear plus cut-off Coulomb 
phase shift and the known Bessel and Neumann functions. 
This phase shift $\overline{\delta}^\rho_\ell$ is
obtained by solving the LS equation for the $T$-matrix in the
momentum space. The nuclear phase shift $\delta^N_\ell$ is then calculated
from the matching condition of the asymptotic waves:
\begin{eqnarray}
{\rm tan}~\delta^N_\ell=-\frac{W\left[F_\ell, u_\ell\right]_\rho
+{\rm tan}~\overline{\delta}^\rho_\ell~W\left[F_\ell, v_\ell\right]_\rho}
{W\left[G_\ell, u_\ell\right]_\rho
+{\rm tan}~\overline{\delta}^\rho_\ell~W\left[G_\ell, v_\ell\right]_\rho}\ ,
\label{eq1-3}
\end{eqnarray}
with the sufficiently large $\rho$.

A recent Coulomb treatment by Deltuva et al. \cite{De05a, De05b, De05c} for
the $pd$ scattering uses a screened Coulomb potential in the form of
\begin{eqnarray}
\omega_\rho(r)=\frac{2 \eta k}{r} e^{-(r/\rho)^n}
\label{eq1-4}
\end{eqnarray}
with $n \sim 4$, and the ``screening and renormalization procedure'', 
which is developed by Alt et al. \cite{Al76,Al80,Al85,Al96,Al02}
The basic ingredient of this approach is the Taylor's 
theorem \cite{Ta74, Se75},
which implies that the phase shift of the screened Coulomb 
potential $\delta^\rho_\ell$ requires the renormalization 
$\delta^\rho_\ell \longrightarrow \sigma_\ell-\zeta^\rho(k)
\quad \hbox{as} \quad \rho \rightarrow \infty$ in the sense 
of distribution, where
\begin{eqnarray}
\zeta^\rho(k)=\frac{1}{2k} \int^\infty_{\frac{1}{2k}}
\omega_\rho(r)~dr
\label{eq1-5}
\end{eqnarray}
is the diverging renormalization phase determined from the
explicit form of the screened Coulomb potential $\omega_\rho(r)$.
Since the relative wave functions with the screened Coulomb potential
always suffer the renormalization of this phase factor, 
the complete $pd$ scattering amplitude is achieved only when
the limit $\rho \rightarrow \infty$ is reached in the
two-potential formula for the scattering amplitude. 
In practice, this limit is taken numerically such that the 
well converged result is obtained.
A problem of this procedure is that the error estimate of the finite
$\rho$ is not possible, and we usually need to take a very large $\rho$ value,
for which solving the Alt-Grassberger-Sandhas 
equation (AGS equation) \cite{AGS67} 
accurately is not easy because of the quasi-singular nature 
of the screened Coulomb potential.  
The convergence of the partial wave decomposition also becomes problematic,
if the Coulomb singularity is so strong.

A final goal of this study is to find an approximate but practical method 
to incorporate
the Coulomb force to the $pd$ elastic scattering, by using a reasonable
magnitude of $\rho$. For this purpose, we incorporate the Vincent and Phatak 
approach \cite{Vi74} to the ``screening and renormalization procedure''.
In this paper, we first consider a simple potential model for the
$\alpha \alpha$ scattering and examine if this approach gives a reasonable
accuracy of the phase shift, using the screened Coulomb potential.
The stability of the nuclear phase shift with respect to the change of
$\rho$ in an appropriate range is examined by $\alpha \alpha$ LS-RGM. 
An application to the $pd$ elastic scattering is briefly discussed.

In the next section, we discuss the sharply cut-off Coulomb problem, 
for which analytic derivation of the cut-off Coulomb wave functions 
is feasible.
The definitions of the pure Coulomb wave functions used in this paper are
gathered in Appendix A.
A general procedure to calculate the nuclear phase shift from
solutions of the LS equations for the two-cluster $T$-matrix
is discussed in $\S$3. In $\S$4, a formulation for the screened 
Coulomb problem is given, 
by paying an attention to new features appearing in the
screened Coulomb potential. An extension to deal with the $pd$ elastic
scattering in the present approach is given in $\S$5.
In $\S$6, numerical performance is examined, first for an exactly solvable
model in the case of Ali-Bodmer's phenomenological $\alpha \alpha$ potential,
secondly for the $\alpha \alpha$ LS-RGM using the Minnesota three-range force,
and finally for the $pd$ elastic scattering using the quark-model baryon-baryon
interaction fss2. In Appendix B, shift functions of various screening functions
are evaluated. The screening function $\alpha^\rho(R)$ for the $pd$ scattering
is derived in Appendix C.
The last section is devoted to a summary and outlook.

\section{Exact solutions of the sharply cut-off Coulomb problem}

In this section, we assume a sharply cut-off Coulomb potential
\eq{eq1-2}
%
%
and consider the pure Coulomb problem as the limit of $\rho \rightarrow \infty$.
The regular solution $\varphi^\rho_\ell(k, r)$ corresponding to the Jost solution
satisfies the following integral equation:\cite{Go64, Ne66, Ta72}
\begin{eqnarray}
\varphi^\rho_\ell(k, r)=\frac{1}{k^{\ell+1}}u_\ell(kr)
+\int^r_0 \CG_{0 \ell} (r, r^\prime; k) \frac{2k\eta}{r^\prime}
\theta(\rho-r^\prime) \varphi^\rho_\ell(k, r^\prime) d\,r^\prime\ .
\label{eq2-2}
\end{eqnarray}
Here, $u_\ell(kr)$ is the Riccati Bessel function and
the Green function $\CG_{0 \ell} (r, r^\prime; k)$ is given by
\begin{eqnarray}
\CG_{0 \ell}(r,r^\prime; k) = \frac{1}{k} \left[
u_\ell(kr)~v_\ell(kr^\prime)-v_\ell(kr)~u_\ell(kr^\prime) \right]
\theta(r-r^\prime)\ ,
\label{eq2-3}
\end{eqnarray}
with $v_\ell(kr)$ being the Riccati Neumann function. 
For $r \leq \rho$, $\varphi^\rho_\ell(k, r)$ is the same as
the regular Coulomb function $\varphi_\ell(k, r)$ given
in \eq{a1}, which implies
\begin{eqnarray}
& & \frac{1}{k^\ell} F^\rho_\ell(k) \psi^\rho_\ell(k, r)
=\frac{1}{k^\ell} F_\ell(k) \psi_\ell(k, r)
\qquad \rightarrow \qquad
\psi^\rho_\ell(k, r)=\frac{F_\ell(k)}{F^\rho_\ell(k)}
\psi_\ell(k, r) \nonumber \\
& & \qquad \hbox{for} \quad r \leq \rho\ ,
\label{eq2-4}
\end{eqnarray}
where $F^\rho_\ell(k)$ is the Jost function of the
sharply cut-off Coulomb potential and $F_\ell(k)$ the Coulomb 
Jost function.
If we use this in the integral equation for
the regular solution $\psi^\rho_\ell(k, r)$,
\begin{eqnarray}
\psi^\rho_\ell(k, r)=\frac{1}{k} u_\ell(kr)
+\langle r|G_{0\ell} \omega_\rho \psi^\rho_\ell \rangle\ ,
\label{eq2-5}
\end{eqnarray}
we obtain 
\begin{eqnarray}
\psi_\ell(k, r)=\frac{F^\rho_\ell(k)}{F_\ell(k)}
\frac{1}{k} u_\ell(kr)
+\langle r|G_{0\ell} \omega_\rho \psi_\ell \rangle
\qquad \hbox{for} \quad r \leq \rho\ ,
\label{eq2-6}
\end{eqnarray}
with $G_{0 \ell}$ being the regular Green function.
The Jost function $F^\rho_\ell(k)$ of the sharply cut-off 
Coulomb potential is calculated from
$F^\rho_\ell(k)=1+k^\ell \langle \omega^{(-)}_\ell
|\omega_\rho|\varphi^\rho_\ell \rangle$ as~\cite{Ha85}
\begin{eqnarray}
F^\rho_\ell(k) & = & -i \eta \frac{\ell !}{\Gamma(\ell+1-i\eta)}
\sum^\ell_{n=0}\frac{(\ell+n)!}{n!}
\sum^{\ell-n}_{m=0} \frac{(-2ik\rho)^m}{m!}
\frac{\Gamma(n+m-i\eta)}{\Gamma(\ell+n+m+1)}
\nonumber \\
& & \times F(n+m-i\eta, \ell+n+m+1, 2ik\rho)\ .
\label{eq2-7}
\end{eqnarray}
In particular, $S$-wave Jost function is very simple:
\begin{eqnarray}
F^\rho_0(k)=F(-i\eta, 1, 2ik\rho)\ .
\label{eq2-8}
\end{eqnarray}
If we use the asymptotic form of the confluent hypergeometrical function 
$F(\alpha, \gamma, z)$  at $|z| \rightarrow \infty$, we obtain
\begin{eqnarray}
\lim_{\rho \rightarrow \infty}(2k\rho)^{-i \eta} F^\rho_\ell(k)=F_\ell(k)
=e^{\pi\eta/2}\frac{\ell !}{\Gamma(\ell+1+i\eta)}\ ,
\label{eq2-9}
\end{eqnarray}
resulting in
\begin{eqnarray}
\lim_{\rho \rightarrow \infty} \frac{F^\rho_\ell(k)}{F_\ell(k)}
(2k\rho)^{-i \eta}=1\ .
\label{eq2-10}
\end{eqnarray}
This relationship yields the limit of \eq{eq2-6} as
\begin{eqnarray}
\psi_\ell(k, r)=\lim_{\rho \rightarrow \infty}
\left\{
\frac{1}{k} u_\ell(kr) (2k\rho)^{i \eta}
+\langle r|G_{0\ell} \omega_\rho \psi_\ell \rangle \right\}\ .
\label{eq2-11}
\end{eqnarray}
Furthermore, \eq{eq2-4} implies
\begin{eqnarray}
\lim_{\rho \rightarrow \infty}
(2k\rho)^{i \eta} \psi^\rho_\ell(k, r)
=\psi_\ell(k, r)
\label{eq2-12}
\end{eqnarray}

The asymptotic behavior of the cut-off Coulomb phase shift
for $\rho \rightarrow \infty$ can be derived from
the non-Coulomb version of \eq{eq1-1},
since Wronskians $W[u_\ell, \psi_\ell]_\rho$ 
and $W[v_\ell, \psi_\ell]_\rho$
with the $\rho \rightarrow \infty$ limit
are analytically calculated. The result is, of course,
\begin{eqnarray}
\delta^\rho_\ell \rightarrow \sigma_\ell - \eta\,{\rm log}\,2k\rho
\qquad \hbox{as} \qquad \rho \rightarrow \infty\ ,
\label{eq2-13}
\end{eqnarray}
with the ambiguity of integral multiples of $\pi$.\cite{Ta74, Se75}

The Jost solution for the sharply cut-off Coulomb potential is defined by
the integral equation
\begin{eqnarray}
f^\rho_\ell(k, r)=\omega^{(+)}_\ell(kr)
+\int^\infty_r g_{0 \ell} (r, r^\prime; k) \frac{2k\eta}{r^\prime}
\theta(\rho-r^\prime) f^\rho_\ell(k, r^\prime) d\,r^\prime\ ,
\label{eq2-14}
\end{eqnarray}
where the Green function is
\begin{eqnarray}
g_{0 \ell}(r,r^\prime; k) = - \frac{1}{k} \left[
u_\ell(kr)~v_\ell(kr^\prime)-v_\ell(kr)~u_\ell(kr^\prime) \right]
\theta(r^\prime-r)\ . 
\label{eq2-15}
\end{eqnarray}
The asymptotic behavior is given by
\begin{eqnarray}
f^\rho_\ell(k, r)=\omega^{(+)}_\ell(kr) \sim e^{i(kr-(\pi/2) \ell)}
\qquad \hbox{for} \quad r \geq \rho\ . 
\label{eq2-16}
\end{eqnarray}
For the Coulomb solutions, we cannot formulate the integral equation,
since the asymptotic behavior is different from \eq{eq2-16}.
However, for $r \leq \rho$, $f^\rho_\ell(k, r)$ can be written as
a linear combination of two independent Coulomb Jost solutions,
$f_\ell(k, r)$ and $f^*_\ell(k, r)$:
\begin{eqnarray}
f^\rho_\ell(k, r)=C^\rho_1 f_\ell(k, r)+C^\rho_2 f^*_\ell(k, r) \qquad
\hbox{for} \quad r \leq \rho\ .
\label{eq2-17}
\end{eqnarray}
The coefficients, $C^\rho_1$ and $C^\rho_2$ are derived by evaluating
the Wronskians $W[f^*_\ell, f^\rho_\ell]$ 
and $W[f_\ell, f^\rho_\ell]$ at $\rho \rightarrow \infty$.
We find
\begin{eqnarray}
C^\rho_1 & = & \left(1-\frac{\eta}{2k\rho}\right) (2k\rho)^{i\eta}\ ,\nonumber \\
C^\rho_2 & = & (-)^{\ell+1}\frac{\eta}{2k\rho}(2k\rho)^{-i\eta} e^{2ik\rho}
\qquad \hbox{as} \quad \rho \rightarrow \infty\ .
\label{eq2-18}
\end{eqnarray}
Thus, if we use the symmetry \eq{a5} for $f^*_\ell(k,r)$, we find
\begin{eqnarray}
f^\rho_\ell(k, r) & = & \left(1-\frac{\eta}{2k\rho}\right) (2k\rho)^{i\eta}
f_\ell(k, r)
-\frac{\eta}{2k\rho} e^{\pi \eta} (2k\rho)^{-i\eta} e^{2ik\rho} f_\ell(-k, r)
\nonumber \\
& \sim & (2k\rho)^{i\eta} f_\ell(k,r) \qquad
\hbox{for} \quad r \leq \rho \rightarrow \infty\ .
\label{eq2-19}
\end{eqnarray}
After all, we have obtained
\begin{eqnarray}
\lim_{\rho \rightarrow \infty}
(2k\rho)^{-i \eta} f^\rho_\ell(k, r)
=f_\ell(k, r)\ .
\label{eq2-20}
\end{eqnarray}
Note that the renormalization phase 
is the complex conjugate of the one appearing in \eq{eq2-12}.
This results in a basic property of the sharply cut-off Coulomb
potential that the Coulomb Green function can be obtained
as the $\rho \rightarrow \infty$ limit of the sharply
cut-off Coulomb Green function. Namely, if we define
\begin{eqnarray}
G^\rho_\ell(r, r^\prime; k) & = &  -\psi^\rho_\ell(k, r_<) f^\rho_\ell(k, r_>)
\nonumber \\
G^C_\ell(r, r^\prime; k) & = &  -\psi_\ell(k, r_<) f_\ell(k, r_>)\ ,
\label{eq2-21}
\end{eqnarray}
then we find
\begin{eqnarray}
\lim_{\rho \rightarrow \infty} G^\rho_\ell(r, r^\prime; k)
= G^C_\ell(r, r^\prime; k)
\qquad \hbox{for} \quad r,~r^\prime \leq \rho \rightarrow \infty\ .
\label{eq2-22}
\end{eqnarray}
This relationship is valid only when the Green functions are
operated on the short-range potentials.

One can derive the Coulomb scattering amplitude from
the scattering amplitude for the sharply cut-off Coulomb potential.
We use the formula for the short range force
\begin{eqnarray}
f^\rho_\ell & = & -\frac{1}{k}\langle u_\ell|\omega_\rho|\psi^\rho_\ell \rangle
=-\frac{1}{k^2}\langle u_\ell|T^\rho_\ell|u_\ell \rangle
= -\frac{1}{k}\frac{\Im m\,F^\rho_\ell(k)}{F^\rho_\ell(k)} \nonumber \\
& = & \frac{1}{2ik}\left(\frac{F^\rho_\ell(k)^*}{F^\rho_\ell(k)}-1\right)\ ,
\label{eq2-23}
\end{eqnarray}
and calculate
\begin{eqnarray}
f_\ell \equiv \lim_{\rho \rightarrow \infty} 
(2k\rho)^{i\eta}~f^\rho_\ell~(2k\rho)^{i\eta}\ .
\label{eq2-24}
\end{eqnarray}
Equation (\ref{eq2-9}) yields for $\rho \rightarrow \infty$
\begin{eqnarray}
f_\ell & \sim & \frac{1}{2ik}\left(\frac{(F^\rho_\ell(k) (2k\rho)^{-i\eta})^*}
{F^\rho_\ell(k) (2k\rho)^{-i\eta}}-(2k\rho)^{2i\eta}\right) \nonumber \\
& = & \frac{1}{2ik}\left(\frac{F_\ell(k)^*}{F_\ell(k)}
-(2k\rho)^{2i\eta}\right)
=\frac{1}{2ik}\left( e^{2i\sigma_\ell}-(2k\rho)^{2i\eta}\right)
\nonumber \\
& = & \frac{1}{2ik}\left( e^{2i\sigma_\ell}-1\right)
- \frac{1}{2ik}\left((2k\rho)^{2i\eta}-1\right)
=f^C_\ell- \frac{1}{2ik}\left((2k\rho)^{2i\eta}-1\right)\ .
\label{eq2-25}
\end{eqnarray}
Here, the last term is $\ell$-independent and contributes only
to $\theta=0$ if we add up over all the partial waves.
Thus, we find
\begin{eqnarray}
f(\theta) & = & \sum^\infty_{\ell=0} (2\ell+1) f_\ell P_\ell(\cos\,\theta)
\nonumber \\
& = & f^C(\theta)-\sum^\infty_{\ell=0} (2\ell+1)\frac{1}{2ik}
\left((2k\rho)^{2i\eta}-1 \right) P_\ell(\cos\,\theta)
\nonumber \\
& = & f^C(\theta) \qquad \hbox{for} \quad \theta \neq 0\ .
\label{eq2-26}
\end{eqnarray}
Here,
\begin{eqnarray}
f^C(\theta) & = & \frac{1}{2ik} \sum^\infty_{\ell=0}
(2\ell+1)\,(e^{2i\sigma_\ell}-1)\,P_\ell(\cos\,\theta) \nonumber \\
& = & -\frac{\eta}{2k\left(\sin\,\frac{\theta}{2}\right)^2}
~e^{-2i\eta\,\log\,\left(\sin\,\frac{\theta}{2}\right)}
\frac{\Gamma(1+i\eta)}{\Gamma(1-i\eta)}
\label{eq2-27}
\end{eqnarray}
is the standard Coulomb scattering amplitude.

We should note that the renormalization phase $(2k\rho)^{i\eta}$ appearing 
in the above equations is nothing but the Taylor's phase factor \eq{eq1-5}.
In fact, we can easily show for \eq{eq1-2}
\begin{eqnarray}
\zeta^\rho(k)=\frac{1}{2k} \int^\infty_{\frac{1}{2k}}
\omega_\rho(r)~dr = \eta \int^\rho_{\frac{1}{2k}}
\frac{1}{r}~d\,r = \eta~{\rm log}\,(2k \rho)\ .
\label{eq2-28}
\end{eqnarray}
Then, the relationship in Eqs.\,(\ref{eq2-12}) and (\ref{eq2-20}) can be written as
\begin{eqnarray}
\lim_{\rho \rightarrow \infty}
e^{i \zeta^\rho(k)} \psi^\rho_\ell(k, r)
=\psi_\ell(k, r)\ \ ,\qquad
\lim_{\rho \rightarrow \infty}
e^{-i \zeta^\rho(k)} f^\rho_\ell(k, r)
=f_\ell(k, r)\ .
\label{eq2-29}
\end{eqnarray}
More basically, different parametrizations of Coulomb functions
in \eq{a9} and a trivial relationship
\begin{eqnarray}
\psi^{(+)}_\ell(k, r)=\frac{1}{k} \Im {\rm m}\,f_\ell(k, r)
+f^C_\ell\,f_\ell(k, r)
\label{eq2-30}
\end{eqnarray}
derived from them are essential.
Since many relations are also valid even for more general 
screened Coulomb functions introduced in $\S4$, we reformulate
the sharply cut-off Coulomb problem in more general form,
using the parametrization of wave functions as
\begin{eqnarray}
\psi^\rho_\ell(k, r) & = & \frac{1}{k} e^{i \delta^\rho_\ell}
~F^\rho_\ell(k, r)\ ,\nonumber \\
\varphi^\rho_\ell(k, r) & = & \frac{1}{k^{\ell+1}} |F^\rho_\ell(k)|
~F^\rho_\ell(k, r)={\rm real}\ ,\nonumber \\
f^\rho_\ell(k, r) & = & e^{-i\delta^\rho_\ell}
\left[ G^\rho_\ell(k,r)+i F^\rho_\ell(k,r)\right]
\ ,\nonumber \\
f^{\rho *}_\ell(k, r) & = & e^{i\delta^\rho_\ell}
\left[ G^\rho_\ell(k,r)-i F^\rho_\ell(k,r)\right]\ .
\label{eq2-31}
\end{eqnarray}
For the sharply cut-off Coulomb force, the basic screened Coulomb
wave functions satisfying
\begin{eqnarray}
\lim_{\rho \rightarrow \infty}
F^\rho_\ell(k, r)=F_\ell(k, r)\ \ ,\qquad
\lim_{\rho \rightarrow \infty}
G^\rho_\ell(k, r)=G_\ell(k, r)\ ,
\label{eq2-32}
\end{eqnarray}
for a fixed $r$ are analytically derived using the regular and irregular Coulomb
wave functions, $F_\ell(k, r)$ and $G_\ell(k, r)$, and various Wronskians
between these wave functions and the free wave functions. They are given by
\begin{eqnarray}
& & F^\rho_\ell(k, r)=\left\{
\begin{array}{ll}
\frac{|F_\ell(k)|}{|F^\rho_\ell(k)|} F_\ell(k, r) & r \leq \rho \\ [2mm]
u_\ell(kr) \cos\,\delta^\rho_\ell
+v_\ell(kr) \sin\,\delta^\rho_\ell & r \geq \rho \\ [2mm]
\end{array} \right. \ ,\nonumber \\ [2mm]
& & G^\rho_\ell(k, r)=\left\{
\begin{array}{ll}
\frac{|F^\rho_\ell(k)|}{|F_\ell(k)|} G_\ell(k, r)
+\frac{1}{k}\left(-W[G_\ell(k, r), u_\ell(kr)]_\rho \sin\,\delta^\rho_\ell
\right. & \\ [2mm]
\qquad \left. +W[G_\ell(k, r), v_\ell(kr)]_\rho \cos\,\delta^\rho_\ell \right)
F_\ell(k, r) & r \leq \rho \\ [2mm]
v_\ell(kr) \cos\,\delta^\rho_\ell - u_\ell(kr) \sin\,\delta^\rho_\ell
& r \geq \rho \ . \\ [2mm]
\end{array} \right.
\label{eq2-33}
\end{eqnarray}
The screened Coulomb wave function $\psi^\rho_\ell(k, r)$ also
has a expression similar to \eq{eq2-30}: 
\begin{eqnarray}
\psi^{\rho(+)}_\ell(k, r)=\frac{1}{k} \Im {\rm m}\,f^\rho_\ell(k, r)
+f^\rho_\ell\,f^\rho_\ell(k, r)\ ,
\label{eq2-34}
\end{eqnarray}
where $f^\rho_\ell=(1/k)e^{i\delta^\rho_\ell}\sin\,\delta^\rho_\ell$.
Since $f^\rho_\ell(k, r)=\omega^{(+)}_\ell (kr)
=v_\ell(kr)+i u_\ell(kr)$ for $r \geq \rho$,
the asymptotic form of \eq{eq2-34} is
\begin{eqnarray}
\psi^{\rho(+)}_\ell(k, r)=\frac{1}{k} u_\ell(kr)
+f^\rho_\ell\,\omega^{(+)}_\ell(kr) \qquad \hbox{for} \quad r \geq \rho\ .
\label{eq2-35}
\end{eqnarray}
We multiply \eq{eq2-34} by $e^{i\zeta^\rho(k)}$ and find
\begin{eqnarray}
\psi^{\rho(+)}_\ell(k, r) e^{i\zeta^\rho(k)}
& = & \frac{1}{k} \Im {\rm m}\,\left\{f^\rho_\ell(k, r)e^{-i\zeta^\rho(k)}\right\}
\nonumber \\
& & +\left[e^{i\zeta^\rho(k)} f^\rho_\ell e^{i\zeta^\rho(k)}
+f^\rho_\eta \right]\,f^\rho_\ell(k, r)e^{-i\zeta^\rho(k)}\ ,
\label{eq2-36}
\end{eqnarray}
where we have set
\begin{eqnarray}
f^\rho_\eta \equiv \frac{1}{2ik}\left( e^{2i\zeta^\rho(k)}-1 \right)
=\frac{1}{2ik}\left( e^{2i\eta\,\log\,(2k\rho)}-1 \right)\ .
\label{eq2-37}
\end{eqnarray}
Here, we take the limit $\rho \rightarrow \infty$  and use 
\eq{eq2-29}. If we compare the resultant expression 
with \eq{eq2-30}, we find the correspondence
\begin{eqnarray}
e^{i\zeta^\rho(k)} f^\rho_\ell e^{i\zeta^\rho(k)} \rightarrow
f^C_\ell - f^\rho_\eta \qquad \hbox{as} \quad \rho \rightarrow \infty\ .
\label{eq2-38}
\end{eqnarray}
In fact, \eq{eq2-38} diverges, but if we add up over all the partial waves,
the second term of \eq{eq2-38} does not contribute except
for $\theta=0$ because $f^\rho_\eta$ is $\ell$-independent.
Thus, the scattering amplitude of the sharply cut-off Coulomb force 
\begin{eqnarray}
f^\rho(\theta)=\sum^\infty_{\ell=0}(2\ell+1) f^\rho_\ell P_\ell(\theta)\ ,
\label{eq2-39}
\end{eqnarray}
satisfies
\begin{eqnarray}
\lim_{\rho \rightarrow \infty}
e^{i\zeta^\rho(k)} f^\rho(\theta) e^{i\zeta^\rho(k)}
=f^C(\theta) \qquad \hbox{for} \quad \theta \neq 0\ .
\label{eq2-40}
\end{eqnarray}

\section{Two-body Coulomb problem}

In this section, we consider a scattering problem for a
two-body Coulomb system consisting of a short-range local 
potential $v(r)$ with the interaction range $a$ and the
Coulomb force $\omega^C(r)=2k\eta/r$. The Schr{\"o}dinger equation
in the configuration space reads 
\begin{eqnarray}
\left[\left(\frac{d}{dr}\right)^2-\frac{\ell (\ell+1)}{r^2}
-v(r)-\frac{2k\eta}{r}+k^2\right]\,\Psi^{(+)}_\ell(r)=0\ ,
\label{eq3-1}
\end{eqnarray}
with the boundary condition
\begin{eqnarray}
\Psi^{(+)}_\ell(r) \sim \frac{1}{k} \Im {\rm m}\,f_\ell(k, r)
+f_\ell\,f_\ell(k, r)\qquad (r \rightarrow \infty)\ .
\label{eq3-2}
\end{eqnarray}
Here, $f_\ell(k, r)$ is the Coulomb Jost solution in \eq{a3}
and the partial-wave scattering amplitude $f_\ell$ 
is expressed as
\begin{eqnarray}
& & f_\ell=f^C_\ell+e^{2i \sigma_\ell} f^N_\ell
=\frac{1}{2ik}(e^{2i(\sigma_\ell+\delta^N_\ell)}-1)\ ,\nonumber \\
& & \hbox{with} \qquad f^N_\ell=\frac{1}{2ik}(e^{2i\delta^N_\ell}-1)\ ,
\label{eq3-3}
\end{eqnarray}
using the nuclear phase shift $\delta^N_\ell$.
In the usual approach, $\delta^N_\ell$ is calculated
from the real regular function $\CF_\ell(k, r)$ for the
Schr{\"o}dinger equation \eq{eq3-1}, which satisfies
the relationship 
\begin{eqnarray}
\Psi^{(+)}_\ell(r)=\frac{1}{k} e^{i(\sigma_\ell+\delta^N_\ell)}
\CF_\ell(k, r)
\label{eq3-4}
\end{eqnarray}
The asymptotic wave of $\CF_\ell(k, r)$ is expressed as
\begin{eqnarray}
\CF_\ell(k, r) & \sim & F_\ell(k, r) \cos\,\delta^N_\ell
+G_\ell(k, r) \sin\,\delta^N_\ell \nonumber \\
& \sim & \sin\,(kr-\eta \log\,2kr-(\pi/2)\ell+\sigma_\ell+\delta^N_\ell)
\qquad (r \rightarrow \infty)\ .
\label{eq3-5}
\end{eqnarray}
%
The nuclear phase shift $\delta^N_\ell$ is then calculated from \eq{eq1-1}
by assigning $\CF_\ell(k, r)$ to $\psi_\ell$ and taking $\rho > a$ large enough.

Similar equations are also valid for the sharply cut-off Coulomb force
$\omega_\rho(r)$ in \eq{eq1-2}. Namely, the Schr{\"o}dinger equation
for this system,
\begin{eqnarray}
\left[\left(\frac{d}{dr}\right)^2-\frac{\ell (\ell+1)}{r^2}
-v(r)-\omega_\rho(r)+k^2\right]\,\Psi^{\rho(+)}_\ell(r)=0\ ,
\label{eq3-6}
\end{eqnarray}
has the asymptotic wave
\begin{eqnarray}
\Psi^{\rho(+)}_\ell(r) & = & \frac{1}{k} \Im {\rm m}\,f^\rho_\ell(k, r)
+\bar{f}^\rho_\ell\,f^\rho_\ell(k, r) \qquad \hbox{for} \quad r \geq a \nonumber \\
& = & \frac{1}{k} u_\ell(kr)
+\bar{f}^\rho_\ell\,\omega^{(+)}_\ell(k, r)\qquad \hbox{for} \quad r \geq \rho \ ,
\label{eq3-7}
\end{eqnarray}
where $f^\rho_\ell(k, r)$ is the Jost solution for $\omega_\rho(r)$.
The scattering amplitude $\bar{f}^\rho_\ell$ in \eq{eq3-7} this time
is parametrized as
\begin{eqnarray}
& & \bar{f}^\rho_\ell=\frac{1}{2ik}(e^{2i \bar{\delta}^\rho_\ell}-1)
=f^\rho_\ell+e^{2i \delta^\rho_\ell} f^{\rho N}_\ell \qquad
\hbox{with} \quad \bar{\delta}^\rho_\ell=\delta^\rho_\ell+\delta^{\rho N}_\ell\ ,
\nonumber \\
& & \hbox{and} \qquad f^{\rho N}_\ell=\frac{1}{2ik}(e^{2i\delta^{\rho N}_\ell}-1)\ ,
\label{eq3-8}
\end{eqnarray}
where $f^\rho_\ell=(1/2ik)(e^{2i\delta^\rho_\ell}-1)$ is the scattering
amplitude for $\omega_\rho(r)$.
Furthermore,we have the relationship
\begin{eqnarray}
& & \Psi^{\rho(+)}_\ell(r)=\frac{1}{k} e^{i \bar{\delta}^\rho_\ell}
\CF^\rho_\ell(k, r)\ ,\nonumber \\
& & \CF^\rho_\ell(k, r)=u_\ell(kr) \cos\,\bar{\delta}^\rho_\ell
+v_\ell(kr) \sin\,\bar{\delta}^\rho_\ell 
\qquad (r \geq \rho)\ .
\label{eq3-9}
\end{eqnarray}
Note that the second equation of \eq{eq3-9} is exact for the sharply
cut-off Coulomb force.
We multiply \eq{eq3-7} by the phase factor $e^{i\zeta^\rho(k)}$ 
and take the limit $\rho \rightarrow \infty$.
Then, a procedure similar to \eq{eq2-36} leads to the
correspondence
\begin{eqnarray}
& & \lim_{\rho \rightarrow \infty}
\left[e^{i\zeta^\rho(k)} \bar{f}^\rho_\ell e^{i\zeta^\rho(k)}
+f^\rho_\eta \right] = f^C_\ell+e^{2i \sigma_\ell} f^N_\ell
\ ,\nonumber \\
& & \lim_{\rho \rightarrow \infty} \Psi^{\rho(+)}_\ell(k, r)e^{i\zeta^\rho(k)}
=\Psi^{(+)}_\ell(k, r) \qquad \hbox{for} \quad r \geq a \ .
\label{eq3-10}
\end{eqnarray}
From \eq{eq3-8}, the nuclear phase shift $\delta^N_\ell$ is obtained through
\begin{eqnarray}
\delta^N_\ell
=\lim_{\rho \rightarrow \infty} \delta^{\rho N}_\ell
=\lim_{\rho \rightarrow \infty} (\bar{\delta}^\rho_\ell-\delta^\rho_\ell)\ .
\label{eq3-11}
\end{eqnarray}
The sharply cut-off Coulomb phase shift $\delta^\rho_\ell$ is calculated from
\begin{eqnarray}
{\rm tan}\,\delta^\rho_\ell
=-\frac{W[F_\ell(k, r), u_\ell(k, r)]_\rho}{W[F_\ell(k, r), v_\ell(k, r)]_\rho}\ .
\label{eq3-12}
\end{eqnarray}
Since $\bar{\delta}^\rho_\ell$ is obtained by solving the potential problem for
$v(r)+\omega_\rho(r)$, Eqs.\,(\ref{eq3-11}) and (\ref{eq3-12}) gives a solution
for the two-body Coulomb problem in the momentum representation, 
using the sharply cut-off Coulomb force.

Another method to derive the nuclear phase shift in the momentum representation is
to use the two-potential formula for the $T$-matrix.
For the short-range potential $v$ and the sharply cut-off (or screened) Coulomb 
potential $\omega_\rho$, we solve the $T$-matrix equation
\begin{eqnarray}
T^\rho=(v+\omega_\rho)+(v+\omega_\rho) G_0 T^\rho\ ,
\label{eq3-13}
\end{eqnarray}
where $G_0=(z-h_0)^{-1}$ with $z=E+i\varepsilon$
is the free Green function with the energy $E$.
We assume the energy factor $(\hbar^2/2\mu)=1$
and set $E=k^2$. Furthermore, the partial wave decomposition
is implicitly assumed and the orbital angular momentum $\ell$
is omitted for typological simplicity.
The kinetic energy operator $h_0$ is, therefore,
$h_0=(d/dr)^2-\ell(\ell+1)/r^2$. 
The two-potential formula for $T^\rho$ is given by
\begin{eqnarray}
& & T^\rho=t_\omeg+(1+t_\omeg G_0) \widetilde{t}_\omeg
(1+G_0 t_\omeg)\ ,\nonumber \\
& & t_\omeg=\omeg+\omeg G_0 t_\omeg=\omeg+\omeg G_\omeg \omeg
\nonumber \\
& & \widetilde{t}_\omeg=v+v G_\omeg \widetilde{t}_\omeg
=v+ v G^\rho v\ ,
\label{eq3-14}
\end{eqnarray}
where $G_\omeg=(z-h_0-\omeg)^{-1}$ and $G^\rho=(z-h_0-v-\omeg)^{-1}$.
To derive the scattering amplitude, we sandwich $T^\rho$ with
the plane wave (with the wave number $k$)
\begin{eqnarray}
|\phi \rangle=\frac{1}{k}|u\rangle\ ,
\label{eq3-15}
\end{eqnarray}
and define
\begin{eqnarray}
|\psi^{\rho(+)} \rangle=|\phi\rangle
+G_0 \omeg |\psi^{\rho(+)} \rangle\ .
\label{eq3-16}
\end{eqnarray}
Then, by using 
\begin{eqnarray}
|\psi^{\rho(+)} \rangle=(1+G_0 t_\omeg) |\phi\rangle
\ \ ,\qquad  
\langle \psi^{\rho(-)} |=\langle \phi|(1+t_\omeg G_0),
\label{eq3-17}
\end{eqnarray}
we find
\begin{eqnarray}
\langle \phi|T^\rho|\phi \rangle
=\langle \phi| t_\omeg |\phi \rangle
+\langle \psi^{\rho(-)}|
\widetilde{t}_\omeg|\psi^{\rho(+)} \rangle\ .
\label{eq3-18}
\end{eqnarray}
In Eqs.\,(\ref{eq3-17}) and (\ref{eq3-18}), 
$\langle  \psi^{\rho(-)}|$ is defined by
$\psi^{(-)}(k, r)=\left(\psi^{(+)}(k, r)\right)^*$.
This equation is essentially equivalent to
the $T$-matrix in the distorted-wave Born approximation 
(DWBA). In fact, if we set
\begin{eqnarray}
\widetilde{t}_\omeg|\psi^{\rho(+)} \rangle
=v(1+G_\omeg \widetilde{t}_\omeg)|\psi^{\rho(+)} \rangle
\equiv v|\Psi^{\rho(+)}\rangle\ ,
\label{eq3-19}
\end{eqnarray}
the on-shell $T$-matrix is expressed as
\begin{eqnarray}
\langle \phi|T^\rho|\phi \rangle
=\langle \phi| \omega_\rho |\psi^{\rho(+)} \rangle
+\langle \psi^{\rho(-)}|v|\Psi^{\rho(+)}\rangle\ .
\label{eq3-20}
\end{eqnarray}
The LS equation for the total wave function
\begin{eqnarray}
|\Psi^{\rho(+)}\rangle=|\psi^{\rho(+)} \rangle
+G_\omeg v |\Psi^{\rho(+)}\rangle
\label{eq3-21}
\end{eqnarray}
is equivalent to Eqs.\,(\ref{eq3-6}) and (\ref{eq3-7}).

Equation (\ref{eq3-18}) gives a starting point 
for the ``screening and renormalization procedure''.
Namely, if we sandwich \eq{eq3-18} with the renormalization
phase $e^{i\zeta^\rho}$ with $\zeta^\rho=\zeta^\rho(k)$,
and take the limit $\rho \rightarrow \infty$,
we find
\begin{eqnarray}
& & \langle \phi|T|\phi \rangle
\equiv \lim_{\rho \rightarrow \infty} 
e^{i\zet} \langle \phi|T^\rho|\phi \rangle e^{i\zet}
\nonumber \\
& & = \lim_{\rho \rightarrow \infty}
e^{i\zet} \langle \phi| t^\rho_\omega |\phi \rangle e^{i\zet}
+\lim_{\rho \rightarrow \infty}
e^{i\zet} \langle \psi^{\rho(-)}|
\widetilde{t}^\rho_\omega|\psi^{\rho(+)} \rangle e^{i\zet}
\nonumber \\
& & = \langle \phi| t_C |\phi \rangle
+\lim_{\rho \rightarrow \infty}
\langle \psi^{(-)}|
\widetilde{t}^\rho_\omega|\psi^{(+)} \rangle \ .
\label{eq3-22}
\end{eqnarray}
Here,
\begin{eqnarray}
|\psi^{(\pm)}\rangle
=\lim_{\rho \rightarrow \infty}
|\psi^{\rho(\pm)}_\omega \rangle e^{\pm i \zet}
\label{eq3-23}
\end{eqnarray}
are the pure Coulomb wave functions.
The first term in \eq{eq3-22} is separated into
the partial-wave Coulomb amplitude $f^C_\ell$ and
$\ell$-independent term from the discussion of the 
preceding section. When all the partial-wave contributions
are added up, the first term becomes the pure Coulomb
amplitude. Actually, the relationship between the
scattering amplitude and the on-shell $T$-matrix yields
\begin{eqnarray}
\langle \bq_f|t_C|\bq_i \rangle
=- \frac{4\pi}{(2\pi)^3}\frac{\hbar^2}{2\mu} f^C(\theta)
=-\frac{\hbar^2}{(2\pi)^2\mu} f^C(\theta)
\label{eq3-24}
\end{eqnarray}
with $|\bq_f|=|\bq_i|=k$.
In the second term of \eq{eq3-22}, $\rho \rightarrow \infty$ limit
can be taken, since the nuclear potential $v$ is short-ranged.
We define $\widetilde{t}$ by the solution of
$\widetilde{t}=v+v G^C \widetilde{t}$, where $G^C
=(z-h_0-2\eta k/r)^{-1}$ is the Coulomb Green function.
Thus, we find
\begin{eqnarray}
& & \lim_{\rho \rightarrow \infty}
\langle \psi^{(-)}|
\widetilde{t}_{\omega_\rho}|\psi^{(+)} \rangle
=\langle \psi^{(-)}|
\widetilde{t}|\psi^{(+)} \rangle\ .
\label{eq3-25}
\end{eqnarray}
To derive this matrix element, we introduce
the total wave function $|\Psi^{(+)}\rangle$ through 
\begin{eqnarray}
\widetilde{t} |\psi^{(+)}\rangle
=v(1+G^C \widetilde{t})|\psi^{(+)}\rangle
=v |\Psi^{(+)}\rangle \ ,
\label{eq3-26}
\end{eqnarray}
which satisfies the LS equation
\begin{eqnarray}
|\Psi^{(+)}\rangle=(1+G^C \widetilde{t})|\psi^{(+)}\rangle
=|\psi^{(+)}\rangle+G^C v |\Psi^{(+)}\rangle\ ,
\label{eq3-27}
\end{eqnarray}
and the Schr{\"o}dinger equation in Eqs.\,(\ref{eq3-1}) and (\ref{eq3-2}).
We should note that \eq{eq3-27} has a solution,
since $v$ is short-ranged.
Here, we introduce a decomposition of the 
partial-wave Green function
\begin{eqnarray}
& & G^C(r, r^\prime; k) = -\psi_\ell(k, r_<) f_\ell(k, r_>)\ ,\nonumber \\
& & G^C=\widetilde{G}^C- |f_\ell \rangle \langle \psi^{(-)}_\ell|
\qquad \hbox{with} \qquad \widetilde{G}^C \rightarrow 0
\qquad \hbox{as} \quad r \rightarrow \infty \ .
\label{eq3-28}
\end{eqnarray}
Then, we find the asymptotic behavior 
\begin{eqnarray}
|\Psi^{(+)}\rangle
& = & |\psi^{(+)}\rangle-|f\rangle \langle \psi^{(-)}| v |\Psi^{(+)}\rangle
+\widetilde{G}^C v |\Psi^{(+)}\rangle \nonumber \\
& \sim & |\psi^{(+)}\rangle-|f\rangle \langle \psi^{(-)}| \widetilde{t}|
\psi^{(+)}\rangle \qquad \hbox{as} \quad r \rightarrow \infty \ .
\label{eq3-29}
\end{eqnarray}
If we use the Wronskians of the Coulomb wave functions
\begin{eqnarray}
& & W[F_\ell, G_\ell]=-k\ ,\nonumber \\
& & W[F_\ell, f_\ell]=-k e^{-i \sigma_\ell}\ ,\nonumber \\
& & W[\psi^{(+)}_\ell, f_\ell]=-1 \ ,
\label{eq3-30}
\end{eqnarray}
derived from \eq{a9}, we obtain
\begin{eqnarray}
& & W[\psi^{(+)}_\ell, \Psi^{(+)}]_{r \rightarrow \infty}
=\langle \psi^{(-)}| \widetilde{t}|\psi^{(+)}\rangle\ ,\nonumber \\
& & W[f_\ell, \Psi^{(+)}]_{r \rightarrow \infty}=1
\label{eq3-31}
\end{eqnarray}
and
\begin{eqnarray}
\langle \psi^{(-)}| \widetilde{t}|\psi^{(+)}\rangle_\ell
=\lim_{r \rightarrow \infty} 
\frac{W[\psi^{(+)}_\ell(k, r), \Psi^{(+)}(k, r)]}
{W[f_\ell(k, r), \Psi^{(+)}(k, r)]} \ .
\label{eq3-32}
\end{eqnarray}
If we further parametrize
\begin{eqnarray}
\langle \psi^{(-)}| \widetilde{t}|\psi^{(+)}\rangle_\ell
=-e^{2i\sigma_\ell} \frac{1}{2ik} \left( e^{2i \delta^N_\ell}-1\right) \ ,
\label{eq3-33}
\end{eqnarray}
\eq{eq3-32} is equivalent to 
\begin{eqnarray}
{\rm tan}~\delta^N_\ell=-\lim_{r \rightarrow \infty} 
\frac{W[F_\ell(k, r), \Psi^{(+)}(k, r)]}
{W[G_\ell(k, r), \Psi^{(+)}(k, r)]} \ .
\label{eq3-34}
\end{eqnarray}
Eventually, we find
\begin{eqnarray}
& & \langle \phi|T|\phi\rangle = -\frac{\hbar^2}{(2\pi)^2\mu} f(\theta)
\ ,\nonumber \\
& &  f(\theta)=f^C(\theta)
+\sum^\infty_{\ell=0} (2\ell+1) e^{2i\sigma_\ell}~f^N_\ell
~P_\ell(\cos\,\theta)\ ,\nonumber \\
& & \hbox{with} \qquad f^N_\ell=\frac{1}{2ik}\left(e^{2i\delta^N_\ell}-1
\right) \ .
\label{eq3-35}
\end{eqnarray}

For practical calculations in the momentum space,
it is much easier to start with the sharply cut-off
Coulomb force from the very beginning. 
We multiply the LS equation in \eq{eq3-21} by
the renormalization phase $e^{i\zet}$
and take the limit $\rho \rightarrow \infty$.
Then, by using Eqs.\,(\ref{eq2-12}) and (\ref{eq2-22}), we obtain
\begin{eqnarray}
\lim_{\rho \rightarrow \infty} |\Psi^{\rho(+)}\rangle e^{i\zet}
=|\psi^{(+)}\rangle+G^C v 
\lim_{\rho \rightarrow \infty}
|\Psi^{\rho(+)}\rangle e^{i\zet}\ .
\label{eq3-36}
\end{eqnarray}
If we compare this with \eq{eq3-27}, we find
\begin{eqnarray}
\lim_{\rho \rightarrow \infty} |\Psi^{\rho(+)}\rangle e^{i\zet}
=|\Psi^{(+)}\rangle \ .
\label{eq3-37}
\end{eqnarray}
If we further use \eq{eq3-37} in ${\rm tan}~\delta^N_\ell$ 
of \eq{eq3-34}, we find
\begin{eqnarray}
{\rm tan}~\delta^N_\ell=-\lim_{r \rightarrow \infty} 
\frac{W[F_\ell(k, r), \Psi^{\rho (+)}_\ell(k, r)]}
{W[G_\ell(k, r), \Psi^{\rho (+)}_\ell(k, r)]}\ ,
\label{eq3-38}
\end{eqnarray}
for sufficiently large $\rho$.
On the other hand, the asymptotic behavior
of the wave functions for the short-range force
yields
\begin{eqnarray}
\Psi^{\rho(+)}_\ell(k, r) & = & \frac{1}{k} e^{i\overline{\delta}^\rho_\ell}
\left\{ u_\ell(kr)~\cos\,\overline{\delta}^\rho_\ell
+v_\ell(kr)~\sin\,\overline{\delta}^\rho_\ell \right\} \nonumber \\
& & \hspace{50mm} \hbox{for} \quad \rho < r \rightarrow \infty\ ,
\label{eq3-39}
\end{eqnarray}
for sufficiently large $\rho$.
Thus, if we calculate Wronskians in \eq{eq3-38} at $r = \rho$,
we obtain
\begin{eqnarray}
{\rm tan}~\delta^N_\ell & = & -\frac{W[F_\ell(k, r), \Psi^{\rho (+)}_\ell(k, r)]
_{r=\rho}}
{W[G_\ell(k, r), \Psi^{\rho (+)}_\ell(k, r)]_{r=\rho}}
\nonumber \\
& = & -\frac{W\left[F_\ell, u_\ell\right]_\rho
+{\rm tan}\,\overline{\delta}^\rho _\ell~W\left[F_\ell, v_\ell\right]_\rho}
{W\left[G_\ell, u_\ell\right]_\rho
+{\rm tan}\,\overline{\delta}^\rho_\ell~W\left[G_\ell, v_\ell\right]_\rho}\ ,
\label{eq3-40}
\end{eqnarray}
which is nothing but \eq{eq1-3}.
After all, if $\overline{\delta}^\rho_\ell$ is calculated in the momentum 
representation, the nuclear phase shift $\delta^N_\ell$ is
obtained through \eq{eq3-40}. The scattering amplitude 
$f_\ell (\theta)$ is calculated from \eq{eq3-35}, using $\delta^N_\ell$.

\section{The screened Coulomb case}

In this section, we will extend the preceding discussion for
the sharply cut-off Coulomb force to a more general screened 
Coulomb force, which is formulated as
\begin{eqnarray}
\omega_\rho(r)=\frac{2k\eta}{r} \alpha_\rho(r)\ ,
\label{eq4-1}
\end{eqnarray}
according to Taylor \cite{Ta74}.
Here, the screening function $\alph$ with 
$1 \geq \alph \geq 0$ is a monotonically decreasing
function of $r$, satisfying
\begin{enumerate}
\item[1)] with $\rho$ fixed, $\alph$ decreases to zero,
faster than $O(r^{-\varepsilon-2})$ ($\varepsilon >0$),
as $r$ approaches to $\infty$, 
\item[2)] with $r$ fixed, $\alph$ appraoches to 1 as 
$\rho$ approaches to $\infty$,
\item[3)] around $r \sim \rho$, there exist sufficiently wide regions
in which $\alph \sim 1$ and $\sim 0$.
\end{enumerate}

\bigskip

In 3) above, we added ``an almost sharply cut-off condition''
in addition to the original conditions 1) and 2) in Ref.\,\citen{Ta74}.
This condition is required if we wish to develop
almost parallel discussion to the sharply cut-off Coulomb
case, as seen below. Note that the sharply cut-off Coulomb
case is included in the above category by taking
$\alph=\theta(\rho-r)$. 

The necessity to relax the sharply cut-off condition is
as follows. First, in the LS-RGM formalism, the longest-range
direct Coulomb potential becomes a screened Coulomb force
as explicitly shown in $\S6.1$ and $\S6.2$. If the cluster 
wave functions are assumed to be standard harmonic-oscillator 
shell-model wave functions, the cut-off function $\alph$ is 
usually expressed by the error function.
Secondly, in the application to the $pd$ elastic scattering,
the asymptotic Hamiltonian involves a screened Coulomb force
which is obtained from the $pp$ Coulomb force by the folding
procedure using a realistic deuteron wave function.
In Ref.\,\citen{De05a}, the same $pp$ screened Coulomb force is
used for the $pd$ screened Coulomb force, but using a more realistic
$pd$ Coulomb potential is certainly desirable to avoid unnecessary
extra distortion of the deuteron in the asymptotic region
by the Coulomb force.
In any case, the screening function $\alph$ should be chosen
most appropriately for each problem, since ``in practice
Coulomb potentials are always screened'' as stated in Ref.\,\citen{Ta74}.
   
For the screened Coulomb force in \eq{eq4-1}, the parametrization 
of screened Coulomb wave functions in \eq{eq2-31} is employed
in the following, but the explicit solutions of $F^\rho_\ell(k,r)$
and $G^\rho_\ell(k, r)$ like in \eq{eq2-33} are no longer available.
In order to extend \eq{eq2-33} to the screened Coulomb case, we
first examine the behavior of the screened Coulomb wave functions
around the origin $r \sim 0$. For the pure Coulomb solutions,
$F_\ell(k, r)$ and $G_\ell(k,r)$, we can easily show that
\begin{eqnarray}
F_\ell(k, r) \sim \frac{1}{|F_\ell(k)|} u_\ell(kr)\ \ ,\qquad
G_\ell(k, r) \sim |F_\ell(k)| v_\ell(kr) \qquad
\hbox{as} \quad r \rightarrow 0\ ,
\label{eq4-2}
\end{eqnarray}
by using the explicit expression of the Coulomb Jost solution 
$f_\ell(k, r)$ in \eq{a3} and the parametrization \eq{a9}.
The corresponding expressions for the screened Coulomb wave functions 
are
\begin{eqnarray}
F^\rho_\ell(k, r) & \sim & \frac{1}{|F^\rho_\ell(k)|} u_\ell(kr) \nonumber \\
G^\rho_\ell(k, r) & \sim & |F^\rho_\ell(k)| v_\ell(kr)
+\frac{1}{|F^\rho_\ell(k)|} A^\rho_\ell(r) u_\ell(kr)
\qquad \hbox{as} \quad r \rightarrow 0\ ,
\label{eq4-3}
\end{eqnarray}
where an extra term including $A^\rho_\ell(r)$ appears 
in the irregular solution $G^\rho_\ell(k, r)$.
The real function $A^\rho_\ell(r)$ is given by 
\begin{eqnarray}
A^\rho_\ell(r) & = & 
-|F^\rho_\ell(k)|\,\frac{1}{k}\,W\left[G^\rho_\ell(k, r), v_\ell(kr) \right]
\nonumber \\
& \sim & 
\left\{ \begin{array}{c}
{\rm log}\,r \\ [2mm]
\frac{1}{r^{2\ell}} \\
\end{array} \right.
\qquad \hbox{for} \quad
\begin{array}{c}
\ell=0 \\ [2mm]
\ell \geq 1 \\
\end{array}
\qquad \hbox{as} \quad r \rightarrow 0\ ,
\label{eq4-4}
\end{eqnarray}
and diverges as $r \rightarrow 0$.
These results are derived by applying the Calogero's
variable phase method \cite{Ca67} to the regular solution
$\varphi^\rho_\ell(k, r)$ and
the Jost solution $f^\rho_\ell(k, r)$.

For practical applications, 
we use the ``almost sharply cut-off condition'' 3)
and assume a screening function satisfying

\bigskip

\begin{enumerate}
\item[3)$^\prime$]
\begin{eqnarray}
\alpha_\rho(r)=\left\{\begin{array}{lll}
1 & \hbox{for} & r < \rho-b=R_{\rm in} \\ [2mm]
0 & \hbox{for} & r > \rho+b=R_{\rm out} \\
\end{array}\right.\ ,
\label{eq4-5}
\end{eqnarray}
\end{enumerate}
with a sufficiently large $\rho \gg b$.
A new parameter $b$ is introduced to make a smooth transition
for the Coulomb force to disappear. To make the pure Coulomb region
available, $R_{\rm in} \gg a$ should also be taken large enough,
compared with the range $a$ of the short-range nuclear force.
By this assumption, we can extend the discussion in the
sharply cut-off Coulomb case, although some modifications are
necessary as seen below.
First we apply Calogero's variable phase method to the
regular function $\varphi^\rho_\ell(k, r)$.
This solution and the pure Coulomb wave function $\varphi_\ell(k, r)$
both satisfy the same integral equation \eq{eq2-2} with $\theta(\rho-r)
\rightarrow \alph$ or 1 for $r < R_{\rm in}$, yielding
\begin{eqnarray}
\varphi^\rho_\ell(k, r)=\varphi_\ell(k, r)\qquad \hbox{for} \quad
r \leq R_{\rm in}\ .
\label{eq4-6}
\end{eqnarray}
If we use the standard relationship
\begin{eqnarray}
\varphi_\ell(k, r)=\frac{1}{k^{\ell+1}} |F_\ell(k)|
F_\ell(k, r)\ ,
\label{eq4-7}
\end{eqnarray}
(see Eqs.\,(\ref{a9}) and (\ref{eq2-31})), 
\eq{eq4-6} implies
\begin{eqnarray}
F^\rho_\ell(k, r)=\frac{|F_\ell(k)|}{|F^\rho_\ell(k)|}
F_\ell(k, r)\qquad \hbox{for} \quad
r \leq R_{\rm in}\ .
\label{eq4-8}
\end{eqnarray}
Here, we can prove
\begin{eqnarray}
\lim_{\rho \rightarrow \infty} \frac{|F_\ell(k)|}{|F^\rho_\ell(k)|}=1\ .
\label{eq4-9}
\end{eqnarray}
Similarly, the screened Coulomb phase shift $\delta^\rho_\ell
=\delta^\rho_\ell(k)$ 
is proved to have the Coulomb limit
\begin{eqnarray}
\lim_{\rho \rightarrow \infty} \left(\delta^\rho_\ell + \zeta^\rho \right)
=\sigma_\ell\ ,
\label{eq4-10}
\end{eqnarray}
where $\zeta^\rho=\zeta^\rho(k)$ is given by \eq{eq1-5} 
for $\omega_\rho(r)$ in \eq{eq4-1}.
For the irregular solution $G^\rho_\ell(k,r)$, the local phase approach
does not work. In this case, we have admixture of the regular solution
$F_\ell(k, r)$ for $r < R_{\rm in}$, which is related to
$A^\rho_\ell(r)$ in \eq{eq4-4}. Summarizing the above discussion,
the explicit results of \eq{eq2-33} in the sharply cut-off Coulomb case
should be modified to
\begin{eqnarray}
& & F^\rho_\ell(k, r)=\frac{1}{a^\rho_\ell} F_\ell(k, r) \ , \nonumber \\ [2mm]
& & G^\rho_\ell(k, r)=a^\rho_\ell G_\ell(k, r)
+A^\rho_\ell F_\ell(k, r)
\qquad \hbox{for} 
\quad r \leq R_{\rm in} \ , \nonumber \\ [2mm]
& & F^\rho_\ell(k, r)=u_\ell(kr) \cos\,\delta^\rho_\ell
+v_\ell(kr) \sin\,\delta^\rho_\ell \ , \nonumber \\ [2mm]
& & G^\rho_\ell(k, r)=v_\ell(kr) \cos\,\delta^\rho_\ell
-u_\ell(kr) \sin\,\delta^\rho_\ell\qquad \hbox{for} 
\quad r \geq R_{\rm out}\ ,\nonumber \\
\label{eq4-11}
\end{eqnarray}
where
\begin{eqnarray}
a^\rho_\ell=\frac{|F^\rho_\ell(k)|}{|F_\ell(k)|}\ \ ,\qquad
\lim_{\rho \rightarrow \infty} a^\rho_\ell=1\ ,
\label{eq4-12}
\end{eqnarray}
and
\begin{eqnarray}
& & \delta^\rho_\ell \rightarrow \sigma_\ell-\zeta^\rho \qquad \hbox{as}
\quad \rho \rightarrow \infty \ ,\nonumber \\ [2mm]
& & \hbox{with} \qquad 
\zeta^\rho = \zeta^\rho(k)
=\frac{1}{2k}\int^{\infty}_{\frac{1}{2k}} \omega_\rho(r)~d\,r\ .
\label{eq4-13}
\end{eqnarray}

We note that, for the pure Coulomb problem, the renormalization of the
screened Coulomb wave functions and the scattering amplitude 
is possible. In particular, Eqs.\,(\ref{eq2-34}) - (\ref{eq2-40}) are all
valid owing to \eq{eq4-13}. However, the renormalization of the irregular
solutions like in \eq{eq2-20} needs a modification, 
since in general $A^\rho_\ell \neq 0$ in \eq{eq4-11}. 
For example, the relationship in \eq{eq2-29} should be modified as
\begin{eqnarray}
& & \lim_{\rho \rightarrow \infty}
\psi^{\rho(+)}_\ell(k, r) e^{i\zeta^\rho}
=\lim_{\rho \rightarrow \infty} \frac{1}{k}
e^{i(\delta^\rho_\ell+\zeta^\rho)}~F^\rho_\ell(k, r)
=\frac{1}{k}e^{i \sigma_\ell}~F_\ell(k, r) \nonumber \\
& & =\psi^{(+)}_\ell(k, r)  \qquad 
(\hbox{for~regular~Coulomb~wave~function})\ ,
\nonumber \\ [2mm]
& & \lim_{\rho \rightarrow \infty}
f^{\rho}_\ell(k, r) e^{-i\zeta^\rho}
=\lim_{\rho \rightarrow \infty}
e^{-i(\delta^\rho_\ell+\zeta^\rho)}
~\left[ G^\rho_\ell(k, r)+i F^\rho_\ell(k, r) \right]
\nonumber \\
& & =e^{-i \sigma_\ell}
~\left[ G_\ell(k, r)+\lim_{\rho \rightarrow \infty} A^\rho_\ell
~F_\ell(k, r)+i F_\ell(k, r) \right]
\nonumber \\
& & = f_\ell(k, r)+A_\ell\,e^{-i\sigma_\ell}~F_\ell(k, r)\ ,
\nonumber \\ [2mm]
& & \lim_{\rho \rightarrow \infty}
\left[e^{i\zeta^\rho} f^\rho_\ell e^{i\zeta^\rho}
+f^\rho_\eta \right]=\frac{1}{2ik}\left(e^{i\sigma_\ell}-1\right)
= f^{\rm C}_\ell \ ,
\label{eq4-14}
\end{eqnarray}
for $r < R_{\rm in}$. Here, we have assumed 
that $A_\ell =\lim_{\rho \rightarrow \infty} A^\rho_\ell$ exists 
for simplicity.
By the same token, the $\rho \rightarrow \infty$ limit 
in \eq{eq2-36} becomes
\begin{eqnarray}
\psi^{(+)}_\ell(k, r)
& = & \frac{1}{k} \Im {\rm m}\,\left\{f_\ell(k, r)
+A_\ell e^{-i\sigma_\ell}~F_\ell(k, r)\right\}
\nonumber \\
& & +f^{\rm C}_\ell \left[ f_\ell(k, r)
+A_\ell e^{-i\sigma_\ell}~F_\ell(k, r)\right]
\qquad \hbox{for} \quad r < R_{\rm in} \ .
\label{eq4-15}
\end{eqnarray}
Here, because $A_\ell$ is real, the contribution
from the terms proportional to $A_\ell$ vanishes as
\begin{eqnarray}
-\frac{1}{k} A_\ell~\sin\,\sigma_\ell~F_\ell(k, r)
+\frac{1}{2ik} \left(e^{2i\sigma_\ell}-1\right)
~A_\ell~e^{-i\sigma_\ell}~F_\ell(k, r)=0\ ,
\label{eq4-16}
\end{eqnarray}
resulting in \eq{eq2-30} again.
It is important to note that this renormalization is
possible only for the regular solution of the pure Coulomb 
problem. Once the nuclear potential is introduced,
we need further renormalization for the magnitude of the
wave function related to $A^\rho_\ell$, 
since the derivation of the regular solution
also requires irregular solution of the screened Coulomb problem.

In order to make the similarity to the sharply cut-off Coulomb case
more transparent, it is convenient to introduce a modified set of
screened Coulomb wave functions by 
\begin{eqnarray}
\widetilde{F}^\rho_\ell(k, r) & = & a^\rho_\ell~F^\rho_\ell(k, r)\ ,
\nonumber \\
\widetilde{G}^\rho_\ell(k, r) & = & \frac{1}{a^\rho_\ell}~G^\rho_\ell(k, r)
-A^\rho_\ell~F^\rho_\ell(k, r)\ .
\label{eq4-17}
\end{eqnarray}
For $r < R_{\rm in}$, these are the pure Coulomb wave functions:
\begin{eqnarray}
\widetilde{F}^\rho_\ell(k, r) = F_\ell(k, r)\ \ ,\qquad
\widetilde{G}^\rho_\ell(k, r) = G_\ell(k, r) \qquad
\hbox{for} \quad r \leq R_{\rm in}\ .
\label{eq4-18}
\end{eqnarray}
However, for $r \geq R_{\rm out}$, \eq{eq4-11} leads to 
\begin{eqnarray}
\widetilde{F}^\rho_\ell(k, r) & = & a^\rho_\ell
\left[ u_\ell(kr) \cos\,\delta^\rho_\ell
+v_\ell(kr) \sin\,\delta^\rho_\ell\right]\ , \nonumber \\ [2mm]
\widetilde{G}^\rho_\ell(k, r) & = & \frac{1}{a^\rho_\ell}
\left[ v_\ell(kr) \cos\,\delta^\rho_\ell
-u_\ell(kr) \sin\,\delta^\rho_\ell \right] 
\nonumber \\
& & -A^\rho_\ell \left[ u_\ell(k, r) \cos\,\delta^\rho_\ell
+v_\ell(k, r) \sin\,\delta^\rho_\ell\right]
\qquad \hbox{for} \quad r \geq R_{\rm out}\ .
\label{eq4-19}
\end{eqnarray}
The Wronskians of these wave functions with the free
scattering solutions in the asymptotic region are given
by
\begin{eqnarray}
\frac{1}{k}W\left[\widetilde{F}^\rho_\ell(k, r), u_\ell(kr)\right]
& = & a^\rho_\ell~\sin\,\delta^\rho_\ell\ ,\nonumber \\ [2mm]
\frac{1}{k}W\left[\widetilde{F}^\rho_\ell(k, r), v_\ell(kr)\right]
& = & -a^\rho_\ell~\cos\,\delta^\rho_\ell\ ,\nonumber \\ [2mm]
\frac{1}{k}W\left[\widetilde{G}^\rho_\ell(k, r), u_\ell(kr)\right]
& = & \frac{1}{a^\rho_\ell}~\cos\,\delta^\rho_\ell
-A^\rho_\ell~\sin\,\delta^\rho_\ell\ ,\nonumber \\ [2mm]
\frac{1}{k}W\left[\widetilde{G}^\rho_\ell(k, r), v_\ell(kr)\right]
& = & \frac{1}{a^\rho_\ell}~\sin\,\delta^\rho_\ell
+A^\rho_\ell~\cos\,\delta^\rho_\ell
\qquad \hbox{for} \quad r \geq R_{\rm out}\ .
\label{eq4-20}
\end{eqnarray}
Let us assume $a \ll R_{\rm in}$ and consider the regular solution of the
Schr{\"o}dinger equation for $v(r)+(2k\eta/r)\alpha_\rho(r)$:
\begin{eqnarray}
\Psi^{\rho (+)}_\ell(r) & = & \widetilde{F}^\rho_\ell(k, r)
~\cos\,\delta^N_\ell
+\widetilde{G}^\rho_\ell(k, r)~\sin\,\delta^N_\ell \qquad
\hbox{for} \quad r >a \ .
\label{eq4-21}
\end{eqnarray}
In the $a < r < R_{\rm in}$ region,
$\delta^N_\ell$ becomes the nuclear phase shift
owing to \eq{eq4-18}. This can be calculated from
\begin{eqnarray}
{\rm tan}\,\delta^N_\ell
=-\frac{W\left[\widetilde{F}^\rho_\ell(k, r), \Psi^{\rho (+)}_\ell(r)\right]}
{W\left[\widetilde{G}^\rho_\ell(k, r), \Psi^{\rho (+)}_\ell(r)\right]}
\qquad \hbox{for} \quad r > a\ .
\label{eq4-22}
\end{eqnarray}
The Wronskians in \eq{eq4-22} can be calculated at any points $r > a$,
since $\widetilde{F}^\rho_\ell(k, r)$, $\widetilde{G}^\rho_\ell(k, r)$
and $\Psi^\rho_\ell(r)$ are all solutions of the Schr{\"o}dinger equation
for the screened Coulomb potential.
In particular, the asymptotic behavior
\begin{eqnarray}
\Psi^{\rho (+)}_\ell(r) = B\,\left[ u_\ell(kr)~\cos\,\overline{\delta}^\rho_\ell
+v_\ell(kr)~\sin\,\overline{\delta}^\rho_\ell \right] 
\qquad \hbox{for} \quad r > R_{\rm out}
\label{eq4-23}
\end{eqnarray}
without the Coulomb force, yields a connection condition
\begin{eqnarray}
\hspace{-10mm}
{\rm tan}\,\delta^N_\ell
=-\frac{W\left[\widetilde{F}^\rho_\ell(k, r), u_\ell(kr)\right]_{R_{\rm out}}
+{\rm tan}\,\overline{\delta}^\rho_\ell
~W\left[\widetilde{F}^\rho_\ell(k, r), v_\ell(kr)\right]_{R_{\rm out}}}
{W\left[\widetilde{G}^\rho_\ell(k, r), u_\ell(kr)\right]_{R_{\rm out}}
+{\rm tan}\,\overline{\delta}^\rho_\ell
~W\left[\widetilde{G}^\rho_\ell(k, r), v_\ell(kr)\right]_{R_{\rm out}}}\ ,
\label{eq4-24}
\end{eqnarray}
which is an extension of \eq{eq3-40} in the sharply cut-off Coulomb case.
The phase shift $\overline{\delta}^\rho_\ell$ is calculated from
the standard procedure to solve $T$-matrix of $v(r)+(2k\eta/r)\alpha_\rho(r)$
in the momentum representation.

To the contrary, we can also recover the asymptotic behavior
of $\Psi^{\rho (+)}_\ell(r)$ in \eq{eq4-23}, 
starting from \eq{eq4-21} and \eq{eq4-24}.
If we use the expressions of Wronskians in \eq{eq4-20},
the connection condition \eq{eq4-24} yields
\begin{eqnarray}
{\rm tan}\,\delta^N_\ell
=\frac{a^\rho_\ell~\sin\,\left(\overline{\delta}^\rho_\ell
-\delta^\rho_\ell \right)}{ \left[ \frac{1}{a^\rho_\ell}
\cos\,\left(\overline{\delta}^\rho_\ell
-\delta^\rho_\ell \right)
+A^\rho_\ell~\sin\,\left( \overline{\delta}^\rho_\ell
-\delta^\rho_\ell \right)\right]}\ .
\label{eq4-25}
\end{eqnarray}
We write this as
\begin{eqnarray}
& & \sin\,\delta^N_\ell = \frac{a^\rho_\ell}{B^\rho_\ell}
~\sin\,\left(\overline{\delta}^\rho_\ell
- \delta^\rho_\ell \right)\ ,\nonumber \\
& & \cos\,\delta^N_\ell = \frac{1}{B^\rho_\ell}
~\left[ \frac{1}{a^\rho_\ell}~\cos\,\left(
\overline{\delta}^\rho_\ell-\delta^\rho_\ell \right)
+ A^\rho_\ell~\sin\,\left(\overline{\delta}^\rho_\ell 
- \delta^\rho_\ell \right)\right]\ ,\nonumber \\
& & B^\rho_\ell = \left\{
\left[ \frac{1}{a^\rho_\ell}~\cos\,\left(
\overline{\delta}^\rho_\ell - \delta^\rho_\ell \right)
+A^\rho_\ell~\sin\,\left( \overline{\delta}^\rho_\ell 
- \delta^\rho_\ell \right)\right]^2
+\left[ a^\rho_\ell~\sin\,\left(\overline{\delta}^\rho_\ell 
- \delta^\rho_\ell \right)\right]^2 \right\}^{\frac{1}{2}}\ .
\nonumber \\
\label{eq4-26}
\end{eqnarray}
If we use this in \eq{eq4-21} for $r > R_{\rm out}$, the asymptotic 
behavior of $\widetilde{F}^\rho_\ell(k, r)$ and $\widetilde{G}^\rho_\ell(k, r)$
in \eq{eq4-19} yields \eq{eq4-23}
with $B=1/B^\rho_\ell$.
In particular, if $v(r)=0$, $\overline{\delta}^\rho_\ell=\delta^\rho_\ell$
in \eq{eq4-25} yields the correct results $\delta^N_\ell=0$.

In fact, $\delta^N_\ell$ in \eq{eq4-21} is $\rho$-dependent: 
$\delta^N_\ell=\delta^{\rho N}_\ell$, and we need to take
the limit $\delta^N_\ell=\lim_{\rho \rightarrow \infty}
\delta^{\rho N}_\ell$.
Furthermore, the present assumption that $\alph=1$ or 0 except for
the interval $\left[R_{\rm in}, R_{\rm out}\right]
=\left[\rho-b, \rho+b\right]$,
is just an approximation. We have to examine the accuracy of
this approximation for the finite $\rho$ on the case-by-case basis.
In practical calculations, we solve $\widetilde{F}^\rho_\ell(k, r)$
and $\widetilde{G}^\rho_\ell(k, r)$ from  $R_{\rm in}$ to $R_{\rm out}$,
by taking the starting values of the pure Coulomb wave functions
$F_\ell(k, R_{\rm in})$ and $G_\ell(k, R_{\rm in})$.
The Wronskians needed in \eq{eq4-24} are calculated numerically.
In the sharply cut-off Coulomb case with $b=0$ 
and $R_{\rm in}=R_{\rm out}=\rho$, this process is unnecessary,
and reduced to \eq{eq3-40}.

The extra term proportional to $A^\rho_\ell$ in \eq{eq4-17} also
affects the relationship of the Green function in \eq{eq2-22}.
To find a new relationship for the screened Coulomb force,
we solve \eq{eq4-17} inversely and express $F^\rho_\ell(k, r)$
and $G^\rho_\ell(k, r)$ as
\begin{eqnarray}
F^\rho_\ell(k, r) & = & \frac{1}{a^\rho_\ell}\,\widetilde{F}^\rho_\ell(k, r)\ ,
\nonumber \\ [2mm]
G^\rho_\ell(k, r) & = & a^\rho_\ell\,\widetilde{G}^\rho_\ell(k, r)
+A^\rho_\ell\,\widetilde{F}^\rho_\ell(k, r)\ .
\label{eq4-27}
\end{eqnarray}
Then the Green function of the screened Coulomb force in \eq{eq2-21}
is expressed for a fixed $\ell$ as
\begin{eqnarray}
G^\rho_{\omega}=\widetilde{G}^\rho_{\omega}
-\frac{1}{k}\frac{1}{a^\rho_\ell} A^\rho_\ell
~|\widetilde{F}^\rho_\ell \rangle \langle \widetilde{F}^\rho_\ell|\ ,
\label{eq4-28}
\end{eqnarray}
with
\begin{eqnarray}
\widetilde{G}^\rho_{\omega}(r, r^\prime; k)
=-\frac{1}{k} \frac{1}{a^\rho_\ell}
\widetilde{F}^\rho_\ell(k, r_{<}) \left[
a^\rho_\ell\,\widetilde{G}^\rho_\ell(k, r_{>})
+i \frac{1}{a^\rho_\ell}\,\widetilde{F}^\rho_\ell(k, r_{>}) \right]\ .
\label{eq4-29}
\end{eqnarray}
For $r,~r^\prime < R_{\rm in}$, the $\rho \rightarrow \infty$ limit
of \eq{eq4-29} yields
\begin{eqnarray}
\hspace{-10mm} \lim_{\rho \rightarrow \infty} \widetilde{G}^\rho_{\omega}(r, r^\prime; k)
= G^C_\ell(r, r^\prime; k)-\frac{1}{k}A_\ell |F_\ell \rangle \langle F_\ell|
\qquad \hbox{for} \quad r,~r^\prime < R_{\rm in} \rightarrow \infty\ .
\label{eq4-30}
\end{eqnarray}
We keep the finite $\rho$ and write \eq{eq3-21} as
\begin{eqnarray}
|\Psi^{\rho(+)}\rangle 
& = & |\psi^{\rho(+)}\rangle
+ \widetilde{G}^\rho_{\omega}~v~|\Psi^{\rho(+)}\rangle 
- \frac{1}{k}\,\frac{1}{a^\rho_\ell} A^\rho_\ell
~|\widetilde{F}^\rho_\ell \rangle \langle \widetilde{F}^\rho_\ell|
v|\Psi^{\rho(+)}\rangle  \nonumber \\
& = & \frac{1}{k}\,e^{i\delta^\rho_\ell} \frac{1}{a^\rho_\ell}
~|\widetilde{F}^\rho_\ell \rangle
\left[1- e^{-i\delta^\rho_\ell} A^\rho_\ell
\langle \widetilde{F}^\rho_\ell|v|\Psi^{\rho(+)}\rangle \right]
+\widetilde{G}^\rho_{\omega}~v~|\Psi^{\rho(+)}\rangle \ . 
\label{eq4-31}
\end{eqnarray}
Here, we define
\begin{eqnarray}
|\Psi^{\rho(+)}\rangle = |\widetilde{\Psi}^{\rho(+)}\rangle
\left[1- e^{-i\delta^\rho_\ell} A^\rho_\ell
~\langle \widetilde{F}^\rho_\ell|v|\Psi^{\rho(+)}\rangle \right]\ .
\label{eq4-32}
\end{eqnarray}
Then, we find
\begin{eqnarray}
|\widetilde{\Psi}^{\rho(+)}\rangle
= \frac{1}{k}\,e^{i\delta^\rho_\ell}\,\frac{1}{a^\rho_\ell}
~|\widetilde{F}^\rho_\ell \rangle
+\widetilde{G}^\rho_{\omega}\,v\,
|\widetilde{\Psi}^{\rho(+)}\rangle \ .
\label{eq4-33}
\end{eqnarray}
Here, we multiply \eq{eq4-33} by $e^{i\zeta^\rho}$ and take a limit
$\rho \rightarrow \infty$ with $r \in [a, R_{\rm in}]$ fixed.
The first term of the right-hand side of \eq{eq4-33} is
$(1/k) e^{i\sigma_\ell}|F_\ell \rangle=|\psi^{(+)}_\ell \rangle$.
In the second term, we further use the decomposition of the Green function
\begin{eqnarray}
& & \widetilde{G}^\rho_{\omega}=\widetilde{g}^{\rm res}_{\omega_\rho}
-\frac{1}{k}\frac{1}{a^\rho_\ell} 
| \left[ a^\rho_\ell~\widetilde{G}^\rho_\ell+i (1/a^\rho_\ell)
~\widetilde{F}^\rho_\ell\right] \rangle 
\langle \widetilde{F}^\rho_\ell |\, \nonumber \\
& & \widetilde{g}^{\rm res}_{\omega_\rho}(r, r^\prime; k)
=-\frac{1}{k} \left\{\widetilde{F}^\rho_\ell(k, r)
~\widetilde{G}^\rho_\ell(k, r^\prime)
-\widetilde{G}^\rho_\ell(k, r)
~\widetilde{F}^\rho_\ell(k, r^\prime)\right\}
~\theta (r^\prime-r) \ ,
\label{eq4-34}
\end{eqnarray}
and find
\begin{eqnarray}
& & \int^\infty_0 d\,r^\prime~\widetilde{G}^\rho_{\omega}(r, r^\prime; k)
v(r^\prime) \widetilde{\Psi}^{\rho(+)}(r^\prime)~e^{i\zeta^\rho}
= \int^\infty_r d\,r^\prime~\widetilde{g}^{\rm res}_{\omega_\rho}(r, r^\prime; k)
v(r^\prime) \widetilde{\Psi}^{\rho(+)}(r^\prime)~e^{i\zeta^\rho} \nonumber \\
& & -\frac{1}{k}\,\frac{1}{a^\rho_\ell} \left[
a^\rho_\ell~\widetilde{G}^\rho_\ell(k, r)
+i \frac{1}{a^\rho_\ell}\,\widetilde{F}^\rho_\ell(k, r) \right]
\int^\infty_0 d\,r^\prime 
~\widetilde{F}^\rho_\ell(k, r^\prime)\,v(r^\prime)
\,\widetilde{\Psi}^{\rho(+)}(r^\prime)~e^{i\zeta^\rho}\ .
\label{eq4-35}
\end{eqnarray}
Here, the first integral in the right-hand side vanishes
since $v(r^\prime)=0$ for $r^\prime > r > a$.
In the second integral, the range of $v(r)$ makes 
$r^\prime < a$ only, so that we can safely replace
$\widetilde{F}^\rho_\ell(k, r^\prime)$ by $F_\ell(k, r^\prime)$.
Thus, we find
\begin{eqnarray}
& & \widetilde{G}^\rho_{\omega}\,v\,
|\widetilde{\Psi}^{\rho(+)}\rangle~e^{i\zeta^\rho}
\sim -\frac{1}{k} |\left[ G_\ell+iF_\ell \right]\rangle
\langle F_\ell|v|\widetilde{\Psi}^{\rho(+)}\rangle~e^{i\zeta^\rho}
\nonumber \\
& & = -\frac{1}{k} e^{i \sigma_\ell} | f_\ell \rangle
\langle F_\ell|v|\widetilde{\Psi}^{\rho(+)}\rangle~e^{i\zeta^\rho}
\nonumber \\
& & =-|f_\ell \rangle \langle \psi^{(-)}|v|
\widetilde{\Psi}^{\rho(+)}\rangle~e^{i\zeta^\rho}
\qquad \hbox{for} \quad \rho \rightarrow \infty \ ,
\label{eq4-36}
\end{eqnarray}
and
\begin{eqnarray}
\lim_{\rho \rightarrow \infty} |\widetilde{\Psi}^{\rho(+)}\rangle
~e^{i\zeta^\rho} & = & |\psi^{(+)}_\ell\rangle
- |f_\ell \rangle~\lim_{\rho \rightarrow \infty} 
\langle \psi^{(-)}|v|
\widetilde{\Psi}^{\rho(+)} \rangle~e^{i\zeta^\rho}
\nonumber \\
& & \hspace{40mm} \qquad \hbox{for} \quad a < r < R_{\rm in}\ .
\label{eq4-37}
\end{eqnarray}
If we compare \eq{eq4-37} with the asymptotic form in the 
exact Coulomb case in \eq{eq3-29},
we find 
\begin{eqnarray}
\lim_{\rho \rightarrow \infty} |\widetilde{\Psi}^{\rho(+)} \rangle
~e^{i\zeta^\rho}=|\Psi^{(+)} \rangle \ .
\label{eq4-38}
\end{eqnarray}
In the matrix element of \eq{eq4-32}, we can also replace 
$\widetilde{F}^\rho_\ell$ by $F_\ell$, since $v(r)$ is short-ranged.
By solving \eq{eq4-32} inversely, we can show that
\begin{eqnarray}
|\Psi^{\rho(+)}\rangle = |\widetilde{\Psi}^{\rho(+)}\rangle
\left[1+ e^{-i\delta^\rho_\ell} A^\rho_\ell
~\langle F_\ell|v|\widetilde{\Psi}^{\rho(+)}\rangle \right]^{-1}\ .
\label{eq4-39}
\end{eqnarray}
If we multiply \eq{eq4-39} by $e^{i\zeta^\rho}$ and
take the limit $\rho \rightarrow \infty$, $\delta^\rho \rightarrow
\sigma_\ell-\zeta^\rho$ yields
\begin{eqnarray}
\lim_{\rho \rightarrow \infty}
|\Psi^{\rho(+)} \rangle e^{i\zeta^\rho}
= |\Psi^{(+)}\rangle
\left[1+ e^{-i\sigma_\ell} A_\ell
~\langle F_\ell|v|\Psi^{(+)} \rangle \right]^{-1}
\label{eq4-40}
\end{eqnarray}
and
\begin{eqnarray}
|\Psi^{(+)} \rangle
=\lim_{\rho \rightarrow \infty} |\Psi^{\rho(+)} \rangle
~e^{i\zeta^\rho}
~\left[1- e^{-i\sigma_\ell} A^\rho_\ell
~\langle F_\ell|v|\Psi^{\rho(+)}\rangle~e^{i\zeta^\rho}\right]^{-1}\ .
\label{eq4-41}
\end{eqnarray}
This expression implies \eq{eq3-37} is no longer valid for the screened
Coulomb force, and we need an extra normalization factor
$\left[1- e^{-i\sigma_\ell} A^\rho_\ell
~\langle F_\ell|v|\Psi^{\rho(+)}\rangle~e^{i\zeta^\rho}\right]^{-1}$.

Finally, we will show that another type of the connection condition,
equivalent to \eq{eq4-24},
is also obtained by considering two types of asymptotic forms of
$|\widetilde{\Psi}^{\rho(+)}\rangle$.
First, the asymptotic form of $|\widetilde{\Psi}^{\rho(+)} \rangle$
for $R_{\rm in} > r \rightarrow \infty$ is 
from Eqs.\,(\ref{eq4-33}) and (\ref{eq4-35})
\begin{eqnarray}
& & |\widetilde{\Psi}^{\rho(+)} \rangle
\sim  \frac{1}{k}\,e^{i\delta^\rho_\ell}\,\frac{1}{a^\rho_\ell}
~|\widetilde{F}^\rho_\ell \rangle
\nonumber \\
& & -\frac{1}{k} \frac{1}{a^\rho_\ell}~\left[
a^\rho_\ell\,\widetilde{G}^\rho_\ell(k, r)
+i \frac{1}{a^\rho_\ell}\,\widetilde{F}^\rho_\ell(k, r) \right]
\langle F_\ell|v|\widetilde{\Psi}^{\rho(+)} \rangle
\quad \hbox{for} \quad r \leq R_{\rm in}\ .
\label{eq4-42}
\end{eqnarray}
The Wronskians at $r \rightarrow \infty$  with $r \leq R_{\rm in}$ are
given by
\begin{eqnarray}
W\left[\widetilde{F}^\rho_\ell, \widetilde{\Psi}^{\rho(+)} \right]
& = & \langle F_\ell|v|\widetilde{\Psi}^{\rho(+)} \rangle
=e^{i\delta^\rho_\ell}\,\frac{1}{a^\rho_\ell}
~\langle F_\ell|v|\widetilde{\Psi}^{\rho(+)}_\ell \rangle
~a^\rho_\ell~e^{-i\delta^\rho_\ell} \ ,\nonumber \\
W\left[\widetilde{G}^\rho_\ell, \widetilde{\Psi}^{\rho(+)} \right]
& = & e^{i\delta^\rho_\ell}\,\frac{1}{a^\rho_\ell}
-i \frac{1}{(a^\rho_\ell)^2} 
\langle F_\ell|v|\widetilde{\Psi}^{\rho(+)} \rangle
\nonumber \\
& = & e^{i\delta^\rho_\ell}\,\frac{1}{a^\rho_\ell}
\left\{ 1-i \langle F_\ell|v|\widetilde{\Psi}^{\rho(+)} \rangle
\frac{1}{a^\rho_\ell}~e^{-i\delta^\rho_\ell} \right\}\ .
\label{eq4-43}
\end{eqnarray}
Thus, if we define $\widetilde{K}^\rho_\ell$ by
\begin{eqnarray}
\widetilde{K}^\rho_\ell~\frac{1}{k} \langle F_\ell|v|
\widetilde{\Psi}^{\rho(+)} \rangle
~a^\rho_\ell\,e^{-i\delta^\rho_\ell}
=1-i \langle F_\ell|v|
\widetilde{\Psi}^{\rho(+)} \rangle \frac{1}{a^\rho_\ell}
\,e^{-i\delta^\rho_\ell} \ ,
\label{eq4-44}
\end{eqnarray}
we obtain
\begin{eqnarray}
\widetilde{K}^\rho_\ell~W\left[ \widetilde{F}^\rho_\ell, 
\widetilde{\Psi}^{\rho(+)} \right]
=k~W\left[ \widetilde{G}^\rho_\ell, 
\widetilde{\Psi}^{\rho(+)} \right]\ .
\label{eq4-45}
\end{eqnarray}
Here, we note that all the wave functions
with tilde satisfy the Schr{\"o}dinger equation for
the screened Coulomb potential for $r > a$, 
so that we can evaluate Wronskians
at any points $r > a$.
If we take the limit $\rho \rightarrow \infty$ 
in \eq{eq4-44}, \eq{eq4-38} yields
\begin{eqnarray}
& & \lim_{\rho \rightarrow \infty} \frac{1}{k} \langle F_\ell|v|
\widetilde{\Psi}^{\rho(+)} \rangle
a^\rho_\ell\,e^{-i\delta^\rho_\ell}
=\frac{1}{k}\langle F_\ell|v|\Psi^{(+)} \rangle e^{-i\sigma_\ell}
=\frac{1}{k}\langle F_\ell|\widetilde{t}_\ell|
\psi^{(+)}_\ell \rangle e^{-i\sigma_\ell}
\nonumber \\
& & = \frac{1}{k^2}\langle F_\ell|\widetilde{t}_\ell|
F_\ell \rangle = e^{-2i\sigma_\ell}
\langle \psi^{(-)}_\ell|\widetilde{t}_\ell|
\psi^{(+)}_\ell \rangle\ .
\label{eq4-46}
\end{eqnarray}
Thus, if we define $\lim_{\rho \rightarrow \infty} 
\widetilde{K}^\rho_\ell=K^N_\ell$, \eq{eq4-44} becomes
\begin{eqnarray}
K^N_\ell~e^{-2i\sigma_\ell}~\langle \psi^{(-)}_\ell|\widetilde{t}_\ell|
\psi^{(+)}_\ell \rangle
=1-i e^{-2i\sigma_\ell}~k~\langle \psi^{(-)}_\ell|\widetilde{t}_\ell|
\psi^{(+)}_\ell \rangle \ .
\label{eq4-47}
\end{eqnarray}
If we further parametrize 
\begin{eqnarray}
\langle \psi^{(-)}_\ell|\widetilde{t}_\ell|
\psi^{(+)}_\ell \rangle
=-e^{2i\sigma_\ell} \frac{1}{2ik} \left( e^{2i\delta^N_\ell}-1 \right) \ ,
\label{eq4-48}
\end{eqnarray}
we find $K^N_\ell=-k~{\rm cot}\,\delta^N_\ell$.

On the other hand, in the region $r \geq R_{\rm out}$,
the Coulomb-free asymptotic wave gives
\begin{eqnarray}
|\Psi^{\rho(+)} \rangle 
=\frac{1}{k} |u_\ell\rangle - |\omega^{(+)}_\ell \rangle
\langle \phi_\ell|T^\rho_\ell|\phi_\ell \rangle
\qquad \hbox{for} \quad r \geq R_{\rm out}\ ,
\label{eq4-49}
\end{eqnarray}
where the $T$-matrix  $T^\rho_\ell$ is defined
in \eq{eq3-13}. If we write \eq{eq4-49} by
the $K$-matrix defined by
\begin{eqnarray}
K^\rho_\ell~\langle \phi_\ell|T^\rho_\ell|\phi_\ell \rangle
=1-i k~ \langle \phi_\ell|T^\rho_\ell|\phi_\ell \rangle\ ,
\label{eq4-50}
\end{eqnarray}
it is expresses as
\begin{eqnarray}
|\Psi^{\rho(+)} \rangle 
=\left[ |u_\ell\rangle K^\rho_\ell - |v_\ell \rangle k \right]
\frac{1}{k} \langle \phi_\ell|T^\rho_\ell|\phi_\ell \rangle
\qquad \hbox{for} \quad r \geq R_{\rm out}\ .
\label{eq4-51}
\end{eqnarray}
We can use this to calculate the Wronskians in \eq{eq4-45}
at $r=R_{\rm out}$, since the difference between 
$\widetilde{\Psi}^{\rho(+)}_\ell$ and $\Psi^{\rho(+)}_\ell$ is
just a normalization.
From these processes, we eventually obtain
\begin{eqnarray}
\widetilde{K}^\rho_\ell \left\{ W[\widetilde{F}^\rho_\ell, 
u_\ell]_{R_{\rm out}} K^\rho_\ell
- k W[\widetilde{F}^\rho_\ell, 
v_\ell]_{R_{\rm out}} \right\}
= k\left\{ W[ \widetilde{G}^\rho_\ell, 
u_\ell]_{R_{\rm out}} K^\rho_\ell
- k W[ \widetilde{G}^\rho_\ell, 
v_\ell ]_{R_{\rm out}} \right\}, \nonumber \\
\label{eq4-52}
\end{eqnarray}
which is equivalent to \eq{eq4-24} since
$K^\rho_\ell=-k~{\rm cot}\,\overline{\delta}^\rho_\ell$
and $\widetilde{K}^\rho_\ell=-k~{\rm cot}\,\delta^{\rho N}_\ell$.

Summarizing this section, we first calculate  
$\langle \phi_\ell|T^\rho_\ell|\phi_\ell \rangle$ from
the LS equation in the momentum representation,
calculate $K^\rho_\ell$ by \eq{eq4-50}, transform to 
$\widetilde{K}^\rho_\ell$ by \eq{eq4-52}, and
take the limit $\lim_{\rho \rightarrow \infty}
\widetilde{K}^\rho_\ell=K^N_\ell$. Then, the nuclear phase shift 
$\delta^N_\ell$ is obtained from $K^N_\ell=-k~{\rm cot}\,\delta^N_\ell$.

\section{ Application to the $pd$ scattering}

Application of the present formalism to the $pd$ scattering is 
not straightforward because of several reasons.
First, the asymptotic $pd$ Coulomb potential suffers 
the strong distortion effect of the deuteron due to the
long-range nature of the Coulomb force.
In the strict three-body treatment of the $nd$ scattering by
the AGS equation, the distortion effect of the deuteron 
is fully taken into account, but only for the short-range force.
Even if we neglect the Coulomb distortion effect by using
the screened Coulomb force, the quasi-singular
nature of this interaction causes the difficulty that
the treatment by the standard
AGS equation eventually breaks down 
at the limit of $\rho \rightarrow \infty$.
To avoid this, a new formulation by the
Coulomb-modified AGS equation was devised. 
However, very singular behavior of the screened Coulomb
wave functions in the momentum representation 
makes it difficult to solve this equation
numerically. Another difficulty lies in the
partial-wave expansion of the AGS equation.
Even in the two-body Coulomb problem,
the partial-wave expansion of the Coulomb
amplitude does not converge in the usual
sense, but converges only as the distribution.
It is therefore attempted to formulate the AGS equation
based on the three-dimensional description of the
two-body $t$-matrix.\cite{Gl09, Gl10, Sk09, Wi09e, Wi09b}
The isospin symmetry breaking by the $T=3/2$ component
should also be taken into account, since the Coulomb
force admixes the different isospins. 
Here, we extend the ``screening and renormalization technique'' 
to incorporate the present approach
and try to find a practical method to deal with
the $pd$ elastic scattering even in an
approximate way.

Let $\omega^\rho(\br; 1, 2)$ be a screened Coulomb force acting
between two nucleons 1 and 2:
\begin{eqnarray}
\omega^\rho(\br; 1, 2)=\frac{e^2}{r} \alpha_\rho(r)
\frac{1+\tau_z(1)}{2} \frac{1+\tau_z(2)}{2}\ .
\label{eq5-1}
\end{eqnarray}
Here, $\br$ is the relative coordinate between the two nucleons.
We use a set of Jacobi coordinates of particles (1-2)+3 
as the standard one and denote it by $\gamma=3$. 
Another relative coordinate is denoted by $\bR$ in this section.
Then, the screened Coulomb potential \eq{eq5-1} is expressed as
$\omega^\rho_\gamma$ with $\gamma=3$.
In the following, we formulate the Coulomb-modified AGS
equation in the isospin representation.
The three-particle symmetric three-body screened Coulomb potential
$\omega^\rho_C=\sum_\alpha \omega^\rho_\alpha$ is given
in the isospin basis as \cite{Be72}
\begin{eqnarray}
\omega^\rho_C=\omega^\rho_\gamma+\CW^\rho_\gamma
+W^\rho_\gamma \qquad \hbox{for} \quad \hbox{}^\forall \gamma\ .
\label{eq5-2}
\end{eqnarray}
Here, $W^{\rho}_{\gamma}$ denotes
the screened Coulomb potential between the nucleon $ \gamma $ and the residual $NN$ pair,
and is a function of the Jacobi coordinate $\bR_\gamma$ between them.
Furthermore, the three-body potential ${\cal W}^{\rho}_{\gamma}$,
which is usually called the polarization potential, \cite{Be72} is defined by
\begin{eqnarray}
{\cal W}^{\rho}_{\gamma}=\sum _{\beta}\left(\bar{\delta}_{\gamma,\beta}\omega^{\rho}_{\beta}
-\delta _{\gamma,\beta}W^{\rho}_{\beta}\right)
=\sum _{\beta}\bar{\delta}_{\gamma,\beta}\omega^{\rho}_{\beta}-W^{\rho}_{\gamma}\ .
\label{eq5-3}
\end{eqnarray}
It should be noted that, for $ppn$ system, either of $ \omega^{\rho}_{\gamma}$ 
or $\CW^\rho_\gamma+W^{\rho}_{\gamma}$ in \eq{eq5-2} is only non-zero.

The two-potential formula for the three-body system is derived for the solutions
of the Coulomb-modified AGS equation.\cite{De05a}
First, the three-body transition operator $U^{\rho}_{\beta,\alpha}$ for the
usual AGS equation is defined through,
\begin{eqnarray}
\CG^{\rho}=\delta _{\beta, \alpha}g^{\rho}_{\alpha}
+g^{\rho}_{\beta}\,U^{\rho}_{\beta,\alpha}
g^{\rho}_{\alpha}\ ,
\label{eq5-4}
\end{eqnarray}
where the full resolvent ${\cal G}^{\rho}$ and the channel 
resolvent $g^{\rho}_{\alpha}$ are defined by
\begin{eqnarray}
{\cal G}^{\rho}=\left(z-H_{0}-\sum _{\alpha} 
v_{\alpha}-\omega^{\rho}_C \right)^{-1}\ ,\qquad
g^{\rho}_{\alpha}=(z-H_{0}-v_{\alpha}-\omega^{\rho}_{\alpha})^{-1}\ ,
\label{eq5-5}
\end{eqnarray}
with $v_{\alpha}$ being the short-range nuclear potential
and $z=E+\varepsilon_d+i 0$ composed of the
incident energy $E$ and the deuteron energy $\varepsilon_d$.
The three-body kinetic-energy operator is expressed 
as $H_0=h_{0 \gamma}+\overline{h}_{0 \gamma}$ for
an arbitrary set of Jacobi coordinate $\gamma$. 
The transition operator $U^{\rho}_{\beta,\alpha}$ satisfies
the AGS equation 
\begin{eqnarray}
U^\rho_{\beta, \alpha}=\bar{\delta}_{\beta, \alpha} G^{-1}_0
+\sum_\sigma \bar{\delta}_{\beta, \sigma}
\,t^\rho_\sigma\,G_0\,U^\rho_{\sigma, \alpha}\ ,
\label{eq5-6}
\end{eqnarray}
where $G^{-1}_0=(z-H_0)^{-1}$ is the free resolvent and basic two-nucleon
$T$-matrix $t^\rho_\sigma$ is generated by solving the LS equation for 
$v_\sigma+\omega^\rho_\sigma$. Namely,
\begin{eqnarray}
t^{\rho}=(v+\omega^{\rho})+(v+\omega^{\rho})\,G_{0}\,t^{\rho}\ .
\label{eq5-7}
\end{eqnarray}
In \eq{eq5-6} and below, we use the usual
convention $\bar{\delta}_{\beta, \alpha}=1-\delta_{\beta, \alpha}$.
The full resolvent ${\cal G}^{\rho}$ can also be decomposed as
\begin{eqnarray}
{\cal G}^{\rho}=\delta _{\beta, \alpha}G^{\rho}_{\alpha}+G^{\rho}_{\beta}
\,\widetilde{U}^{\rho}_{\beta,\alpha} G^{\rho}_{\alpha}\ ,
\label{eq5-8}
\end{eqnarray}
using another resolvent $G^{\rho}_{\alpha}$ defined by
\begin{eqnarray}
G^{\rho}_{\alpha}=\left(z-H_{0}-v_{\alpha}-\omega^{\rho}_{\alpha}-W^{\rho}_{\alpha}
\right)^{-1}\ ,
\label{eq5-9}
\end{eqnarray}
The operator $\widetilde{U}^{\rho}_{\beta,\alpha}$ satisfies the Coulomb-modified 
AGS equation: \cite{De05a}
\begin{eqnarray}
\widetilde{U}^{\rho}_{\beta,\alpha}=\bar{\delta}_{\beta, \alpha}
\left( (G^\rho_{\alpha})^{-1}+v_{\alpha}\right)
+\delta _{\beta, \alpha}{\cal W}^{\rho}_{\alpha}
+\sum _{\sigma}\left(\bar{\delta}_{\beta,\sigma}v_{\sigma}
+\delta _{\beta,\sigma}{\cal W}^{\rho}_{\sigma}\right)
G^\rho_{\sigma}\widetilde{U}^\rho_{\sigma,\alpha}\ .
\label{eq5-10}
\end{eqnarray}
From the relationship between $g^{\rho}_{\alpha}$ and $G^{\rho}_{\alpha}$,
the operator $ \widetilde{U}^{\rho}_{\beta,\alpha}$ is related 
to $U^{\rho}_{\beta,\alpha}$ through
\begin{eqnarray}
U^{\rho}_{\beta, \alpha}=\delta _{\beta, \alpha}
T^{\rho}_{\alpha}+(1+T^{\rho}_{\beta}g^{\rho}_{\beta})
\widetilde{U}^{\rho}_{\beta,\alpha}(1+g^{\rho}_{\alpha}T^{\rho}_{\alpha})\ ,
\label{eq5-11}
\end{eqnarray}
where the screened Coulomb $T$-matrix $T^{\rho}_{\alpha}$ for 
the $pd$ scattering is obtained from $W^{\rho}_{\alpha}$ through
\begin{eqnarray}
T^{\rho}_{\alpha}=W^{\rho}_{\alpha}+W^{\rho}_{\alpha}
\,g^{\rho}_{\alpha}\,T^{\rho}_{\alpha}
=W^{\rho}_{\alpha}+W^{\rho}_{\alpha}\,G^{\rho}_{\alpha}\,W^{\rho}_{\alpha}\ .
\label{eq5-12}
\end{eqnarray}
Equation (\ref{eq5-11}) is the two-potential formula 
for the three-body system.
The Coulomb-distorted asymptotic wave function 
is defined by $|\psi^{\rho(+)}_{\alpha}\rangle
=(1+g^{\rho}_{\alpha}T^{\rho}_{\alpha})|\phi_{\alpha}\rangle$
from the channel wave function $|\phi_{\alpha}\rangle
=|\bq_{0 \alpha},\psi^{d}_{\alpha} \rangle $.
From this definition and \eq{eq5-12}, we obtain
\begin{eqnarray}
|\psi^{\rho(+)}_{\alpha}\rangle 
= |\phi _{\alpha}\rangle+g^{\rho}_{\alpha}W^{\rho}_{\alpha}
|\psi^{\rho(+)}_{\alpha} \rangle \ .
\label{eq5-13}
\end{eqnarray}
We define $|\psi^{\rho(-)}_{\alpha}\rangle $ as 
the complex conjugate of $|\psi^{\rho(+)}_{\alpha}\rangle$
and find
\begin{eqnarray}
\langle \phi _{\beta}|U^{\rho}_{\beta,\alpha}|\phi _{\alpha} \rangle=
\delta _{\beta,\alpha}\langle \phi _{\alpha}|T^{\rho}_{\alpha}
|\phi _{\alpha}\rangle+\langle \psi^{\rho(-)}_{\beta}
|\widetilde{U}^{\rho}_{\beta, \alpha}|\psi^{\rho(+)}_{\alpha}\rangle \ .
\label{eq5-14}
\end{eqnarray}
%

We can separate the deuteron part in \eq{eq5-13} and we obtain
\begin{eqnarray}
|\psi^{\rho(+)}_{\alpha}\rangle 
& = & |\chi^{\rho(+)}_{\alpha},\psi^{d}_{\alpha}\rangle \ ,\nonumber \\
|\chi^{\rho(+)}_{\alpha}\rangle & = & |\bq_{0 \alpha}\rangle
+(E_{\alpha}+i0-\bar{h}_{0 \alpha})^{-1}W^{\rho}_{\alpha}
|\chi^{\rho(+)}_{\alpha}\rangle \ ,
\label{eq5-15}
\end{eqnarray}
where $E_{\alpha}$ is the incident energy in the $ \alpha $-channel 
and the deuteron wave function $|\psi^{d}_{\alpha}\rangle $ satisfies 
\begin{eqnarray}
(\varepsilon _{d}-h_{0 \alpha}-v_{\alpha}-\omega^{\rho}_{\alpha})
|\psi^{d}_{\alpha} \rangle = 0 \ .
\label{eq5-16}
\end{eqnarray}
Note that $ \omega^{\rho}$ does not actually contribute in \eq{eq5-16},
since the isospin of the deuteron is zero.
From Eq.~(\ref{eq5-15}), we find
\begin{eqnarray}
(G^{\rho}_{\alpha})^{-1}|\psi^{\rho(+)}_{\alpha}\rangle
=(E_{\alpha}-\bar{h}_{0 \alpha}-W^{\rho}_{\alpha})|
\chi^{\rho(+)}_{\alpha},\psi^{d}_{\alpha}\rangle=0 \ .
\label{eq5-17}
\end{eqnarray}
For three identical particles in the isospin formalism,
a transition operator to the channel $ \gamma $, $ \widetilde{U}^{\rho}_{\gamma}$,
is defined through
\begin{eqnarray}
\sum _{\alpha}\widetilde{U}^{\rho}_{\gamma, \alpha}|\psi^{\rho(+)}_{\alpha}\rangle
\equiv \widetilde{U}^{\rho}_{\gamma}
|\psi^{\rho(+)}_{\gamma}\rangle \ .
\label{eq5-18}
\end{eqnarray}
We assume $ \gamma $ to be the standard coordinate system $\gamma$ = 3
and abbreviate the subscript $\gamma$. Then, we obtain from 
Eqs.~(\ref{eq5-10}) and (\ref{eq5-17})
the Coulomb-modified AGS equation for three identical particles:
\begin{eqnarray}
\widetilde{U}^{\rho}|\psi^{\rho(+)}\rangle=(Pv+{\cal W}^{\rho})|\psi^{\rho(+)}\rangle
+(Pv+{\cal W}^{\rho})G^{\rho}\widetilde{U}^{\rho}|\psi^{\rho(+)}\rangle \ ,
\label{eq5-19}
\end{eqnarray}
where $G^{\rho}=(z-H_{0}-v-\omega^{\rho}-W^{\rho})^{-1}$
and $P=P_{(12)}P_{(23)}+P_{(13)}P_{(23)}$ is the permutation
operator for the rearrangement.
In \eq{eq5-19}, we set $ \widetilde{U}^{\rho}|\psi^{\rho(+)}\rangle=(Pv+{\cal W}^{\rho})
|\Psi^{\rho(+)}\rangle $ and obtain
\begin{eqnarray}
|\Psi^{\rho(+)}\rangle=|\psi^{\rho(+)}\rangle+G^{\rho}(Pv+{\cal W}^{\rho})
|\Psi^{\rho(+)}\rangle\ .
\label{eq5-20}
\end{eqnarray}
Here, $|\Psi^{\rho(+)}\rangle $ is the total wave function for the screened 
Coulomb problem and is related to the total wave function for the full 
Coulomb problem $|\Psi^{(+)}\rangle $ through
\footnote{Strictly speaking, this relationship is valid only for 
the sharply cutoff Coulomb potential.
For general screened Coulomb potentials, 
an extra finite normalization factor like in \eq{eq4-32} is necessary
for $|{\Psi}^{\rho(+)}\rangle$.
The following relations are all valid by modifying 
$|{\Psi}^{\rho(+)}\rangle$ to $|\widetilde{\Psi}^{\rho(+)}\rangle$.}
\begin{eqnarray}
\lim _{\rho \rightarrow \infty}|{\Psi}^{\rho(+)}\rangle\,e^{i \zeta^{\rho}}
=|\Psi^{(+)}\rangle
\label{eq5-21}
\end{eqnarray}
with a shift function $ \zeta^{\rho}$. 
The shift function $ \zeta^{\rho}=\zeta^{\rho}(k)$ is defined by
\begin{eqnarray}
\zeta^{\rho}(q_{0})=\frac{1}{2q_{0}}\int^{\infty}_{\frac{1}{2q_{0}}}
W^{\rho}(R)~d R \ ,
\label{eq5-22}
\end{eqnarray}
where $q_{0}$ is the wave number between the incident proton and the deuteron
in the center-of-mass (cm) system.
The ``screening and renormalization procedure'' \cite{De05a} converts 
\eq{eq5-14} to its full Coulomb correspondence
\begin{eqnarray}
\langle \phi|U^{C}|\phi \rangle=\langle \phi|T^{C}|\phi \rangle+
\langle \psi^{C(-)}|\widetilde{U}^{C}|\psi^{C(+)}\rangle \ .
\label{eq5-23}
\end{eqnarray}

Equation (\ref{eq5-20}) is the distorted-wave version of
\begin{eqnarray}
|\Psi^{\rho(+)}\rangle=|\phi \rangle+g^{\rho}P(v+\omega^{\rho})
|\Psi^{\rho(+)}\rangle \ ,
\label{eq5-24}
\end{eqnarray}
which can be derived similarly from the AGS equation in \eq{eq5-6}
by assigning $U^\rho |\phi \rangle
=P(v+\omega^\rho)|\Psi^{\rho(+)}\rangle$.
In fact, if we note that $|\Psi^{\rho(+)}\rangle$ is
three-nucleon antisymmetric,
we can easily derive \eq{eq5-19} from \eq{eq5-24} by using
$P(v+\omega^{\rho})=Pv+{\cal W}^{\rho}+W^{\rho}$.
On the other hand, the Faddeev components $|\Psi^{\rho} \rangle$,
satisfying $|\Psi^{\rho(+)}\rangle=(1+P)|\Psi^{\rho} \rangle$,
can be derived by setting $G_{0}U^\rho|\phi \rangle
=P|\Psi^{\rho} \rangle $ in the AGS equation:
\begin{eqnarray}
|\Psi^{\rho} \rangle=|\phi \rangle +G_{0}t^{\rho}P|\Psi^{\rho}\rangle \ .
\label{eq5-25}
\end{eqnarray}
In the isospin formalism for the total isospin $T=1/2$ state,
we use the effective $T$-matrix $t^{I=1}=(2/3)t^{\rho}_{pp}+(1/3)t_{np}$ for 
the isospin 1 $NN$ channel. \cite{Wi91b}

Instead of using the ``screening and renormalization'' procedure,
we use an extension of Vincent and Phatak procedure~\cite{Vi74} 
of the two-cluster Coulomb problem,
which is equivalent to the ``screening and renormalization procedure'' 
in the limit of $\rho \rightarrow \infty$.
The scattering amplitude is obtained by imposing a connection condition 
on the $K$-matrix\,\footnote{Here, the $K$-matrix is defined by the form
of $K_\ell(k)=(1/k)\,\hbox{tan}\,\delta_{\ell}(k)$ for the
on-shell matrix elements.}
$K^{\rho}_{\alpha,\beta}\equiv(Z^{-1})_{\alpha,\beta}
-\langle \phi _{\alpha}|X^{\rho}|\phi _{\beta}\rangle$
for the $pd$ scattering,\cite{ndscat1} which is derived from
the two different asymptotic forms of the total wave function
in Eqs.~(\ref{eq5-20}) and (\ref{eq5-24}).
From here on, the subscripts $\alpha$, $\beta$, etc. specify 
the channel quantum numbers.
We define a reduced wave function $\Phi^{\rho(+)}_{\alpha,\gamma}(R)
\equiv \langle R, \psi^{d}_{\alpha}|\Psi^{\rho(+)}_{\gamma}\rangle $.
The asymptotic form for the wave function Eq~(\ref{eq5-24}) is
without a constant normalization factor
\begin{eqnarray}
\Phi^{\rho(+)}_{\alpha,\gamma}(R)\sim u_{\alpha}(q_{0}R)
~K^{\rho}_{\alpha,\gamma}
-c\,v_{\alpha}(q_{0}R)~\delta _{\alpha,\gamma} 
\qquad \hbox{for} \qquad R > R_{\rm out}\ ,
\label{eq5-26}
\end{eqnarray}
where $c=q_0 (\pi/2) (4 M_N/3\hbar^2)$ with $M_N$ being the nucleon mass.
For the total wave function \eq{eq5-20}, the asymptotic form is
\begin{eqnarray}
\Phi^{\rho(+)}_{\alpha,\gamma}(R) &\sim& \frac{1}{q_{0}}\sum _{\beta}\left\{ 
\widetilde{F}^{\rho}_{\alpha}(q_{0},R)\,\widetilde{K}^{\rho}_{\alpha, \beta}
-c\,\widetilde{G}^{\rho}_{\alpha}(q_{0},R)
\,\delta _{\alpha,\beta}\right\} \nonumber \\
&& \times \frac{1}{q_{0}}\langle F_{\beta},\psi^{d}_{\beta}|
(Pv+{\cal W}^{\rho})|\Psi^{\rho(+)}_{\gamma}\rangle
\qquad \hbox{for} \qquad R > a \ ,
\label{eq5-27}
\end{eqnarray}
where $a$ is the range of the nuclear force.
Here, $ \widetilde{F}^{\rho}_{\alpha}$ and 
$ \widetilde{G}^{\rho}_{\alpha}$ are
the screened Coulomb wave functions defined in \eq{eq4-17}.
In the inside region $R<R_{\rm in}$, $ \widetilde{F}_{\alpha}$ 
and $ \widetilde{G}_{\alpha}$ are equal 
to $F_{\alpha}$ and $G_{\alpha}$, respectively.
The connection condition for $\Phi^{\rho(+)}_{\alpha,\gamma}(R)$ 
at $R=R_{\rm out}$ is written in terms of Wronskians:
\begin{eqnarray}
& & \sum _{\beta}\widetilde{K}^{\rho}_{\alpha, \beta}
\left\{W[\widetilde{F}^{\rho}_{\beta},u_{\beta}]_{R_{\rm out}}
\,K^{\rho}_{\beta,\gamma}-W[\widetilde{F}^{\rho}_{\beta},
v_{\beta}]_{R_{\rm out}}
\,c\,\delta _{\beta, \gamma}\right \} \nonumber \\
& & = c \left\{W[\widetilde{G}^{\rho}_{\alpha},u_{\alpha}]_{R_{\rm out}}
\,K^{\rho}_{\alpha,\gamma}
-W[\widetilde{G}^{\rho}_{\alpha},v_{\alpha}]_{R_{\rm out}}
\,c\,\delta _{\alpha,\gamma}\right\}\ .
\label{eq5-28}
\end{eqnarray}
Matrix elements $ \widetilde{U}^{\rho}_{\beta,\gamma}$, defined by
\begin{eqnarray}
\sum _{\beta}\left[\widetilde{K}^{\rho}_{\alpha,\beta}
+i\,c\,\delta _{\alpha,\beta}\right]
\widetilde{U}^{\rho}_{\beta, \gamma}=\delta _{\alpha, \gamma}
\label{eq5-29}
\end{eqnarray}
in the limit of $\rho \rightarrow \infty$, are related 
to $ \langle \psi^{\rho(-)}_{\beta}|\widetilde{U}^{\rho}
|\psi^{\rho(+)}_{\gamma}\rangle $ through
\begin{eqnarray}
\langle \psi^{\rho(-)}_{\beta}|\widetilde{U}^{\rho}|\psi^{\rho(+)}_{\gamma}\rangle
=e^{i(\sigma _{\beta}+\sigma _{\gamma})}\,\widetilde{U}^{\rho}_{\beta, \gamma}\ .
\label{eq5-30}
\end{eqnarray}
Here, $ \sigma _{\beta}$ and $ \sigma _{\gamma}$ are the Coulomb 
phase shifts in the channels $ \beta $ and $ \gamma $, respectively.
The scattering amplitude $f^{N,\rho}_{\beta,\gamma}$ is obtained 
from $ \widetilde{U}^{\rho}_{\beta, \gamma}$ through
\begin{eqnarray}
f^{N,\rho}_{\beta, \gamma}=-\frac{\pi}{2}\frac{4M_{N}}{3 \hbar^{2}}
\widetilde{U}^{\rho}_{\beta, \gamma} \ .
\label{eq5-31}
\end{eqnarray}
In the channel-spin representation, the full scattering amplitude is written as
\begin{eqnarray}
& & f^{\rho}_{S'_{c}S'_{cz},S_{c}S_{cz}}(\widehat \bq_{f},\widehat{\bq}_{i})
=\delta _{S'_{c},S_{c}}
\delta _{S'_{cz},S_{cz}}f^{C}(\theta)+4 \pi \sum _{\ell'\ell JJ_{z}}
e^{i(\sigma _{\ell'}+\sigma _{\ell})}
~f^{NJ,\rho}_{(\ell'S'_{c}),(\ell S_{c})}\nonumber \\
& & \times \sum _{m'}\langle \ell'm'S'_{c}S'_{cz}|JJ_{z}\rangle
\,Y_{\ell'm'}(\widehat{\bq}_{f})
\sum _{m}\langle \ell mS_{c}S_{cz}|JJ_{z}\rangle\, 
Y^{*}_{\ell m}(\widehat{\bq}_{i})\ ,
\label{eq5-32}
\end{eqnarray}
for a sufficiently large $ \rho $.

\section{Numerical performance}

\subsection{Comparison with the exact solutions for the
Ali-Bodmer $\alpha \alpha$ potential}

Ali-Bodmer $\alpha \alpha$ potential is a simple phenomenological potential 
which reproduces the results of the phase-shift analysis for
the $\alpha \alpha$ scattering up to $E_{\rm cm} \sim 15$ MeV. 
The angular-momentum-dependent version called 
Ali-Bodmer d (ABd) has the explicit form
\begin{eqnarray}
V^{\rm ABd}_{\alpha \alpha}(r)
=V_1~e^{-\eta_1 r^2}+V_2~e^{-\eta_2 r^2}+\frac{4e^2}{r} {\rm erf}\,(\beta r)\ ,
\label{eq6-1}
\end{eqnarray}
with the parameters $\eta_1=0.7^2~\hbox{fm}^{-2}$,
$\eta_2=0.475^2~\hbox{fm}^{-2}$, $V_2=-130~\hbox{MeV}$ and
\begin{eqnarray}
& & V_1=\left\{
\begin{array}{ll}
500~\hbox{MeV} \qquad \hbox{for} \qquad S \\ [2mm]
320~\hbox{MeV} \qquad \hbox{for} \qquad D \\ [2mm]
0 \qquad \hbox{for} \qquad \ell \geq 4 \\ [2mm]
\end{array}
\right. \ ,
\nonumber \\
& & \beta=\frac{\sqrt{3}}{2 \times 1.44}=0.6014\cdots~\hbox{fm}^{-1}\ .
\label{eq6-2}
\end{eqnarray}
In \eq{eq6-1}, erf($x$) stands for the error function defined by
${\rm erf} (x)=(2/\sqrt{\pi})\int^x_0 e^{-t^2}~d\,t$.
Since this potential model is exactly solvable by the 
Runge-Kutta-Gill (RKG) method, it is suitable
to test the accuracy of the Coulomb approach developed 
in this paper.
With the assignment $\alpha=4e^2$, the error function-type 
Coulomb force
\begin{eqnarray}
V_{\rm D}(r) = \frac{\alpha}{r} {\rm erf}\,(\beta r)
\label{eq6-3}
\end{eqnarray}
in \eq{eq6-1} is the direct potential of the $\alpha$-$\alpha$ RGM.
When a simple $(0s)^4$ harmonic-oscillator shell-model wave function
with the width parameter $\nu$ is assumed for the $\alpha$-cluster,
the parameter $\beta$ is expressed as
\begin{eqnarray}
\beta=\sqrt{\frac{\nu}{\left(1-\frac{1}{2\mu}\right)}}=2\sqrt{\nu/3}\ ,
\label{eq6-4}
\end{eqnarray}
where $\mu=4\cdot 4/(4+4)=2$ is the reduced mass number of the $\alpha \alpha$
system. On the other hand, the rms radius of the $\alpha$-cluster with
$A=4$ is given by
\begin{eqnarray}
r_\alpha=\sqrt{\langle r^2\rangle_\alpha}=\sqrt{\frac{3}{4}
\left(1-\frac{1}{A}\right)\frac{1}{\nu}}=\frac{3}{4}
\frac{1}{\sqrt{\nu}}\ ,
\label{eq6-5}
\end{eqnarray}
without the proton size effect, so that $\beta$ is related to $r_\alpha$
through
\begin{eqnarray}
\beta=\frac{\sqrt{3}}{2 \cdot r_\alpha}\ .
\label{eq6-6}
\end{eqnarray}
In ABd, $r_\alpha=1.44$ fm is assumed, corresponding to 
$\nu=0.271~\hbox{fm}^{-2}$.

In the momentum representation, we use the sharply cut-off Coulomb
force at the nucleon level. The corresponding direct $\alpha \alpha$ 
potential is given by
\begin{eqnarray}
V^\rho_{\rm D}(r) = \frac{\alpha}{r} \left\{
{\rm erf}\,(\beta r) - \frac{1}{2}\left[ {\rm erf}\,(\beta (r+\rho))
+{\rm erf}\,(\beta (r-\rho))\right]\right\}\ .
\label{eq6-7}
\end{eqnarray}
If we use this screened Coulomb potential in \eq{eq6-1}, we find
\begin{eqnarray}
V^{\rho}_{\alpha \alpha}(r)
=V_1~e^{-\eta_1 r^2}+V_2~e^{-\eta_2 r^2}+V^\rho_{\rm D}(r)\ .
\label{eq6-8}
\end{eqnarray}
Here, we separate $V^\rho_{\rm D}(r)$ into
\begin{eqnarray}
V^\rho_{\rm D}(r) & = & \frac{\alpha}{r}
\left\{ \left[{\rm erf}\,(\beta r)-1\right]
+1-\frac{1}{2}\left[ {\rm erf}\,(\beta (r+\rho))
+{\rm erf}\,(\beta (r-\rho))\right]\right\}
\nonumber \\
& = & -\frac{\alpha}{r}\left[1-{\rm erf}\,(\beta r)\right]
+\frac{\alpha}{r} \alpha_\rho(r)\ ,
\label{eq6-9}
\end{eqnarray}
and set
\begin{eqnarray}
\alpha_\rho(r) = 1 - \frac{1}{2}\left[ {\rm erf}\,(\beta (r+\rho))
+{\rm erf}\,(\beta (r-\rho))\right]\ .
\label{eq6-10}
\end{eqnarray}
Then, the $\alpha \alpha$ potential which should be used in the
momentum representation becomes
\begin{eqnarray}
& & V^{\rho}_{\alpha \alpha}(r)
=V(r)+\frac{\alpha}{r}\alpha_\rho(r)
\nonumber \\
& & \hbox{with} \qquad
V(r)=V_1~e^{-\eta_1 r^2}+V_2~e^{-\eta_2 r^2}+\CW(r)\ .
\label{eq6-11}
\end{eqnarray}
Here, $\CW(r)=-(\alpha/r) \left[1-{\rm erf}\,(\beta r)\right]$ 
is the short-range attraction originating from the Coulomb potential.
In fact, the asymptotic expansion of the error function yields
%
%
%
\begin{eqnarray}
\CW(r)=\frac{\alpha}{r}\left[{\rm erf}\,(\beta r)-1\right]
\sim -\frac{\alpha}{r} e^{-(\beta r)^2} \sum^\infty_{n=0}
(-)^n \frac{(2n-1)!!}{2^{n+1}}\left( \frac{1}{\beta r}\right)^{2n+1}\ .
\label{eq6-12}
\end{eqnarray}
We find that $\CW(r)$ is sufficiently small around $(\beta r)^2 \sim 16$; namely,
$r \sim 4/\beta \sim 7$ fm. (Actually, even around $\sim 4$ fm, 
as seen in Fig.\,\ref{fig2} below.)

%
\begin{figure}[t]
\begin{minipage}{0.49\textwidth}
\centerline{\includegraphics[width=\columnwidth]{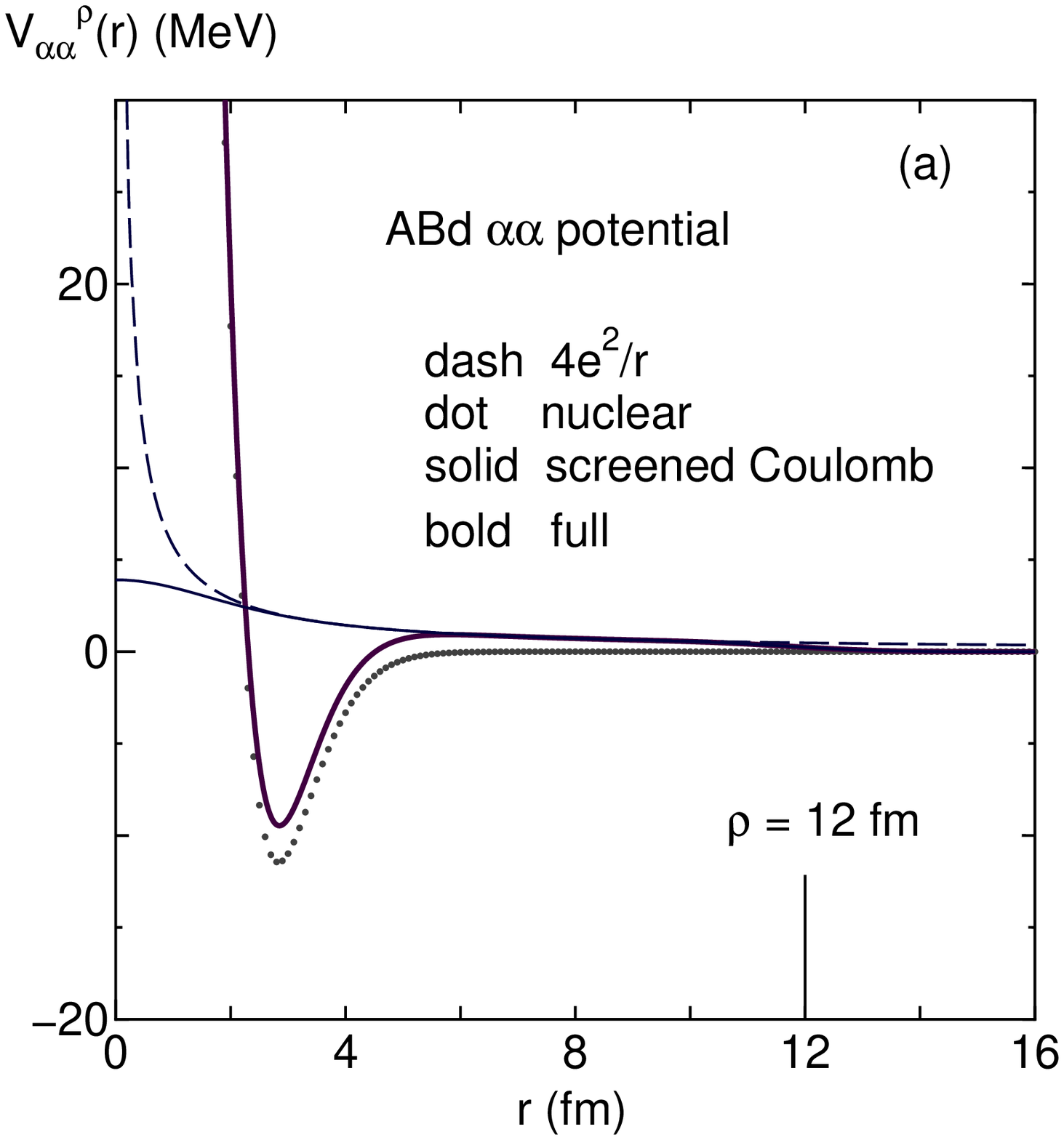}}
\end{minipage}~%
\hfill~%
\begin{minipage}{0.49\textwidth}
\centerline{\includegraphics[width=\columnwidth]{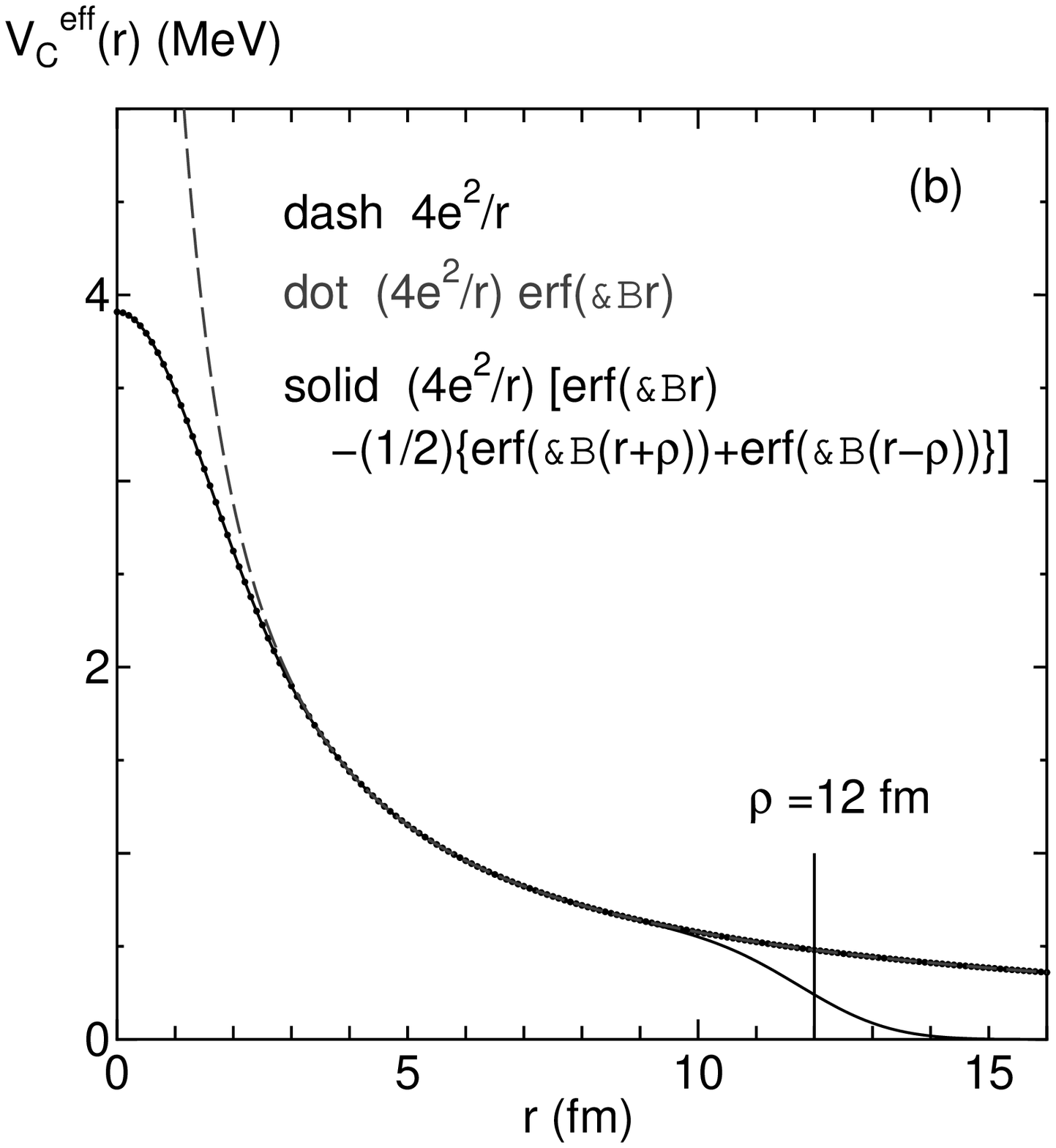}}
\end{minipage}
\caption{
(a): $S$-wave Ali-Bodmer potential ABd
with the screened Coulomb force (solid curve) 
for the $\alpha \alpha$ system.
The dashed curve denotes the simple
Coulomb potential $V_C(r)=4e^2/r$ and the dotted curve
the nuclear part. The full potential is shown 
in the bold solid curve.
The cut-off radius of the sharply cut-off Coulomb force
at the nucleon level is assumed to be $\rho=12$ fm.
(b): The enlarged profiles of (a) for the various Coulomb
potentials. The solid curve denotes the screened Coulomb
direct potential \protect\eq{eq6-7} with the error function form.
\label{fig1}}
\end{figure}

We illustrate in Fig.\,\ref{fig1}(a) the $S$-wave Ali-Bodmer potential
ABd and in (b) the enlarged profiles of various types of
Coulomb potentials. The cut-off function $\alpha_\rho(r)$ in \eq{eq6-10}
for the cut-off Coulomb radius $\rho=12$ fm
and the short-range Coulomb potential $\CW(r)$ in \eq{eq6-12} are
shown in Fig.\,\ref{fig2}. We find that $\alpha_\rho(r)$ satisfies
the conditions 1) - 3) of the screened Coulomb potential.
In particular, the much stringent condition $3)^\prime$ in \eq{eq4-5}
is also satisfied with the smoothness parameter $b \sim 3~\hbox{fm}$.  
If we take $b=6$ fm, the deviation of $\alpha_\rho(r)$ 
from 1 (or 0) at $R_{\rm in}=\rho-b=6$ fm
(or at $R_{\rm out}=\rho+b=18$ fm) is less than $10^{-6}$.
Note that this kind of a rapid transition from 1 to 0 is not
achieved in the standard screening functions in the form of
$\alpha_\rho(r)=e^{-(r/\rho)^n}$, unless $n$ is taken to be
very large like $n \geq 20$. In this sense, our screened Coulomb
potential is a small deviation from the sharply cut-off Coulomb
potential, which is probably related to the smallness of
the limit $\lim_{\rho \rightarrow \infty} A^\rho_\ell \sim 0$ if it exists.
This property must also be related to the small deviation of the 
shift function $\zeta^\rho(k)$ in \eq{eq4-13} from 
$\eta {\rm log}\,(2k\rho)$, which is the result of the sharply cut-off Coulomb 
potential in \eq{eq2-28}.
We will show in Appendix B that the screening function
$\alpha_\rho(r)$ in \eq{eq6-10} satisfies the limit
\begin{eqnarray}
\zeta^\rho(k) \rightarrow \eta~{\rm log}\,(2k\rho)
\qquad \hbox{as} \quad \rho \rightarrow \infty \ ,
\label{eq6-13}
\end{eqnarray}
in contrast to the $\alpha_\rho(r)=e^{-(r/\rho)^n}$ case.
In the latter case, the right-hand side of \eq{eq6-13} contains 
an extra constant term $-(\eta/n)\gamma$ with $\gamma$ being
the Euler constant. (See \eq{b3}.)

\begin{figure}[htb]
\centerline{\includegraphics[width=0.45\columnwidth]{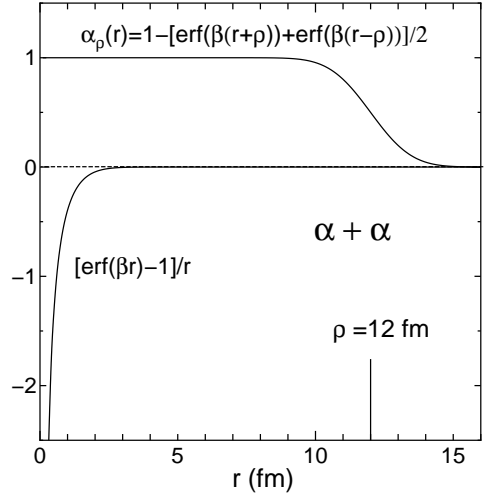}}
\caption{
The cut-off function $\alpha_\rho(r)$ in \eq{eq6-10}
with the cut-off Coulomb radius $\rho=12$ fm
and the short-range Coulomb potential $\CW(r)/\alpha
=\left[\hbox{erf}(\beta r)-1\right]/r$ in \eq{eq6-12} 
for the $\alpha \alpha$ screened Coulomb potential
(in the unit of $\hbox{fm}^{-1}$).
\label{fig2}}
\end{figure}

First, we have neglected the nuclear potential $V_1=0$ and $V_2=0$
in \eq{eq6-1} and compared the nuclear phase shifts between the
present method and the direct method using \eq{eq1-1}. 
In the direct method, the relative wave function $\psi_\ell(r)$ 
in \eq{eq1-1} is solved from $r=0$ to $R_{\rm out}=12+6=18$ fm
by the RKG method and smoothly connected to 
a linear combination of the pure Coulomb wave functions
at $r=R_{\rm out}$. Since we are using the error function Coulomb,
the nuclear phase shift does not become zero, In the $S$-wave,
$\delta^N_0$ increases from 0 to $11.088^\circ$, when the energy
increases up to $E_{\rm cm}=15$ MeV. Similarly, $\delta^N_2 = 0.473^\circ$
and $\delta^N_4 = 0.013^\circ$ at $E_{\rm cm}=15$ MeV.
In the momentum-space approach, we first solve the LS equation 
and calculate $\delta^\rho_\ell$ (which is the screened
Coulomb phase shift) by assuming $\rho=12$ fm.
The phase shift is then transformed to $\delta^N_\ell$ through
the connection condition \eq{eq4-24}. Here, we assumed $b=6$ fm,
and $\widetilde{F}^\rho_\ell(k, r)$, $\widetilde{G}^\rho_\ell(k, r)$
are calculated from $R_{\rm in}=12-6=6$ fm to $R_{\rm out}=12+6=18$ fm,
also by the RKG method, with the pure Coulomb values 
at $R_{\rm in}=6$ fm as the starting values.
The results by these two different methods, of course, agree to each other
completely within the numerical accuracy less than $0.001^\circ$.
Next, we switch on $V_1$ and $V_2$ and repeated the same calculations.
The result is shown in Table \ref{table1}.
For each incident energy, the first row indicates solutions 
obtained by the RKG method, and the second row those in the
momentum-space approach. Only different figures from the first row
are shown. In the left-hand side, the final results of $\delta^N_\ell$
are compared. In the right-hand side, 
the phase shifts $\overline{\delta}^\rho_\ell$ directly obtained from
the LS equation (before the transformation) are also
compared. We find that, in the lowest energy $E_{\rm cm}=1$ MeV,
a difference of $0.005^\circ$ exists both in $\delta^N_\ell$
and $\overline{\delta}^\rho_\ell$. This is probably the inaccuracy
of solving the LS equation in the low energies.
For other energies, the difference is less than $0.001^\circ$,
and the agreement of the results obtained by our method 
with the exact solutions is quite satisfactory.

\begin{table}[htb]
\caption{Comparison of the $\alpha \alpha$ nuclear
phase shifts ($\delta^N_\ell$) of the ABd potential
with the direct method.
For each cm energy $E_{\rm cm}$, the first row
indicates solutions by the RKG method,
connected at $R_{\rm out}=18$ fm by \eq{eq1-1}.
The second row stands for the solutions by the present
momentum-space approach.
Only different figures from the first row
are shown. In the left-hand side, the final results of $\delta^N_\ell$
are compared. In the right-hand side, 
the phase shifts $\overline{\delta}^\rho_\ell$ directly obtained from
the LS equation (before the transformation) are also
compared. 
The cut-off Coulomb radius $\rho$ is
chosen to be $\rho=12$ fm and
the smoothness parameter in \eq{eq4-5} is $b=6$ fm.
$(\hbar^2/M_N)=41.786~\hbox{MeV}\cdot \hbox{fm}^2$
and $e^2=1.44~\hbox{MeV}\cdot \hbox{fm}$ are used.
}
\label{table1}
\begin{center}
\renewcommand{\arraystretch}{1.1}
\setlength{\tabcolsep}{4mm}
\begin{tabular}{r|rrr|rrr}
\hline
$E_{\rm cm}$ & \multicolumn{3}{c|}{$\delta^N_\ell$}
& \multicolumn{3}{c}{$\overline{\delta}^\rho_\ell$} \\
(MeV) & $S$ & $D$ & $G$ & $S$ & $D$ & $G$ \\
\hline
 1 & 147.021 &   0.485 &  0.000 &  40.403 & 152.767 &  178.017 \\
   &      16 &         &        &     398 &         &          \\ 
 2 & 110.751 &   9.736 &  0.008 &  10.133 & 141.754 &  168.315 \\
   &       2 &       5 &        &       4 &         &          \\
 3 &  85.082 &  66.577 &  0.065 & 176.887 &  22.894 &  158.138 \\
   &         &       6 &        &         &       3 &          \\
 4 &  65.251 & 109.426 &  0.261 & 165.105 &  65.200 &  153.102 \\ 
   &         &         &        &         &         &          \\
 5 &  49.031 & 115.142 &  0.748 & 153.868 &  73.736 &  153.013 \\
   &         &         &        &         &         &          \\
 6 &  35.271 & 113.611 &  1.756 & 144.160 &  73.663 &  154.882 \\
   &         &         &        &         &         &          \\
 7 &  23.299 & 110.142 &  3.661 & 135.833 &  71.241 &  157.196 \\
   &         &         &        &         &         &          \\
 8 &  12.689 & 106.077 &  7.140 & 128.379 &  68.336 &  160.738 \\
   &         &         &        &         &         &          \\
 9 &   3.153 & 101.875 & 13.575 & 121.482 &  65.314 &  167.386 \\
   &         &         &        &         &         &          \\
10 &  $-5.513$ & 97.719 &  26.095 & 115.039 & 62.204 &   0.375 \\
   &           &        &         &         &        &         \\
11 & $-13.462$ & 93.685 &  50.814 & 109.027 & 59.060 &  25.569 \\
   &           &        &         &         &        &         \\
12 & $-20.807$ & 89.805 &  86.674 & 103.419 & 55.958 &  61.717 \\
   &           &        &         &         &        &         \\
13 & $-27.637$ & 86.086 & 113.376 &  98.168 & 52.954 &  88.789 \\
   &           &        &         &         &        &       8 \\
14 & $-34.024$ & 82.526 & 127.388 &  93.223 & 50.067 & 103.178 \\
   &           &        &         &         &        &         \\
15 & $-40.025$ & 79.119 & 134.966 &  88.538 & 47.293 & 111.085 \\
   &           &        &         &         &      2 &         \\
\hline
\end{tabular}
\end{center}
\end{table}

\subsection{$\alpha \alpha$ Lippmann-Schwinger RGM by the Minnesota 
three-range force}

\begin{figure}[t]
\begin{minipage}{0.45\textwidth}
\centerline{\includegraphics[width=\columnwidth]{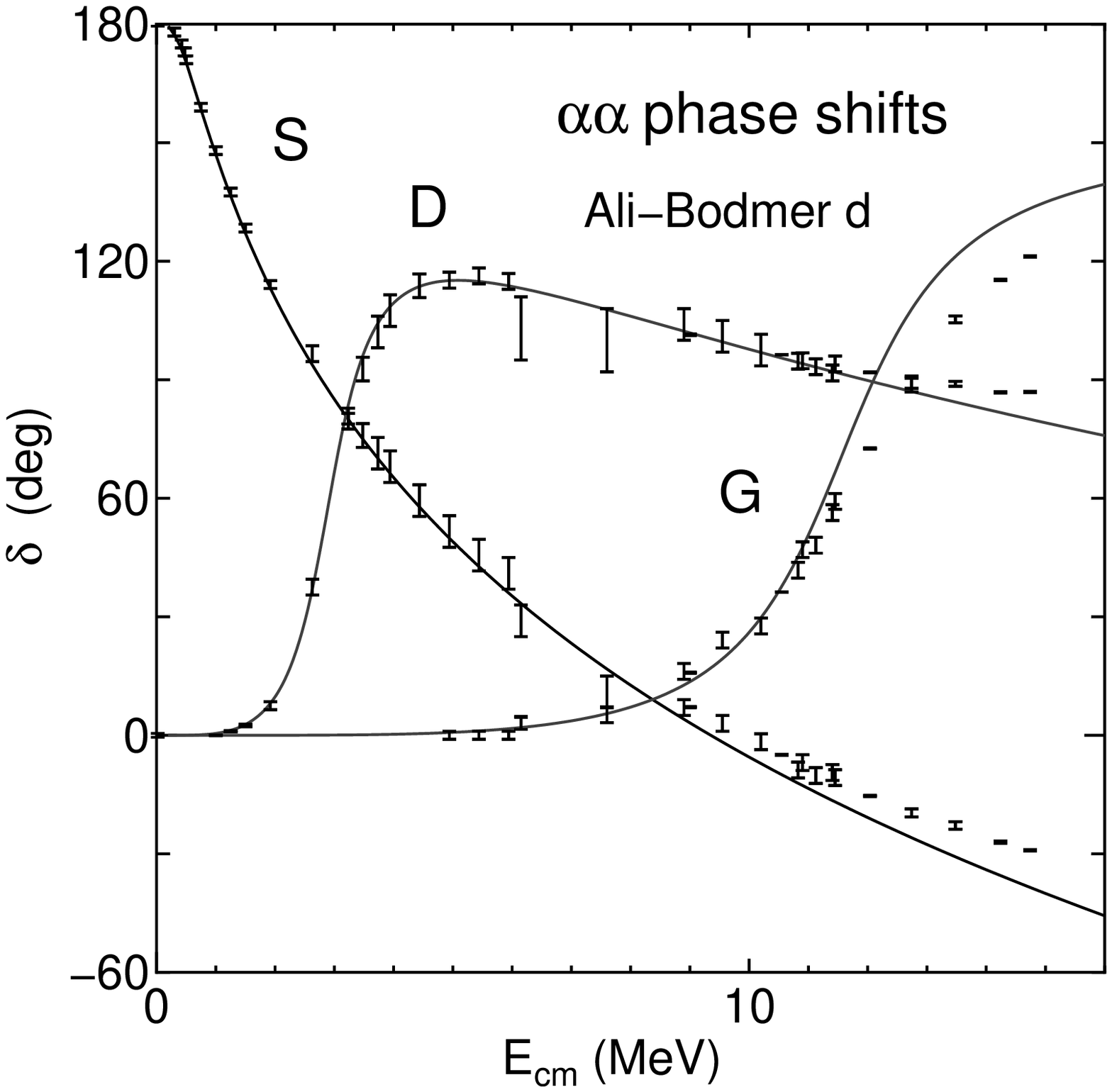}}
\end{minipage}~%
\hfill~%
\begin{minipage}{0.45\textwidth}
\centerline{\includegraphics[width=\columnwidth]{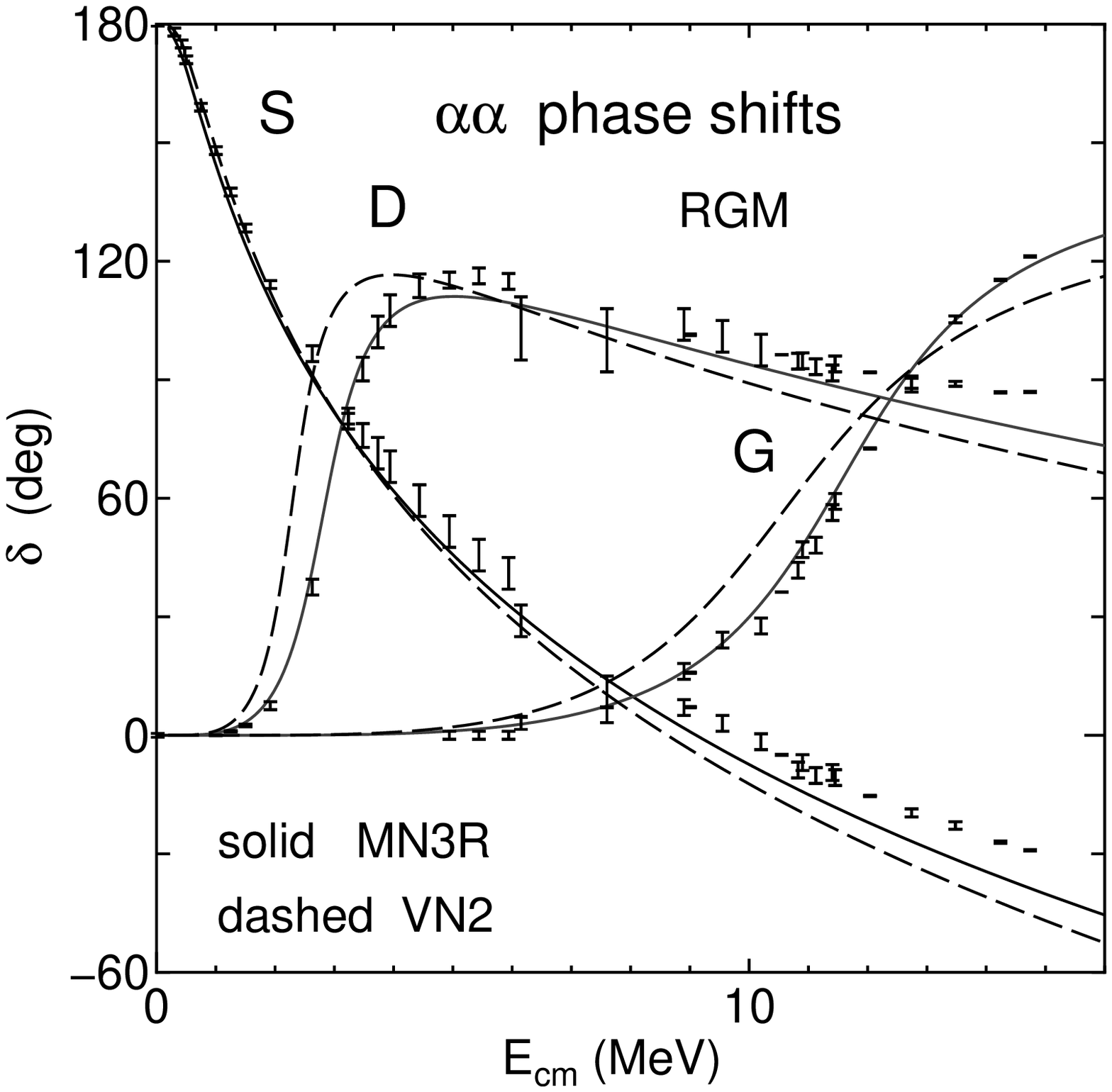}}
\end{minipage}
\caption{(a): $S$-, $D$- and $G$-wave $\alpha \alpha$ phase shifts
predicted by the Ali-Bodmer potential ABd.
(b): The same as (a), but for $\alpha \alpha$ RGM by
Volkov No.\,2 (VN2) with $m=0.59$ and $\nu=0.275~\hbox{fm}^{-1}$ 
(dashed curves) and by Minnesota three-range (MN3R) potentials 
with $u=0.94687$ and $\nu=0.257~\hbox{fm}^{-1}$ (solid curves), 
the latter result is in better agreement with experiment.
\label{fig3}}
\end{figure}

As a more complex system, we apply the present method to the
$\alpha \alpha$ LS-RGM, using the Minnesota
three-range force. In this calculation, we solve the RGM
equation in the momentum space. All the Born kernels including
the direct term and the RGM exchange kernels for the sharply
cut-off Coulomb force between two protons are analytically calculated.
For example, the direct Born kernels of the error function Coulomb
potential in \eq{eq6-3} and the screened Coulomb potential
in \eq{eq6-7} are given by
\begin{eqnarray}
\hspace{-10mm} M^{CL}_D(\bq_f, \bq_i) & = & 
\langle e^{i\bm{q}_f \cdot \bm{r}} \vert
\frac{4e^2}{r}{\rm erf}\,(\beta r) \vert e^{i\bm{q}_i \cdot \bm{r}} \rangle
=4e^2\,\frac{4\pi}{\bm{k}^2}\,e^{-\frac{1}{4}\left(\frac{\bm{k}}{\beta}\right)^2}
\ ,\nonumber \\
\hspace{-10mm} M^{\rho CL}_D(\bq_f, \bq_i) & = & 
\langle e^{i\bm{q}_f \cdot \bm{r}} \vert
V^\rho_{\rm D}(r)\vert e^{i\bm{q}_i \cdot \bm{r}} \rangle
=4e^2\,2\pi \rho^2 \left(\frac{\sin\,\frac{\bm{k} \rho}{2}}{\frac{\bm{k} \rho}{2}} 
\right)^2 \,e^{-\frac{1}{4}\left(\frac{\bm{k}}{\beta}\right)^2}\ ,\hfill
\label{eq6-14}
\end{eqnarray}
where $\bk=\bq_f-\bq_i$.
Note that $M^{CL}_D(\bq_f, \bq_i)$ involves the Coulomb singularity
at $|\bq_f|=|\bq_i|$, while $M^{\rho CL}_D(\bq_f, \bq_i)$ does
not have such a singularity.
A numerical challenge is the angular momentum projection of this kernel.
We have used a standard Gauss-Legendre integration quadrature, taking
many discretization points. We can check the accuracy of this numerical
integration by examining the redundancy condition 
of the Pauli forbidden states for the $S$- and $D$-waves.
Various cut-off Coulomb parameters are chosen 
from $\rho=8$ fm to 16 fm, with $b=6$ fm fixed,
The modified Coulomb wave functions are therefore solved 
from $R_{\rm in}=\rho-6~\hbox{fm}$
to $R_{\rm out}=\rho+6~\hbox{fm}$.
In Table 2, we list the variation of the nuclear phase shifts,
depending on the choice of $\rho$. We find that the results are 
quite stable in this appropriate range of $\rho$. 
We show in Fig.\,\ref{fig3}(a) the $\alpha \alpha$ phase shifts
predicted by Ali-Bodmer d potential and in Fig.\,\ref{fig3}(b)
the results by the LS-RGM using the Minnesota 
three-range force and the Volkov No.\,2 two-range force.


\begin{table}[htb]
\caption{Cut-off radius ($\rho$) dependence of the nuclear phase 
shifts $\delta_\ell$ for the $\alpha \alpha$ Lippmann-Schwinger RGM. 
The Minnesota three-range force with $u=0.94687$ 
and $\nu=0.257~\hbox{fm}^{-2}$ are used.
}
\label{table2}
\begin{center}
\renewcommand{\arraystretch}{1.1}
\setlength{\tabcolsep}{5mm}
\begin{tabular}{rr|rrrrr}
\hline
\multicolumn{2}{c|}{$E_{\rm cm}$ (MeV)} & \multicolumn{5}{c}{$\rho$ (fm)} \\
\multicolumn{2}{c|}{} & 8 \qquad & 10 \qquad & 12 \qquad & 14 \qquad 
& 16 \qquad \\
\hline
   &    1.000  &   144.448   &  144.450  &   144.450  &   144.451  
   &   144.450   \\
   &    2.000  &   107.561   &  107.563  &   107.563  &   107.563  
   &   107.563   \\
   &    3.000  &    81.689   &   81.690  &    81.690  &    81.690  
   &    81.690   \\
   &    4.000  &    61.847   &   61.848  &    61.848  &    61.848  
   &    61.848   \\
$\hbox{}^1S_0$ &    5.000    &   45.740  &    45.741  &    45.741  
   &    45.741 & 45.741 \\
   &    6.000  &    32.176   &   32.177  &    32.177  &    32.177  
   &    32.177   \\
   &    8.000  &    10.145   &   10.146  &    10.146  &    10.146  
   &    10.146   \\
   &     0.933   \\
   &   10.000  &   $-7.395$  &  $-7.394$ &   $-7.394$ &   $-7.394$ 
   &   $-7.394$  \\
   &   12.000  &  $-21.988$  & $-21.987$ &  $-21.987$ &  $-21.987$ 
   &  $-21.987$  \\
   &   15.000  &  $-40.163$  & $-40.163$ &  $-40.163$ &  $-40.163$ 
   &  $-40.163$  \\
\hline
   &    1.000  &    0.589    &    0.590  &   0.590   &    0.590  
   &   0.587  \\
   &    2.000  &   11.641    &   11.644  &  11.644   &   11.644  
   &  11.645  \\
   &    3.000  &   70.120    &   70.134  &  70.134   &   70.134  
   &  70.134  \\
   &    4.000  &  106.361    &  106.365  & 106.365   &  106.365  
   & 106.365  \\
$\hbox{}^1D_2$ &    5.000    &  111.080  &  111.081  & 111.081   
   &  111.081  & 111.081  \\
   &    6.000  &  109.344    &  109.345  & 109.345   &  109.345  
   & 109.345  \\
   &    8.000  &  101.886    &  101.887  & 101.887   &  101.887  
   & 101.887  \\
   &   10.000  &   93.853    &   93.854  &  93.854   &   93.854  
   &  93.854  \\
   &   12.000  &   86.367    &   86.368  &  86.368   &   86.368  
   &  86.368  \\
   &   15.000  &   76.354    &   76.355  &  76.355   &   76.355  
   &  76.355  \\
\hline
   &    1.000  &    0.000  &    0.000 &    0.000  &    0.000  &   0.000  \\
   &    2.000  &    0.013  &    0.013 &    0.013  &    0.013  &   0.014  \\
   &    3.000  &    0.102  &    0.102 &    0.102  &    0.102  &   0.102  \\
   &    4.000  &    0.398  &    0.399 &    0.399  &    0.399  &   0.399  \\
$\hbox{}^1G_4$ &    5.000  &    1.107 &    1.108  &    1.108  &    1.108  
   &  1.108  \\
   &    6.000  &    2.522  &    2.523 &    2.523  &    2.523  &   2.523  \\
   &    8.000  &    9.514  &    9.517 &    9.517  &    9.517  &   9.517  \\
   &   10.000  &   29.924  &   29.928 &   29.928  &   29.928  &  29.928  \\
   &   12.000  &   75.818  &   75.820 &   75.820  &   75.820  &  75.820  \\
   &   15.000  &  120.744  &  120.745 &  120.745  &  120.745  & 120.745  \\
\hline
\end{tabular}
\end{center}
\end{table}
%

%

\subsection{$pd$ elastic scattering}

\begin{figure}[t]
\begin{minipage}{0.45\textwidth}
\centerline{\includegraphics[width=\columnwidth]{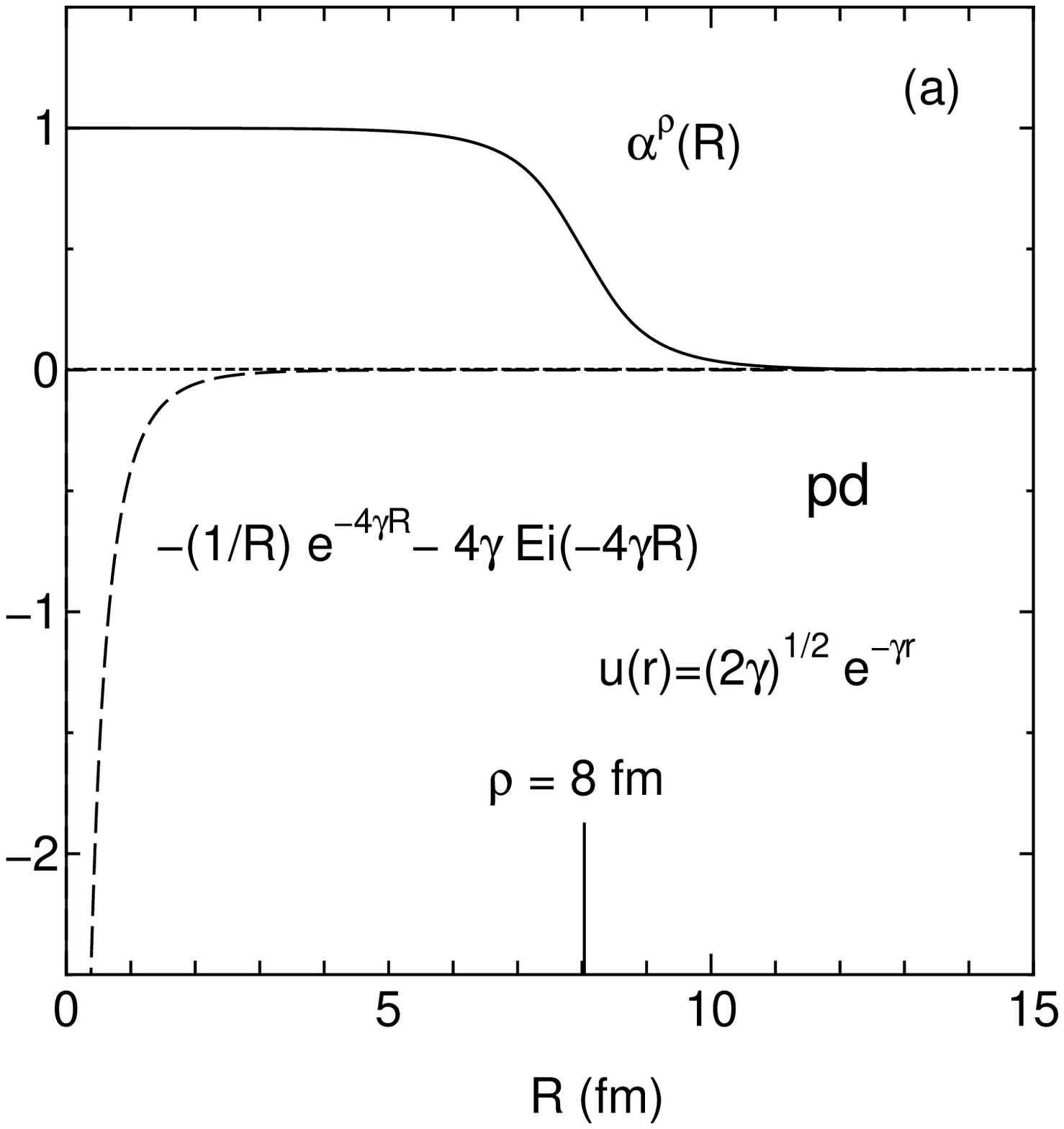}}
\end{minipage}~%
\hfill~%
\begin{minipage}{0.45\textwidth}
\centerline{\includegraphics[width=\columnwidth]{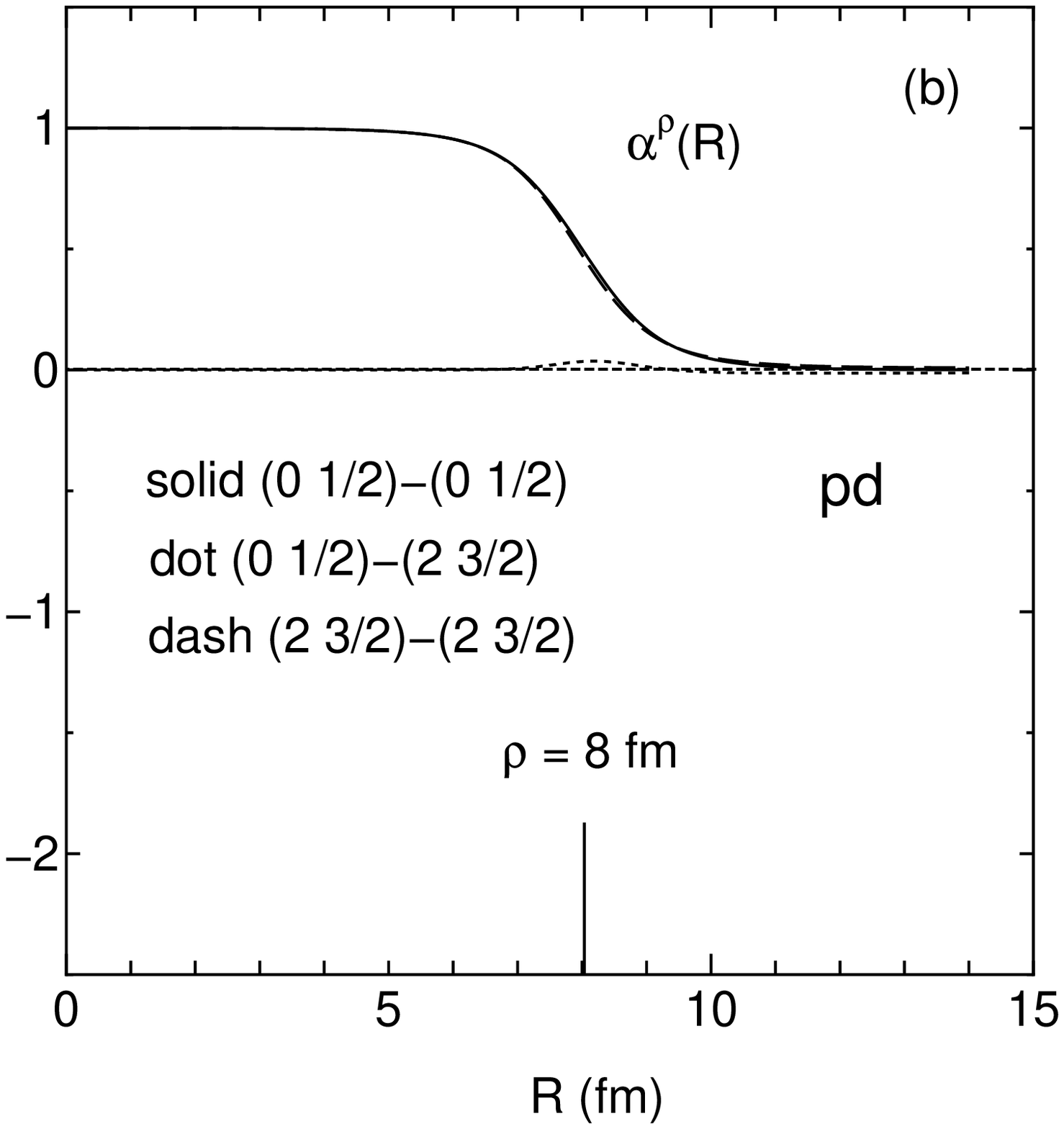}}
\end{minipage}
\caption{(a) The screening function $\alpha^\rho(R)$ (solid curve) 
and the short-range potential $\CW(R)/e^2$ (dashed curve) (in the
unit of $\hbox{fm}^{-1}$), given in \eq{c7} for the
$pd$ scattering. A simple $S$-wave deuteron wave function
$u(r)=\sqrt{2\gamma}\,e^{-\gamma r}$ is used. 
The Coulomb cut-off radius is $\rho=8$ fm.
(b) The realistic $\alpha^\rho(R)$ for the $pd$ scattering
in the simplest deuteron channel with $J^\pi=1/2^+$.
In the ($\ell S_c)=(0~1/2)$ (solid curve) and (2~3/2) (dashed curve) 
diagonal channels, curves are almost identical. 
The off-diagonal $\alpha^\rho(R)$
with ($\ell S_c$)-($\ell^\prime S^\prime_c$)=(0~1/2)-(2~3/2) is
very small. Here, $\ell$ is the relative angular-momentum between
$p$ and $d$ and $S_c$ is the channel spin. 
}
\label{fig4}
\end{figure}

As in the case of the $\alpha \alpha$ scattering 
discussed in the preceding subsections,
the screening function $\alpha^\rho(R)$ for the $pd$ elastic 
scattering should be derived in a consistent way with the
screened Coulomb potential between two protons in \eq{eq5-1}.
In our application of the quark-model baryon-baryon interaction
fss2 to the $pd$ elastic scattering in Ref.\,\citen{ndscat2},
the sharply cut-off Coulomb force is introduced at the quark level
in the form of $(1/r_{qq})\theta (\rho-r_{qq})$, 
where $r_{qq}$ is the relative distance between two quarks.
The proton-proton ($pp$) potential $ \omega^{\rho}(r)$ is obtained 
by folding it with the $(3q)$-$(3q)$ internal wave function, resulting in
\begin{eqnarray}
\omega^\rho(r; 1, 2) & = & \frac{e^2}{r} 
\left\{~{\rm erf}\left(\frac{\sqrt{3}}{2}\frac{r}{b}\right)
-\frac{1}{2}\left[{\rm erf}\left(\frac{\sqrt{3}}{2}\frac{r+\rho}{b}\right)
+{\rm erf}\left(\frac{\sqrt{3}}{2}\frac{r-\rho}{b}\right)\right]\right\}
\nonumber \\
& & \times \frac{1+\tau_z(1)}{2} \frac{1+\tau_z(2)}{2}\ .
\label{eq6-15}
\end{eqnarray}
where $r$ is the distance between the two protons, $r=|\br_{12}|
=|\bx_1-\bx_2|$ and $b$ is 
the harmonic-oscillator range parameter of the $(3q)$-clusters. 
Note that this screened Coulomb potential for the two protons is not equal
to \eq{eq5-1} with a mere change of $\alpha^\rho(r)$ to \eq{eq6-10}
(with a trivial modification $\beta \rightarrow (\sqrt{3}/2b)$), 
but also contains the contributions
from the short-range Coulomb potential in \eq{eq6-12}.
We calculate the $pd$ screened Coulomb potential 
by further folding the $pp$ potential in \eq{eq6-15}
with the deuteron wave function $\langle \br; 1,2|\psi _{d}\rangle $:
\begin{eqnarray}
V^{\rho C}_{pd}(R)
=\langle \psi^{d}|\omega^{\rho}(|\bR+\br/2|; 2,3)|\psi^{d}\rangle
+\langle \psi^{d}|\omega^{\rho}(|\bR-\br/2|; 3,1)|
\psi^{d}\rangle \ ,
\label{eq6-16}
\end{eqnarray}
where $\bR=\bx_3-(\bx_1+\bx_2)/2$ is the relative coordinate between 
the center-of-mass of the deuteron and the proton. 
This calculation is made in Appendix C.
We assign the long-range part of $V^{\rho C}_{pd}(R)$ in \eq{c2}
to $W^\rho(R)$ in \eq{eq5-2},
and parametrize it as $W^\rho(R)=(e^2/R) \alpha^\rho(R)$.
The screening function $\alpha^\rho(R)$ is numerically calculated
by using Eqs.\,(\ref{c15}) - (\ref{c17}) and
the momentum-space deuteron wave function expanded
in the dipole form factors \cite{Fu02a}.
Here, we only show in Fig.\,\ref{fig4} the profiles of 
the screening function $\alpha^\rho(R)$
and the short-range Coulomb potential 
(the polarization potential) $\CW(R)$ for 
the simplest deuteron channel with $J^\pi=1/2^+$.
%
We find that the cut-off behavior around $R \sim \rho$ is
fairly sharp even in $\rho \sim 8$ fm.
The short-range Coulomb potential $\CW(R)$ is $\rho$-independent
as shown in \eq{c5}.
The coupling potential $W^\rho(R)$ between different channel-spin
states, $(\ell S_c) \neq (\ell^\prime S^\prime_c)$, is very small.
We therefore neglect this and solve the screened Coulomb problem
only by using the diagonal part of $(\ell S_c)$, in order to generate
the regular and irregular screened Coulomb wave functions
for the connection condition.

\begin{table}[b]
\caption{Real parts of the nuclear eigenphase shifts for the $Nd$ elastic 
scattering at $E_N=65$ MeV. The $nd$ phase shifts with no Coulomb force 
and the $pd$ phase shifts including the cut-off Coulomb
force with $\rho=8$, 16 and 20 fm are listed. 
For $\rho=8$ fm (before), the eigenphase shifts before 
transformation in \protect\eq{eq5-28} are also shown.
The maximum total angular-momentum of the two-nucleon subsystem
is $I_{\rm max}=4$, and the momentum discretization 
points $n=n_1\hbox{-}n_2\hbox{-}n_3=6\hbox{-}6\hbox{-}5$ are used
in the definition shown in Ref.\,\citen{ndscat1}.  
}
\label{table3}
\begin{center}
\renewcommand{\arraystretch}{1.2}
\setlength{\tabcolsep}{5mm}
\begin{tabular}{r|r|r|rrr}
\hline
$\hbox{}^{2S+1}\ell_J$ & no Coulomb & \multicolumn{4}{c}{with~Coulomb} \\
\cline{3-6}
&  & $\rho=8$ fm & $\rho=8$ fm & $\rho=16$ fm & $\rho=20$ fm \\
&  &  (before)   &      &       &       \\
\hline
$\hbox{}^2S_{1/2}$   &  26.84  &  24.38  &  28.70  &  28.99  &  29.01  \\
$\hbox{}^4D_{1/2}$   & $-7.25$ & $-9.41$ & $-7.20$ & $-6.92$ & $-6.76$ \\
$\hbox{}^2P_{1/2}$   & $-0.44$ & $-2.51$ & $-0.04$ &   0.35  &   0.45  \\
$\hbox{}^4P_{1/2}$   &  24.28  &  21.83  &  24.76  &  24.98  &  24.99  \\
$\hbox{}^4S_{3/2}$   &  32.11  &  31.23  &  33.79  &  34.25  &  34.63  \\
$\hbox{}^2D_{3/2}$   &   8.74  &   6.87  &   9.15  &   9.35  &   9.51  \\
$\hbox{}^4D_{3/2}$   & $-5.49$ & $-7.84$ & $-5.32$ & $-5.01$ & $-5.11$ \\
$\hbox{}^4P_{3/2}$   &  24.98  &  22.53  &  25.32  &  25.84  &  25.72  \\
$\hbox{}^2P_{3/2}$   &   6.73  &   4.60  &   7.16  &   7.38  &   7.47  \\
$\hbox{}^4F_{3/2}$   & $-1.05$ & $-2.81$ & $-0.86$ & $-0.63$ & $-0.55$ \\
\hline
\end{tabular}
\end{center}
\end{table}

\begin{figure}[t]
\begin{center}
\begin{minipage}{0.48\textwidth}
\begin{center}
\includegraphics[width=\columnwidth]{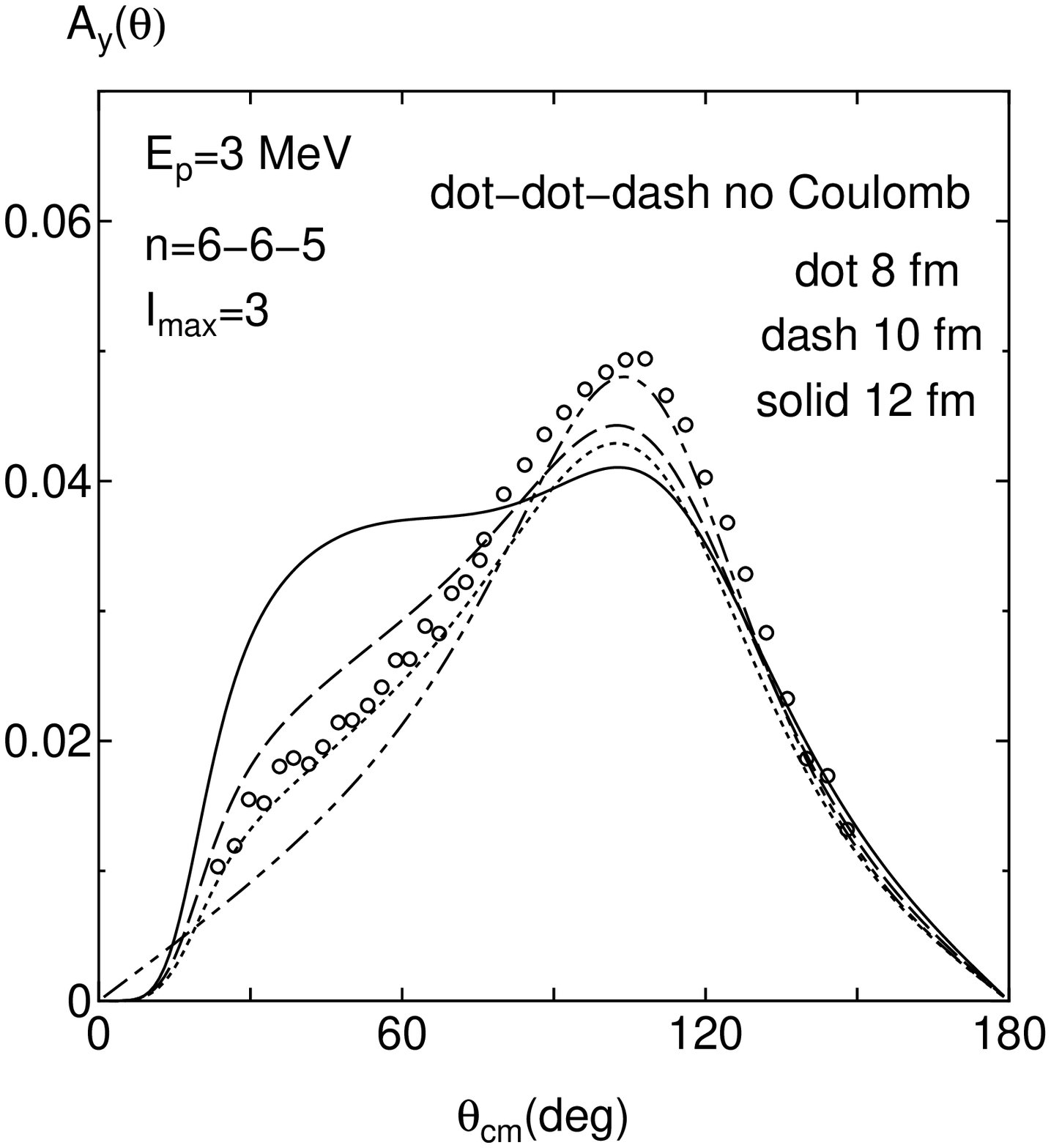}
\end{center}
\end{minipage}~%
\hfill~%
\begin{minipage}{0.48\textwidth}
\begin{center}
\includegraphics[width=\columnwidth]{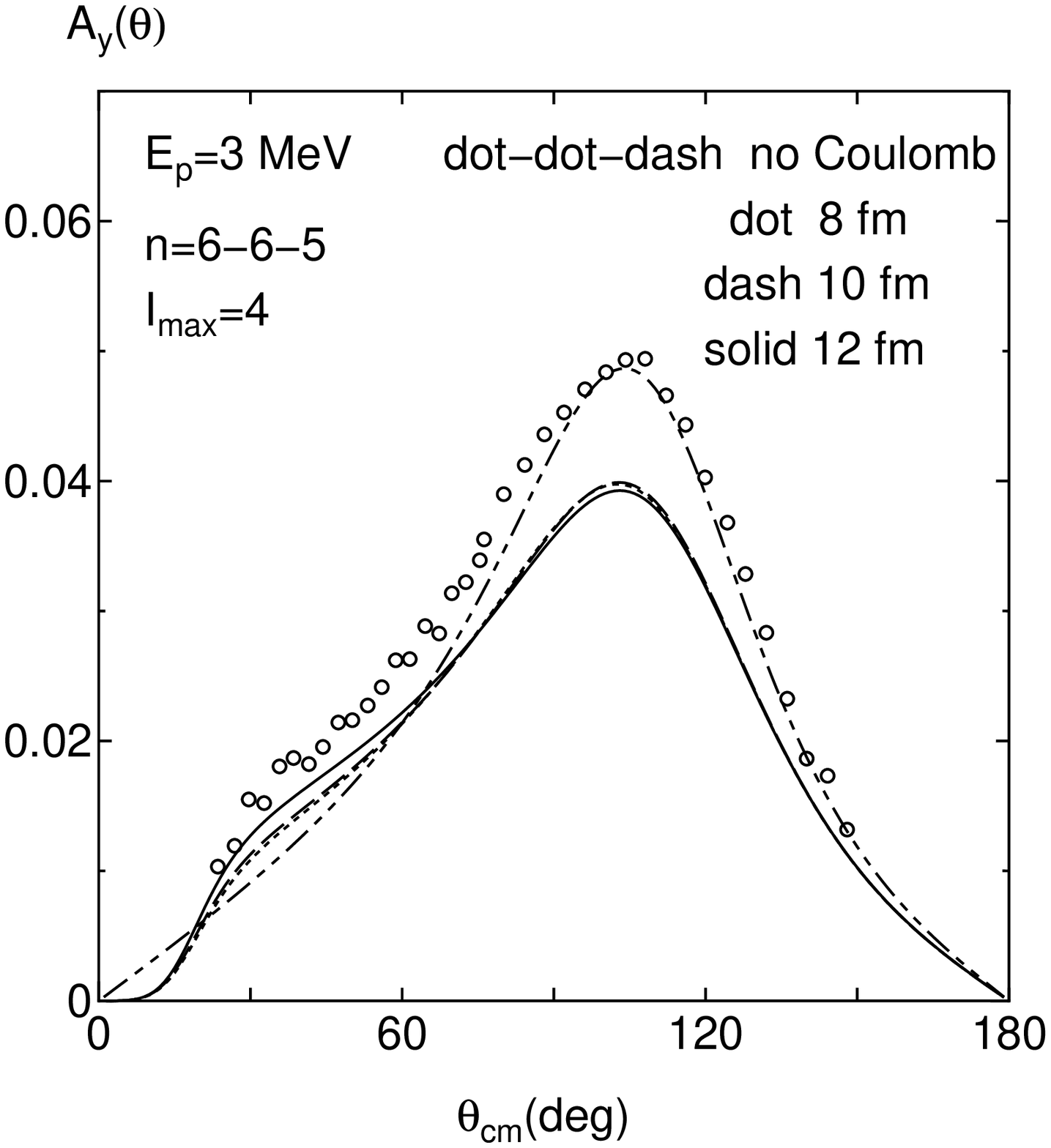}
\end{center}
\end{minipage}
\caption{Cut-off radius dependence of the proton analyzing power
for the $pd$ elastic scattering at $E_p=3$ MeV. The results of
no Coulomb case, $\rho=8$, 10, and 12 fm are shown
by the dot-dot-dashed, dotted, dashed, and
solid curves, respectively. 
The left panel shows the results of $I_{\rm max}=3$ and the right
panel of $I_{\rm max}=4$. 
The $pd$ experimental data from Ref.\,\citen{Sa94} are also shown by circles.
\label{fig5}}
\end{center}
\end{figure}

Some typical eigenphase shifts of the $E_p=65$ MeV $pd$ scattering 
with the Coulomb cut-off radius $\rho=8$, 16 and 20 fm are 
listed in Table \ref{table3} for the $S$ and $P$ waves. 
Here, we have assumed the maximum total angular-momentum 
of the two-nucleon subsystem, $I_{\rm max}=4$.
The real parts of the eigenphase shifts are only given
for simplicity. 
We find that the inclusion of the cut-off Coulomb 
force gives an apparent repulsive effect, namely, the $S$-wave and $P$-wave
eigenphase shifts are $-0.9^\circ \sim -2.5^\circ$ 
($-2.5^\circ \sim -3.1^\circ$) more repulsive than in the no Coulomb case, 
if $\rho=8$ fm ($\rho=16$ fm) is assumed.
The transformation by the connection condition in \eq{eq5-28} gives 
an attractive effect to make the resultant eigenphase shifts rather 
close to the no Coulomb case. As long as the low partial waves such as
the $S$ and $P$ waves are concerned, the final results of the nuclear
eigenphase shifts are rather stable within the fluctuation of less than
0.8$\hbox{}^\circ$. We have calculated $pd$ differential cross sections
and other polarization observables, using various $\rho$ values.
The results by $\rho=8$ fm is quite reasonable, but if we take larger
values like $\rho=16$ and 20 fm, we have found that unpleasant oscillations
develop in all the observables. The origin of the oscillations is traced back
to the high partial waves, in which the restriction of $I_{\rm max}=4$ is 
too severe.
Since we are using the channel-spin formalism, 
the total angular momentum $J^\pi$ of the three nucleon system 
is achieved by the angular-momentum coupling $(\ell S_c)J$, 
where the channel spin $S_c$ is constructed from $(I\frac{1}{2})S_c$.
For a large $J^\pi$, a large contribution of the Coulomb force from the
large relative orbital angular momentum of the two-proton subsystem is
not fully taken into account, since the magnitude of $S_c$ is restricted
by $I_{\rm max}=4$. To demonstrate this situation, we show in Fig. \ref{fig5} 
the $\rho$-dependence of the nucleon analyzing power for the 3 MeV $pd$
scattering, calculated with $I_{\rm max}=3$ and $I_{\rm max}=4$.
In the forward angular region with $\theta_{\rm cm} < 90^\circ$,
we find that unpleasant bump structure develops as $\rho$ increases 
from 8 fm to 12 fm, when $I_{\rm max}=3$ is used. However, such 
enhancement is strongly suppressed when $I_{\rm max}=4$ is used.
This demonstrates very clearly that two-nucleon partial waves
should be included up to sufficiently higher values to obtain the 
well converged results, if the screened Coulomb force 
is incorporated into the standard AGS equations.

%
\begin{figure}[htb]
\begin{center}
\begin{minipage}{0.48\textwidth}
\includegraphics[angle=0,width=52mm]
{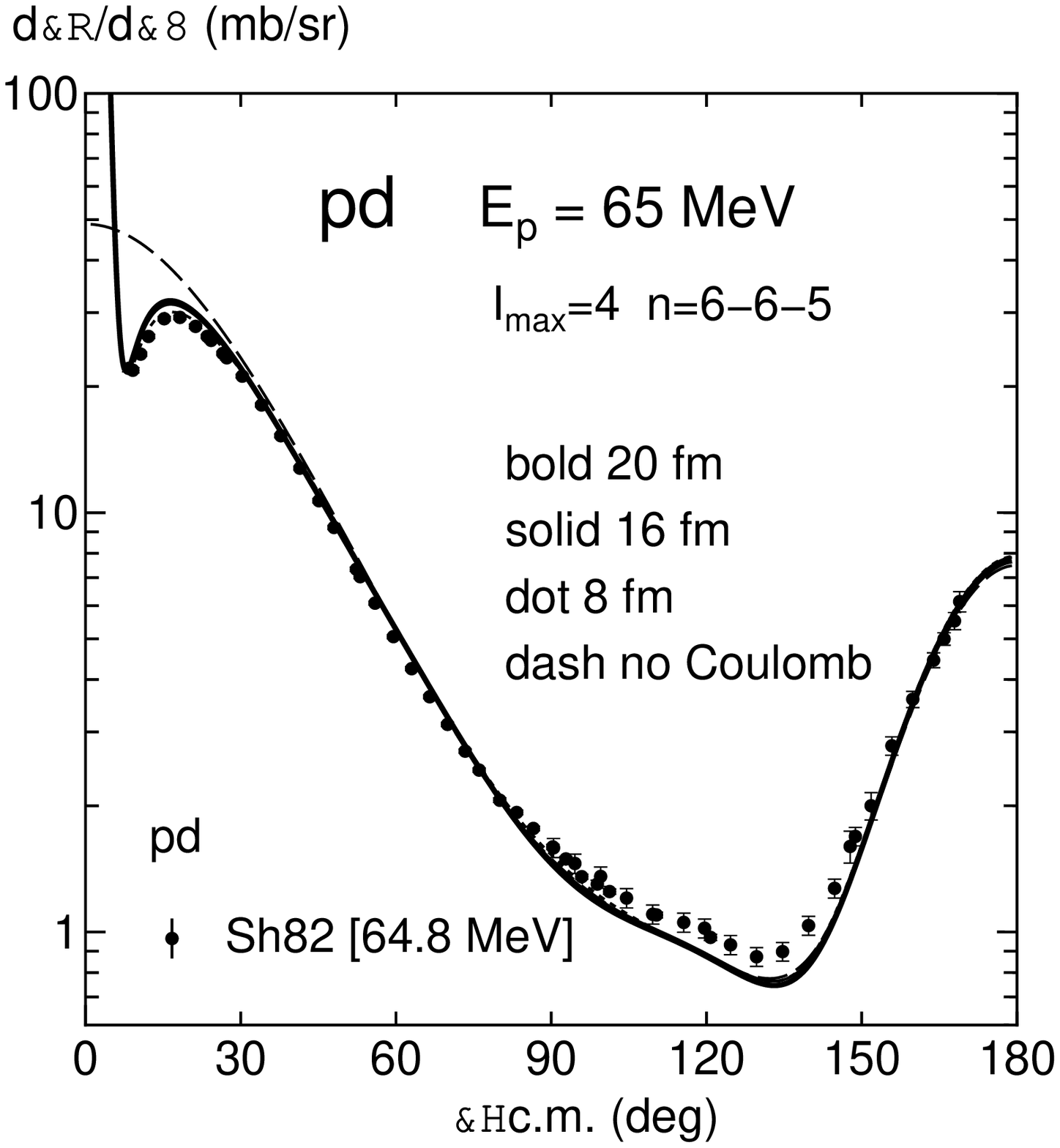}

\includegraphics[angle=0,width=52mm]
{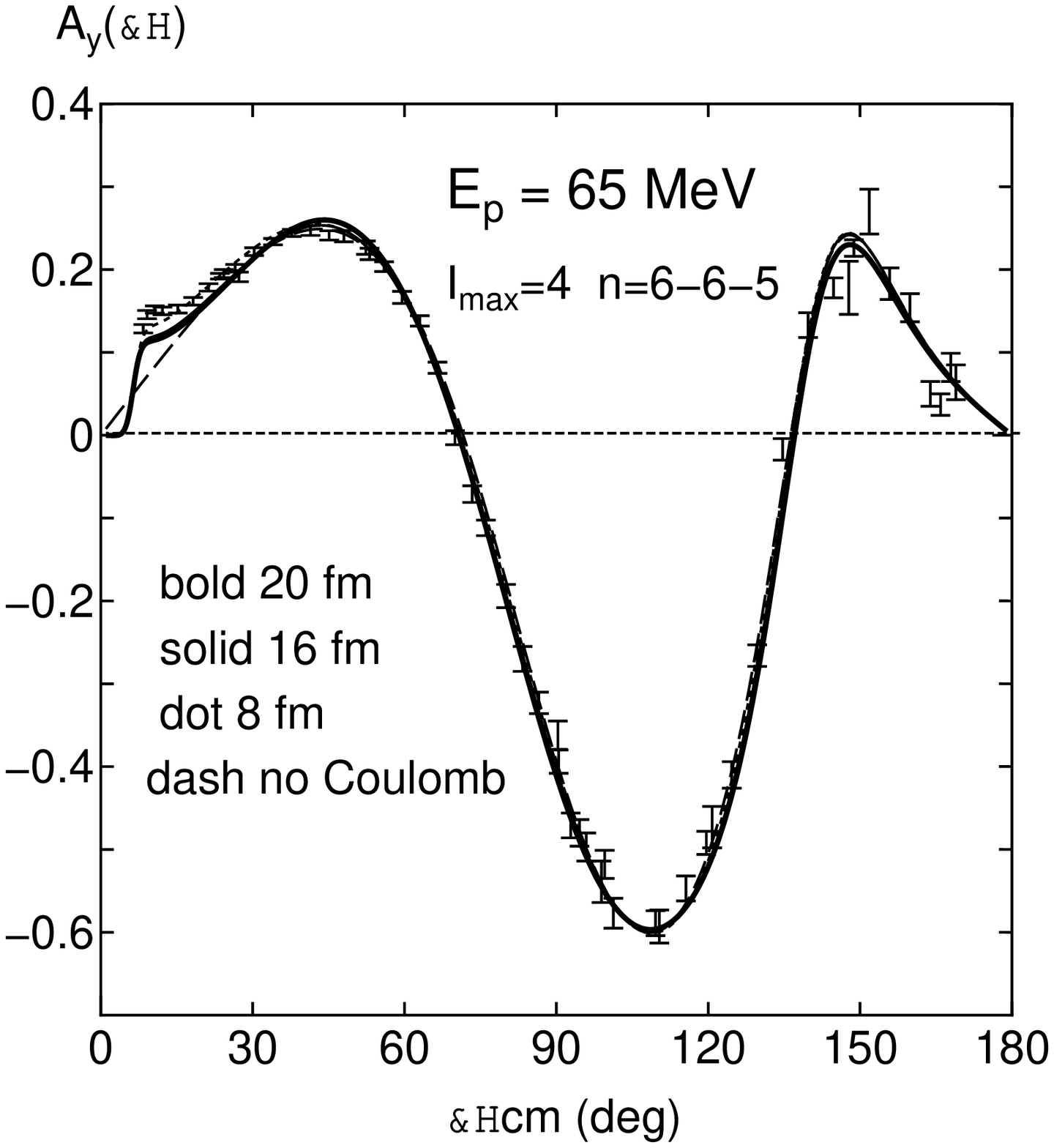}

\includegraphics[angle=0,width=52mm]
{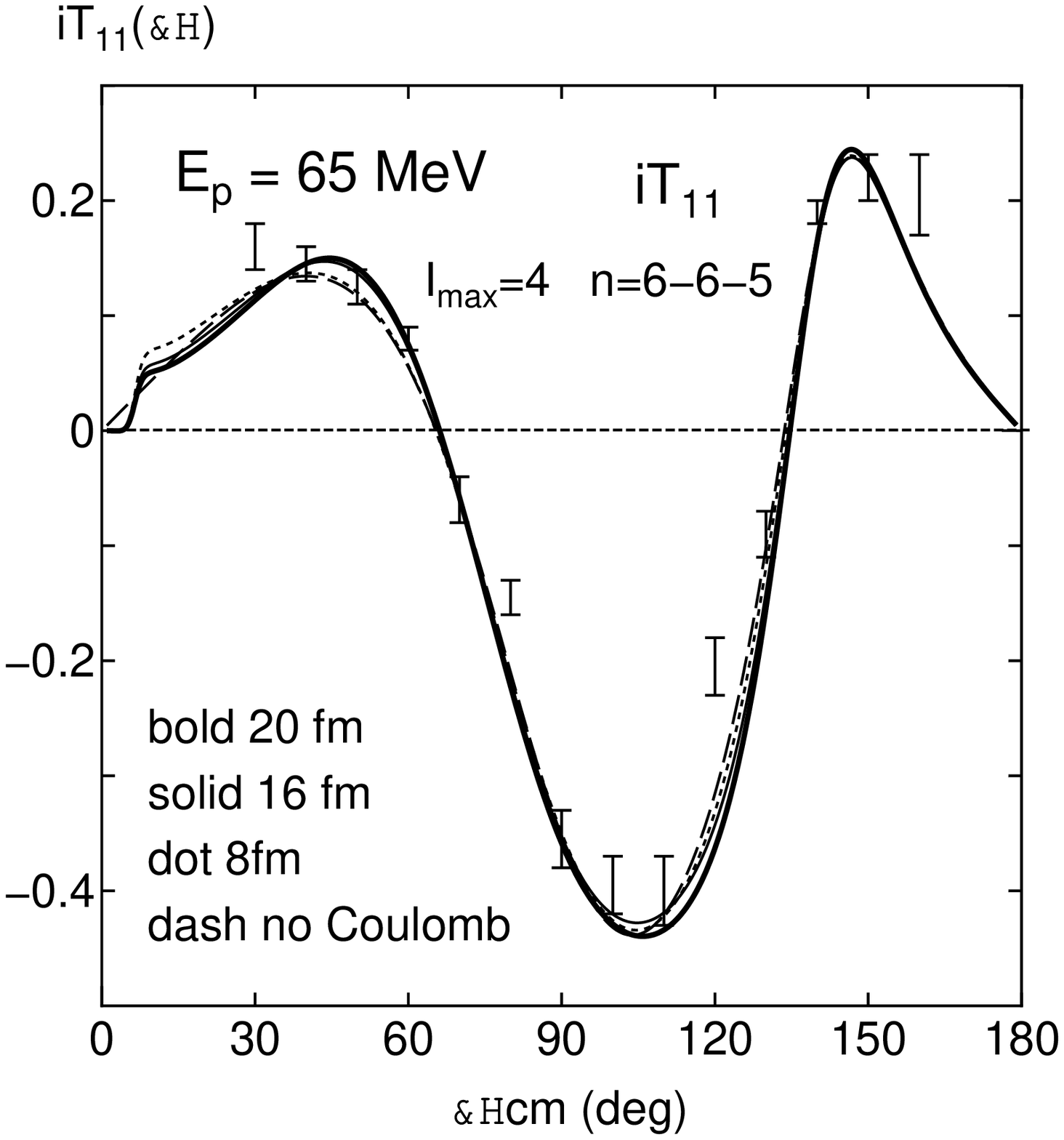}
\end{minipage}~%
\hfill~%
\begin{minipage}{0.48\textwidth}
\includegraphics[angle=0,width=52mm]
{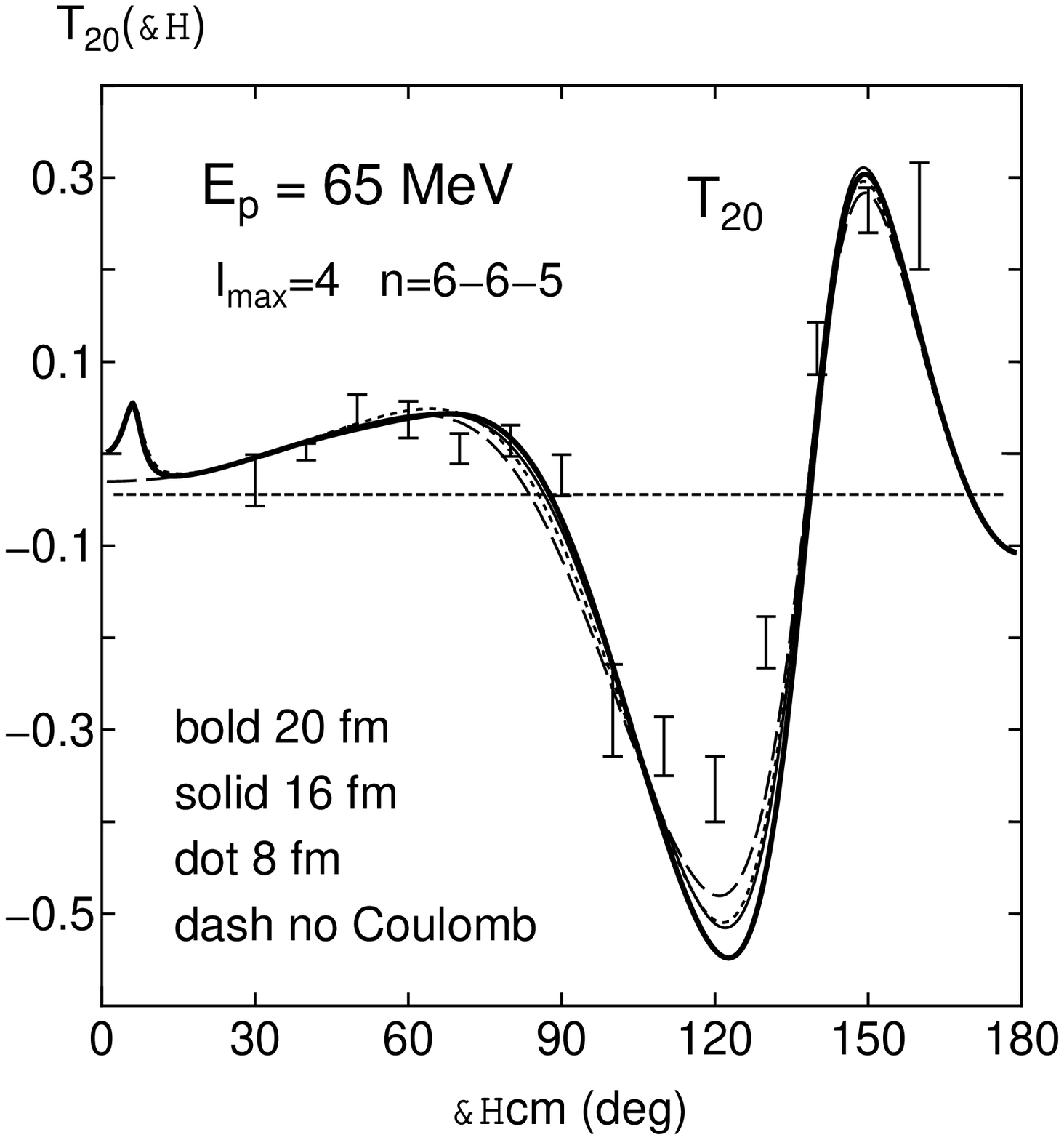}

\includegraphics[angle=0,width=52mm]
{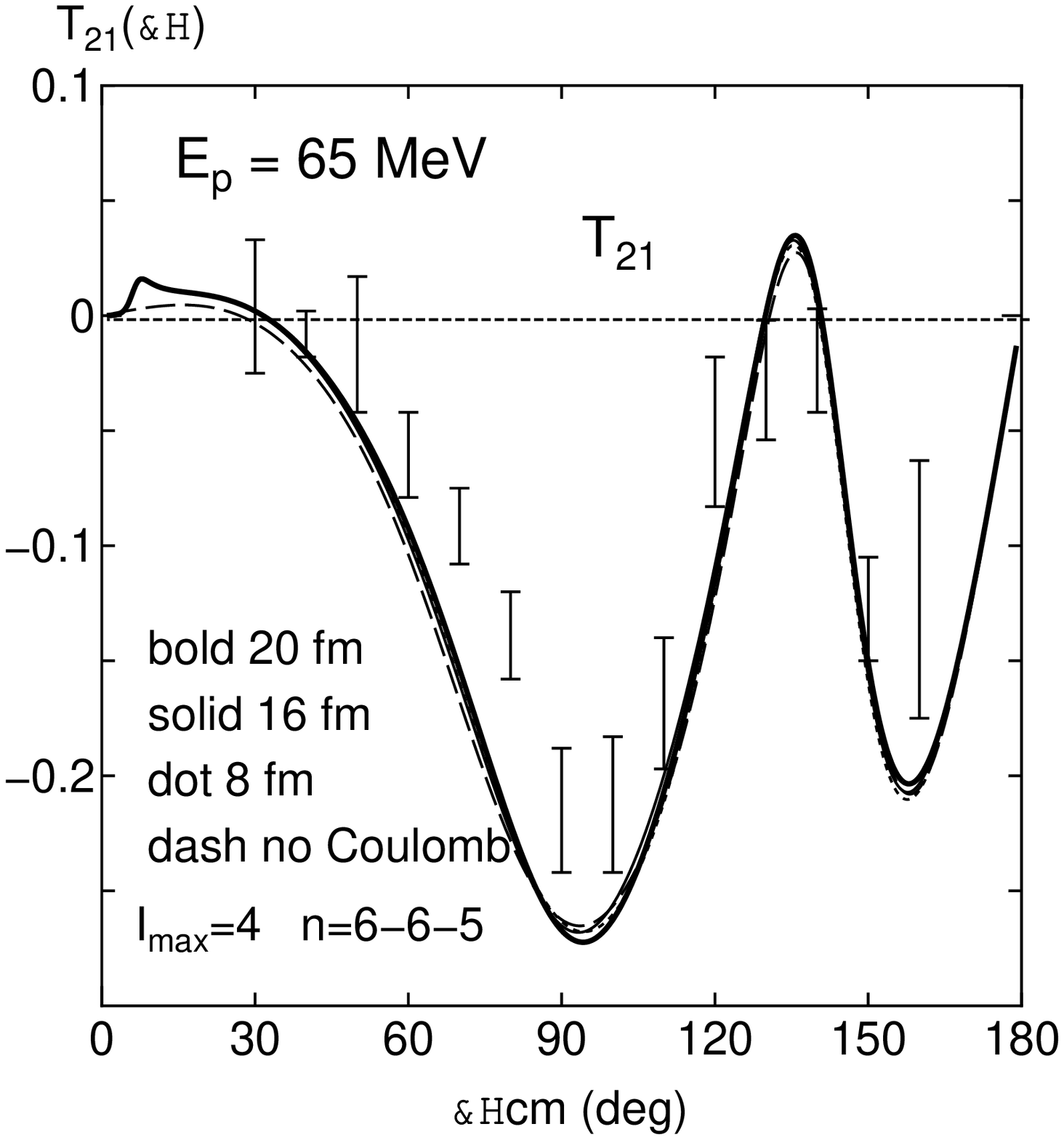}

\includegraphics[angle=0,width=54mm]
{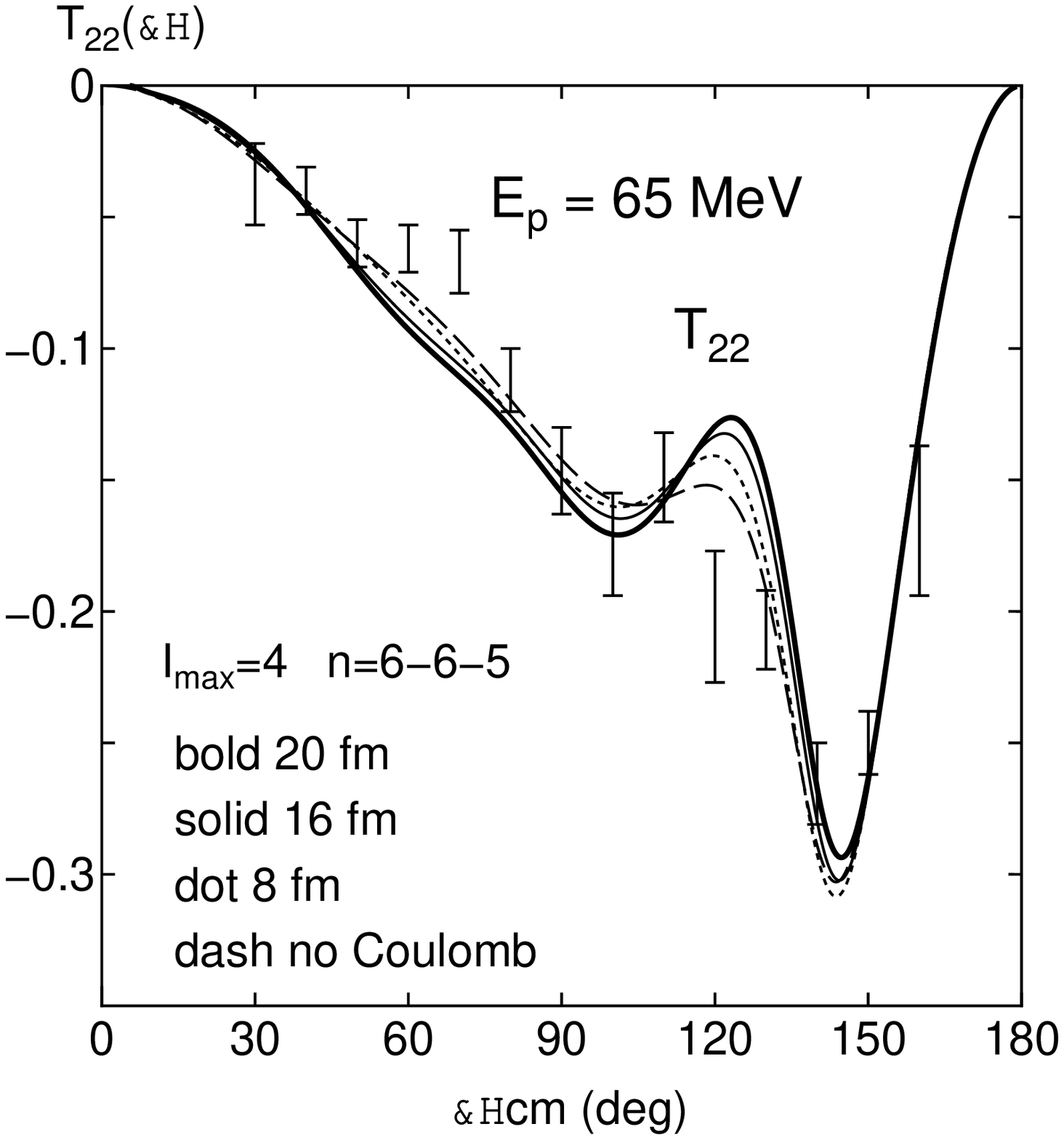}
\end{minipage}
\end{center}
\caption{
$pd$ differential cross sections ($d \sigma/d \Omega$),
analyzing power ($A_y(\theta)$) of the proton, and 
vector-type ($iT_{11}(\theta)$) and 
tensor-type ($T_{2m}(\theta)$) analyzing powers 
of the deuteron at $E_p=65$ MeV.
The results with no Coulomb case, $\rho=8$, 16, and 20fm
are shown by the dashed, dotted, solid, and
bold-solid curves, respectively. These curves almost overlap
with each other, except for the forward nuclear-Coulomb
interference region. The screened Coulomb force is neglected
for higher partial waves with $J^\pi \geq 11/2^+$.
The experimental data are taken from Ref.\,\citen{Sh82}
for $d \sigma/d \Omega$ and $A_y(\theta)$, and 
from Ref.\,\citen{Wi93} for $iT_{11}(\theta)$ 
and $T_{2m}(\theta)$.
}
\label{fig6}       
\end{figure}

Since the calculation with $I_{\rm max}=6$ and more 
is not presently possible because of the computer resources, 
here we propose to cut the Coulomb force for higher $J^\pi$ values 
and use a simple ``Coulomb externally corrected approximation'', 
in which the $nd$ eigenphase shifts are directly used
for the nuclear phase shifts \cite{Do82}.
Figure \ref{fig6} shows the $pd$ differential cross sections and
some polarization observables at $E_p=65$ MeV, 
calculated by neglecting the Coulomb force for $J^\pi \geq 11/2^+$. 
We find that the results with $\rho=8$, 16 and 20 fm
are very similar, although some difference is seen in $T_{22}(\theta)$.
The results with $\rho=8$ fm are almost the same as the full calculation
including the Coulomb force to all the partial waves.  

\clearpage

\section{Summary and outlook}

In the present work we have proposed a practical method to deal
with the Coulomb problem in the momentum space.
Although the standard procedure to deal with the Coulomb force
in two-body systems is formulated in the configuration space, 
the extension of such an approach to three-body systems is 
not trivial.\cite{Is09}
Here, we have reformulated the momentum-space approach
of the two-cluster systems based on the essential idea
of the ``screening and renormalization procedure'',
which is recently used in the standard formulation of
the AGS equations for the $pd$ scattering
in the momentum representation. \cite{De05a, De05b, De05c}
In this approach, the screened Coulomb force with
a cut-off parameter $\rho$ is introduced to the
basic equations as if it is a part of the short-range
nuclear force. The two-potential formula for the
short-range potentials is used to generate the scattering
amplitude. The pure Coulomb results are
reproduced by taking $\rho \rightarrow \infty$ limit,
based on the Taylor's formula \cite{Ta74,Se75} for the
phase renormalization of the asymptotic wave 
functions of the screened Coulomb potential.
The central issue in this approach is if one can reproduce
the exact Coulomb results by taking a finite $\rho$.
Since the quasi-singular nature of the screened Coulomb force
becomes stronger for larger $\rho$, it is essential that
one can reproduce the almost $exact$ results with a reasonable
choice of $\rho$.

To achieve this, we propose to extend the 
Vincent and Phatak approach \cite{Vi74}, which is originally
formulated for the sharply cut-off Coulomb problems.
When a sharply cut-off Coulomb force 
with the cut-off radius $\rho$ is introduced at the level of
constituent particles, two-cluster direct potential of the 
Coulomb force becomes in general a local screened Coulomb potential
implemented with the short-range Coulomb force.
The screening function $\alpha^\rho(r)$ is determined by the 
properties of the cluster wave functions, and
involves a smoothness parameter $b$ related
to the size of clusters.
In practice, $b$ satisfies $b \ll \rho$, which is
an additional condition to the Taylor's properties \cite{Ta74}
of screening functions. We find that this
condition is necessary to make the present treatment work well.
We pay attention to the existence of two different types of 
asymptotic waves contained in the screened
Coulomb wave functions. The first one is 
the approximate Coulomb wave for the relative
distance of two clusters, $r$, smaller than $R_{\rm in}=\rho-b$,
and the other is the free (no-Coulomb) wave in the longer range
region, $r > R_{\rm out}=\rho+b$.
The asymptotic Hamiltonian composed of the screened Coulomb force
allows us to calculate constant Wronskians of this Hamiltonian 
in either region. Using this property, we can extend the standard 
procedure of matching conditions for asymptotic waves 
to the screened Coulomb potential.

We should note that the renormalization property of the screened 
Coulomb wave functions is more involved than in the sharply
cut-off Coulomb case. In particular, the irregular function
of the screened Coulomb potential in general contains an admixture 
of the regular solution even in the $\rho \rightarrow \infty$ limit.
As the result, the limit of the Green function for
the screened Coulomb potential is not reduced to the
Coulomb Green function. This requires an extra renormalization
of the regular wave function for the problem of the short-range 
nuclear potential plus the screened Coulomb potential. 
This renormalization factor, however, does not affect 
the final expression of the connection condition, 
since it is given by the ratio of Wronskians.

We have applied this method first to an exactly solvable model 
of the $\alpha \alpha$ scattering with the Ali-Bodmer potential 
and confirmed that essentially $ exact$ phase shifts
are reproduced, using a finite $\rho$.
The stability of nuclear phase shifts with respect to the change
of $\rho$ in some appropriate range
is demonstrated by using the $\alpha \alpha$ Lippmann-Schwinger 
RGM with the Minnesota three-range force.
In the application to the $pd$ elastic scattering, some dependence 
on the choice of $\rho$ remains although the essential features
of the nuclear and Coulomb interference in forward angles
are reproduced not only for the differential cross sections
but also for the deuteron tensor analyzing powers.

We have to admit that the completely satisfactory Coulomb treatment
of the three-body system is still beyond the way.
First, the stability of $\rho$ in the case of the above $pd$ elastic 
scattering is not completely realized.
We have examined all the observables for the $pd$ elastic scattering
in the energy range $E_p \leq 65$ MeV, and found that
the present choice $\rho \sim 8$ - 9 fm is a reasonable choice to reproduce
almost all the experimental data.\cite{ndscat2}
The forward behavior of the vector analyzing power $A_y(\theta)$ 
for the proton and $iT_{11}(\theta)$ for the deuteron is not consistently 
achieved in the low-energy region, using a unique $\rho$.
Choosing much larger $\rho$ around $\rho \sim 16$ - 20 fm is almost 
prohibited since the solution of the AGS equation becomes very singular
and the partial waves included in the actual calculations are
restricted by the hardware.
Another problem is the treatment of the Coulomb force 
in the breakup processes.\cite{ndscat3}
The phase renormalization for the observed two protons at the
final stage is not trivial because of the exchange breakup 
amplitude. We probably need to solve the Coulomb-modified
AGS equations in spite of the very singular nature of
the screened Coulomb wave functions in the momentum representation.
Finally, we mention that the Coulomb treatment
of three charged particles like the three $\alpha$ system 
is a big challenge, since the the asymptotic behavior of
the three charged particles is not $a~prior$ known.

\section*{Acknowledgements}

This work was supported by a Grant-in-Aid for Scientific
Research (C) from the Japan Society for the Promotion
of Science (JSPS) (Grant No.~23540302), and
by a Grant-in-Aid for the Global COE Program
``The Next Generation of Physics, Spun from Universality and Emergence'' 
from the Ministry of Education, Culture, Sports, Science 
and Technology (MEXT) of Japan.
It was also supported by core-stage backup subsidies from Kyoto University.
The numerical calculations were carried out on SR16000 at YITP 
in Kyoto University and on the high-performance computing system
Intel Xeon X5680 at RCNP in Osaka University.

\appendix

\section{Definition of the Coulomb wave functions}

The usual regular solution $\psi_\ell(k, r)$ and the regular solution
corresponding to the Jost solution, $\varphi_\ell(k, r)$, for the Coulomb 
problem are defined by the confluent hypergeometrical functions through  
\begin{eqnarray}
\varphi_\ell(k, r) & = & \frac{1}{k^\ell} F_\ell(k) \psi_\ell(k, r)
\nonumber \\
& = & \frac{r^{\ell+1}}{(2\ell+1)!!} e^{ikr} F(\ell+1+i\eta, 2\ell+2,
-2ikr) \nonumber \\
& = & \frac{r^{\ell+1}}{(2\ell+1)!!} e^{-ikr} F(\ell+1-i\eta, 2\ell+2,
2ikr) = {\rm real} \nonumber \\
& \sim & \frac{1}{k^{\ell+1}} |F_\ell(k)| \sin\,\left(kr-\eta {\rm log}\,2kr
-\frac{\pi}{2}\ell+\sigma_\ell \right) \qquad (r \rightarrow \infty)\ .
\label{a1}
\end{eqnarray}
Here, $\eta$ is the Sommerfeld parameter and $F_\ell(k)$ is the Coulomb Jost function

\begin{eqnarray}
F_\ell(k)=e^{\frac{\pi}{2}\eta}\frac{\ell !}{\Gamma(\ell+1+i\eta)}\ ,
\label{a2}
\end{eqnarray}
which can be obtained by comparing the behavior at the origin
between $\psi_\ell(k, r)$ and $\varphi_\ell(k, r)$.
The Jost solution of the Coulomb problem is defined by the irregular
solution with the asymptotic 
behavior $f_\ell(k, r) \sim e^{i(kr-\eta {\rm log}\,2kr-\pi\ell/2)}$ 
for $r \rightarrow \infty$.
More explicitly, it is given by
\begin{eqnarray}
f_\ell(k, r) & = & (-i)^\ell (2kr)^{-i\eta}e^{ikr}
G(\ell+1+i\eta, -\ell+i\eta, 2ikr) \nonumber \\
& = & i (-)^{\ell+1} e^{\pi\eta/2}(2kr)^{\ell+1}
e^{ikr} \Psi(\ell+1+i\eta, 2\ell+2, -2ikr) \nonumber \\
& \sim & e^{i(kr-\eta {\rm log}\,2kr-\pi\ell/2)}
\qquad (r \rightarrow \infty)\ .
\label{a3}
\end{eqnarray}
Here, $G(\alpha, \beta, z)$ and
$\Psi(\alpha, \gamma, z)$ are irregular solutions
of the confluent hypergeometric functions defined in
Refs.\,\citen{LL} and \citen{Le}, respectively,
and they are related to each other by
\begin{eqnarray}
\Psi(\alpha, \gamma, z)=z^{-\alpha} G(\alpha, \alpha-\gamma+1, -z)\ .
\label{a4}
\end{eqnarray}
The symmetries of the Jost solution and the Jost function are given by
\begin{eqnarray}
f^*_\ell(k,r)=(-)^\ell e^{\pi\eta}f_\ell(-k,r)\ \ ,\qquad
F^*_\ell(k)=e^{\pi\eta}F_\ell(-k)\ ,
\label{a5}
\end{eqnarray}
with the Coulomb factor $e^{\pi\eta}$. They satisfy
the usual definition of the Jost function
\begin{eqnarray}
F_\ell(k)=\lim_{r \rightarrow 0} \frac{(kr)^\ell}{(2\ell-1)!!}
f_\ell(k,r)\ ,
\label{a6}
\end{eqnarray}
and the relationship 
\begin{eqnarray}
\varphi_\ell(k, r)=\frac{1}{2ik^{\ell+1}}
\left\{ F^*_\ell(k) f_\ell(k, r)-F_\ell(k) f^*_\ell(k, r)\right\}\ ,
\label{a7}
\end{eqnarray}
for real $k$.

The usual Coulomb wave functions are defined
as the real functions satisfying the asymptotic behavior
\begin{eqnarray}
F_\ell(k, r) & \sim & \sin\,\left(kr-\eta \hbox{log}\,2kr
-\frac{\pi}{2}\ell+\sigma_\ell \right)\ ,\nonumber \\
G_\ell(k, r) & \sim & \cos\,\left(kr-\eta \hbox{log}\,2kr
-\frac{\pi}{2}\ell+\sigma_\ell \right)\ ,
\label{a8}
\end{eqnarray}
for $r \rightarrow \infty$.
These Coulomb wave functions are related to each other through
\begin{eqnarray}
\psi_\ell(k, r) & = & \frac{1}{k} e^{i \sigma_\ell} F_\ell(k, r)
\ ,\nonumber \\
\varphi_\ell(k, r) & = & \frac{1}{k^{\ell+1}} |F_\ell(k)|
~F_\ell(k, r)={\rm real}\ ,\nonumber \\
f_\ell(k, r) & = & e^{-i\sigma_\ell}\left[ G_\ell(k,r)+i F_\ell(k,r)\right]
\ ,\nonumber \\
f^*_\ell(k, r) & = & e^{i\sigma_\ell}\left[ G_\ell(k,r)-i F_\ell(k,r)\right]
\label{a9}
\end{eqnarray}
The relationship with the 
usual ``incident plane wave + outgoing (or incoming) spherical wave''
is $\psi^{(+)}_\ell(k, r)=\psi_\ell(k, r)$ 
and $\psi^{(-)}_\ell(k, r)=\psi^*_\ell(k, r)$.
This implies that
\begin{eqnarray}
\psi^{(+)}_\ell(k, r) & = & \psi_\ell(k, r)
\sim \frac{1}{k} e^{i\sigma_\ell} 
\sin\,\left(kr-\eta {\rm log}\,2kr
-\frac{\pi}{2}\ell+\sigma_\ell \right)\ ,\nonumber \\
& \sim & \frac{1}{k} \sin\,\left(kr-\eta {\rm log}\,2kr
-\frac{\pi}{2}\ell \right)
+f^C_\ell e^{i(kr-\eta {\rm log}\,2kr-(\pi/2)\ell)}\ , \nonumber \\
\psi^{(-)}_\ell(k, r) & = & \psi^*_\ell(k, r)
\sim \frac{1}{k} e^{-i\sigma_\ell} 
\sin\,\left(kr-\eta {\rm log}\,2kr
-\frac{\pi}{2}\ell+\sigma_\ell \right)\ ,\nonumber \\
& \sim & \frac{1}{k} \sin\,\left(kr-\eta \hbox{log}\,2kr
-\frac{\pi}{2}\ell \right)
+{f^C_\ell}^* e^{-i(kr-\eta {\rm log}\,2kr-(\pi/2)\ell)}\ .
\label{a10}
\end{eqnarray}
Here, $f^C_\ell=(1/2ik)(e^{2i\sigma_\ell}-1)$ is the Coulomb partial-wave
amplitude, and
\begin{eqnarray}
e^{2i\sigma_\ell}=\frac{\Gamma(\ell+1+i\eta)}{\Gamma(\ell+1-i\eta)}
\ \ ,\qquad
|F_\ell(k)|=\left[\frac{e^{2\pi \eta}-1}{2\pi \eta} \prod^\ell_{n=1}
\frac{n^2}{n^2+\eta^2}\right]^{\frac{1}{2}}\ .
\label{a11}
\end{eqnarray}

\section{Shift function of various screening functions}

In this appendix, we calculate the shift function
\begin{eqnarray}
\zeta^\rho(k) \equiv \frac{1}{2k} \int^\infty_{\frac{1}{2k}}
\frac{2k\eta}{r} \alpha_\rho(r)~d\,r
=\eta \int^\infty_{\frac{1}{2k}}
\frac{1}{r} \alpha_\rho(r)~d\,r\ ,
\label{b1}
\end{eqnarray}
appearing in \eq{eq1-5} for
various screening functions $\alpha^\rho(r)$ and
evaluate the no-screening limit $\rho \rightarrow \infty$.
When the screening is $\alpha_\rho(r)=e^{-(r/\rho)^n}$,
we can write an analytic expression 
\begin{eqnarray}
\hspace{-10mm} \zeta^\rho(k)=\eta \int^\infty_{\frac{1}{2k}} \frac{1}{r}
~e^{-\left(\frac{r}{\rho}\right)^n} d\,r
=\eta~{\rm log}\,(2k\rho)
-\frac{\eta}{n}\gamma
+\frac{\eta}{n} \sum^\infty_{r=1} \frac{(-)^r}{r~r!}
\left(\frac{1}{2k\rho}\right)^{nr}\ ,
\label{b2}
\end{eqnarray}
which leads to 
\begin{eqnarray}
\zeta^\rho(k) \rightarrow \eta~{\rm log}\,(2k\rho)
-\frac{\eta}{n}\gamma
\qquad \hbox{as} \quad \rho \rightarrow \infty\ .
\label{b3}
\end{eqnarray}
Here, $\gamma$ is the Euler constant.
On the other hand, the screening functions 
with more sharp transitions like 
$3)^\prime$ in \eq{eq4-5} seem to have 
no constant term like \eq{eq6-13}
in the limit of $\rho \rightarrow \infty$.
We will show this for the error function
screening in \eq{eq6-10}. The proof for the
exponential screening function in \eq{c7}
is also carried out similarly.

In order to prove \eq{eq6-13},
we separate the $r$ integral in \eq{b1}
into three pieces as
\begin{eqnarray}
\zeta^\rho(k) & = & \eta \int^\rho_{\frac{1}{2k}}
\frac{1}{r}~d\,r
-\eta \int^\rho_{\frac{1}{2k}}
\frac{1}{r} \left(1-\alpha_\rho(r)\right)~d\,r
+\eta \int^\infty_\rho
\frac{1}{r} \alpha_\rho(r)~d\,r \nonumber \\
& = & \eta~{\rm log}\,(2k\rho)-I_1(\rho)+I_2(\rho)\ .
\label{b4}
\end{eqnarray}
First, the positive integral $I_2(\rho)$ is estimated by
\begin{eqnarray}
I_2(\rho) < \frac{\eta}{\rho} \int^\infty_\rho
\alpha_\rho(r)~d\,r \ ,
\label{b5}
\end{eqnarray}
so that we only need to evaluate the integral
over $\alpha_\rho(r)$.
For the error function screening, the expression
\begin{eqnarray}
\alpha_\rho(r)=\frac{1}{\sqrt{\pi}}
\left\{ \int^\infty_{\beta (r-\rho)} e^{-t^2}~d\,t
+\int^\infty_{\beta (r+\rho)} e^{-t^2}~d\,t \right\}
\label{b6}
\end{eqnarray}
yields
\begin{eqnarray}
\int^\infty_\rho \alpha_\rho(r)~d\,r=\frac{1}{\sqrt{\pi}}
\frac{1}{2\beta}\left(1+e^{-(2\beta \rho)^2}\right)
-\frac{\rho}{2} \left( 1-{\rm erf}\,(2\beta \rho) \right)\ .
\label{b7}
\end{eqnarray}
We therefore find
\begin{eqnarray}
\hspace{-10mm} 
I_2(\rho) < \frac{1}{\sqrt{\pi}}
\frac{\eta}{2\beta \rho}\left(1+e^{-(2\beta \rho)^2}\right)
-\frac{\eta}{2} \left( 1-{\rm erf}\,(2\beta \rho) \right)
\longrightarrow 0 \quad \hbox{as} \quad \rho \rightarrow \infty\ .
\label{b8}
\end{eqnarray}
In order to evaluate $I_1(\rho)$, we use
\begin{eqnarray}
\alpha_\rho(r)=1-\frac{1}{\sqrt{\pi}}
\int^{\beta (\rho+r)}_{\beta (\rho-r)} e^{-t^2}~d\,t
\label{b9}
\end{eqnarray}
derived from \eq{b6}, and express it as
\begin{eqnarray}
I_1(\rho)=\frac{\eta}{\sqrt{\pi}} \int^\rho_{\frac{1}{2k}}
\frac{1}{r} \left(
\int^{\beta (\rho+r)}_{\beta (\rho-r)} e^{-t^2}~d\,t \right)~d\,r\ .
\label{b10}
\end{eqnarray}
Here, we change the integral variable from $r$ to $x$ by $r=\rho x$
and obtain
\begin{eqnarray}
I_1(\rho) & = & \frac{\eta}{\sqrt{\pi}} \int^1_{\varepsilon}
\frac{1}{x} \left(
\int^{\alpha (1+x)}_{\alpha (1-x)} e^{-t^2}~d\,t \right)~d\,x\ ,
\label{b11}
\end{eqnarray}
with $\alpha = \beta \rho$ and $\varepsilon= \frac{1}{2k\rho}$.
We consider the upper bound $\frac{\eta}{\sqrt{\pi}} 
\widetilde{I}_1(\alpha) > I_1(\rho)$ with
\begin{eqnarray}
\widetilde{I}_1(\alpha)=\int^1_{0}
\frac{1}{x} \left(
\int^{\alpha (1+x)}_{\alpha (1-x)} e^{-t^2}~d\,t \right)~d\,x \ .
\label{b12}
\end{eqnarray}
We separate the integral interval $[0, 1]$ into $[0, 1-\delta]$
and $[1-\delta, 1]$ with a small positive $\delta > 0$.
Then, we find
\begin{eqnarray}
\widetilde{I}_1(\alpha) & = & \int^{1-\delta}_{0}
\frac{1}{x} \left(
\int^{\alpha (1+x)}_{\alpha (1-x)} e^{-t^2}~d\,t \right)~d\,x
\nonumber \\
& & +\int^{1}_{1-\delta}
\frac{1}{x} \left(
\int^{\alpha (1+x)}_{\alpha (1-x)} e^{-t^2}~d\,t \right)~d\,x\ .
\label{b13}
\end{eqnarray}
Here, the first term is bounded by $2 \alpha 
e^{-(\alpha \delta)^2} (1-\delta)$. As to the second term,
we change the integral variable from $x$ to $y$ by $x=1-y$ 
and find
\begin{eqnarray}
2\hbox{-nd~term} & = & \int^\delta_0 \frac{1}{1-y}
\left( \int^{\alpha(2-y)}_{\alpha y}
e^{-t^2}~d\,t \right)~d\,y
\nonumber \\
& < & \frac{1}{1-\delta} \int^\delta_0 
\left( \int^\infty_0 e^{-t^2}~d\,t \right)~d\,y
=\frac{\delta}{1-\delta}\frac{\sqrt{\pi}}{2}\ .
\label{b14}
\end{eqnarray}
Thus, we obtain
\begin{eqnarray}
0 \leq \widetilde{I}_1(\alpha) \leq
2 \alpha e^{-(\alpha \delta)^2} (1-\delta)
+\frac{\delta}{1-\delta}\frac{\sqrt{\pi}}{2}\ .
\label{b15}
\end{eqnarray}
First, we take the limit $\alpha \rightarrow \infty$ 
in \eq{b15} and obtain
\begin{eqnarray}
0 \leq \lim_{\alpha \rightarrow \infty}
\widetilde{I}_1(\alpha) \leq \frac{\delta}{1-\delta}
\frac{\sqrt{\pi}}{2}\ .
\label{b16}
\end{eqnarray}
Since we can take $\delta > 0$ arbitrary small,
we eventually find
\begin{eqnarray}
I_1(\rho) \leq \frac{\eta}{\sqrt{\pi}} \widetilde{I}_1(\alpha)
\longrightarrow 0 \qquad \hbox{as} \quad \rho \rightarrow \infty\ .
\label{b17}
\end{eqnarray}

\section{Screening function $\alpha^\rho(R)$
for the $dp$ scattering}

In this appendix, we derive the screening function $\alpha^\rho(R)$
for the $pd$ scattering, starting from
the screened Coulomb function $\omega^\rho(r; 1, 2)$ in \eq{eq6-15}
for the $pp$ system of the quark-model baryon-baryon interaction.
We first note that the $\rho \rightarrow \infty$ limit,
$\omega=\lim_{\rho \rightarrow \infty} \omega^\rho$, is
an error function Coulomb, which satisfies
\begin{eqnarray}
\langle \psi_d|(P \omega)| \psi_d\rangle
\sim \frac{e^2}{R}
\qquad \hbox{for} \quad R \rightarrow \infty\ ,
\label{c1}
\end{eqnarray}
where $\langle 1, 2|\psi_d \rangle$ is the deuteron wave function
and $P$ is the rearrangement permutation operator $P=P_{(12)}P_{(23)}
+P_{(13)}P_{(23)}$.
We follow the procedure similar to the $\alpha \alpha$ case
in \eq{eq6-9} and separate the folding $pd$ potential 
in \eq{eq6-16} for the screened Coulomb force 
into the long-range and short-range parts:
\begin{eqnarray}
V^{\rho C}_{pd}(R)
& = & \langle \psi_d|(P \omega^\rho)| \psi_d\rangle
=\langle \psi_d|(P \omega^\rho)-(P\omega)| \psi_d\rangle
+\langle \psi_d|(P\omega)| \psi_d\rangle
\nonumber \\
& = & \left(\frac{e^2}{R}-\langle \psi_d|(P \omega)
-(P\omega^\rho)| \psi_d\rangle \right)
+\left(-\frac{e^2}{R}+\langle \psi_d|(P\omega)| \psi_d\rangle
\right)
\nonumber \\
& = & \frac{e^2}{R} \alpha^\rho(R)+ \CW(R)
=W^\rho(R)+\CW(R)\ .
\label{c2}
\end{eqnarray}
Here, the screening function $\alpha^\rho(R)$ and
the short-range Coulomb potential $\CW(R)$ is given by
\begin{eqnarray}
\alpha^\rho(R) & = & 1-\frac{R}{e^2} \langle \psi_d|(P \omega)
-(P\omega^\rho)|\psi_d \rangle\ ,\nonumber \\
\CW(R) & = & \langle \psi_d|(P\omega)-\frac{e^2}{R}| \psi_d \rangle\ .
\label{c3}
\end{eqnarray}
On the other hand, the exchange term \eq{eq5-2} in the three-body model space
yields the matrix element
\begin{eqnarray}
V^{\rho C}_{pd}(R)=\langle \psi_d|\CW^\rho+W^\rho| \psi_d\rangle
=\langle \psi_d|\CW^\rho| \psi_d\rangle+W^\rho(R)\ .
\label{c4}
\end{eqnarray}
We therefore find that the deuteron matrix element
of the polarization potential is $\rho$-independent:
\begin{eqnarray}
\langle \psi_d|\CW^\rho| \psi_d\rangle=\CW(R)\ .
\label{c5}
\end{eqnarray}

We first assume the sharply cut-off Coulomb force
\eq{eq5-1} with $\alpha_\rho(r)=\theta(\rho-r)$ for
the two protons, and examine the screening property
discussed in $\S 4$ by using available analytic expressions.
This is possible, if we further neglect the $D$-state 
component of the deuteron wave function and assume that
the spatial part of the $S$-wave component
is given by a simple exponential function
$u(r)=\sqrt{2\gamma}\,e^{-\gamma r}$.
In this case, the folding potential is
expressed in terms of the integral exponential function
defined by
\begin{eqnarray}
{\rm Ei}(-x) & = & -\int^\infty_x \frac{e^{-t}}{t}~d\,t
\nonumber \\
& = & {\rm log}\,x+\gamma-x+\frac{x^2}{2\cdot 2!}
- \cdots + \frac{(-x)^r}{r\cdot r!}-\cdots <0 \quad (\hbox{for} \quad x > 0)
\nonumber \\
& \sim & e^{-x} \sum^\infty_{n=1}(-)^n \frac{(n-1)!}{x^n}
\qquad (\hbox{asymptotic expansion})\ .
\label{c6}
\end{eqnarray}
We find
\begin{eqnarray}
& & V^{\rho C}_{pd}(R) = \CW(R)+\frac{e^2}{R}\,\alpha_\rho(R) 
\ ,\nonumber \\ [2mm]
& & \CW(R) = V^C_{pd}(R)-\frac{e^2}{R}
=-\frac{e^2}{R} e^{-4\gamma R}-4 \gamma e^2~{\rm Ei}(-4\gamma R) \nonumber \\
& & \sim - e^{-4\gamma R} \frac{e^2}{4\gamma R^2}
\left(1-\frac{2!}{4\gamma R}+\frac{3!}{(4\gamma R)^2}-\cdots \right)\ ,
\nonumber \\ [2mm]
& & \alpha_\rho(R) = 1-2\gamma \int^\infty_\rho d\,r
~\left[~{\rm Ei}(-4\gamma(r+R))-{\rm Ei}(-4\gamma|r-R|)~\right] \nonumber \\
& & = -2\gamma~\left[~\int^\infty_{R-\rho}d\,r~{\rm Ei}(-4\gamma r)
+\int^\infty_{R+\rho}d\,r~{\rm Ei}(-4\gamma r)~\right]
\qquad \hbox{for} \quad R \geq \rho \ .
\label{c7}
\end{eqnarray}
Here, $V^C_{pd}(R)=\lim_{\rho \rightarrow \infty}
V^{\rho C}_{pd}(R)=\langle \psi_d|(P \omega)| \psi_d\rangle$.
In order to derive the last expression of \eq{c7},
we use the relationship
\begin{eqnarray}
(-4\gamma)\int^\infty_0 d\,r~{\rm Ei}(-4\gamma r)=1\ ,
\label{c8}
\end{eqnarray}
which is obtained by exchanging the integration order.
From here, we can obtain
\begin{eqnarray}
\hspace{-10mm}
(-2\gamma)\int^\infty_\rho d\,r~{\rm Ei}(-4\gamma |r-R|)
=1+2\gamma \int^\infty_{R-\rho} d\,r~{\rm Ei}(-4\gamma r)
\qquad \hbox{for} \quad R \geq \rho \ .
\label{c9}
\end{eqnarray}
The asymptotic form of $\CW(R)$ in \eq{c7} is due to
\begin{eqnarray}
\hspace{-10mm} (-4\gamma)~{\rm Ei}(-4\gamma r)
\sim \frac{1}{r} e^{-4\gamma r}
\left(1-\frac{1}{4\gamma r}+\frac{2!}{(4\gamma r)^2}- \cdots \right)
\qquad \hbox{as} \quad r \rightarrow \infty\ .
\label{c10}
\end{eqnarray}
If we further use this in the last expression of \eq{c7},
we find
\begin{eqnarray}
\alpha_\rho(R) & \sim & -\frac{1}{2}
\left[~{\rm Ei}(-4\gamma (R-\rho))+{\rm Ei}(-4\gamma (R+\rho))~\right]
\nonumber \\
& \sim & -{\rm Ei}(-4\gamma R)
\rightarrow 0 \qquad \hbox{as} \quad R \rightarrow \infty\ .
\label{c11}
\end{eqnarray}
With $R$ fixed, we can show $\lim_{\rho \rightarrow \infty} 
\alpha_\rho(R)=1$ as follows.
First, \eq{c9} and some calculations yield
\begin{eqnarray}
& & \frac{2\gamma}{R}\int^\infty_0 d\,r~\left[~{\rm Ei}(-4\gamma (r+R))
-{\rm Ei}(-4\gamma |r-R|)~\right]
\nonumber \\
& & =\frac{1}{R}+\frac{4\gamma}{R}\int^\infty_R d\,r~{\rm Ei}(-4\gamma r)
=\frac{1}{R}\left(1-e^{-4\gamma R}\right)
-4\gamma~{\rm Ei}(-4\gamma R)\ .
\label{c12}
\end{eqnarray}
Thus, \eq{c10} gives
\begin{eqnarray}
& & 2\gamma~\int^\infty_0 d\,r~\left[~{\rm Ei}(-4\gamma (r+R))
-{\rm Ei}(-4\gamma |r-R|)~\right]
\nonumber \\
& & \sim 1-e^{-4\gamma R} \frac{1}{4\gamma R}
\left(1-\frac{2!}{4\gamma R} + \cdots \right)
\qquad \hbox{as} \quad R \rightarrow \infty\ .
\label{c13}
\end{eqnarray}
Here, because of $r+R \geq |r-R|$, the integrand of \eq{c13}
is always positive.
Furthermore, the integral from 0 to $\rho$ in \eq{c13} is 
the monotonically increasing function of $\rho$ and 
the limit $\rho \rightarrow \infty$ exists.
We therefore find
\begin{eqnarray}
\lim_{\rho \rightarrow \infty}
2\gamma~\int^\infty_\rho d\,r~\left[~{\rm Ei}(-4\gamma (r+R))
-{\rm Ei}(-4\gamma |r-R|)~\right]=0\ .
\label{c14}
\end{eqnarray}
After all, we find that the screening function $\alpha^\rho(R)$
satisfies the condition 1) - 3) at least in this simplest case.
If we calculate the shift function $\zeta^\rho(k)$ using 
$\alpha^\rho(R)$ in \eq{c7}, we find the same result \eq{eq6-13};
namely, there is no constant term as in the sharply cut-off Coulomb
case.

The calculation of $\alpha^\rho(R)$ using the screened Coulomb potential 
in \eq{eq6-15} and the realistic deuteron wave function by the
quark-model baryon-baryon interaction is rather involved.
We here show only the final result for the numerical calculations.
The screening function $\alpha^\rho(R)$ in the channel-spin formalism
is given by
\begin{eqnarray}
\alpha^J_{(\ell S_c), (\ell^\prime S^\prime_c)}(R)
=\delta_{\ell, \ell^\prime} \delta_{S_c, S^\prime_c}
-\sum_{\lambda, \lambda^\prime=0,2} \sum_{\kappa=0,2,4}
f^\kappa_{\lambda \lambda^\prime}(R)
~g^{\lambda \lambda^\prime \kappa J}_{(\ell S_c),
(\ell^\prime S^\prime_c)}\ ,
\label{c15}
\end{eqnarray}
where the kinematical factor 
$g^{\lambda \lambda^\prime \kappa J}_{(\ell S_c), 
(\ell^\prime S^\prime_c)}$ is given by
\begin{eqnarray}
 & & \hspace{-10mm} g^{\lambda \lambda^\prime \kappa J}
_{(\ell S_c), (\ell^\prime S^\prime_c)}
=(-)^{S_c+S^\prime_c+1} 3 \widehat{S}_c \widehat{S}^\prime_c
\widehat{\lambda} \widehat{\ell}
\widehat{\lambda}^\prime \widehat{\ell}^\prime
\langle \lambda 0 \lambda^\prime 0|\kappa 0 \rangle
\langle \ell 0 \ell^\prime 0|\kappa 0 \rangle
\nonumber \\
& & \hspace{-10mm} \times \sum_S (2S+1)
\left\{ \begin{array}{ccc}
\H & 1 & S_c \\
\lambda & S & 1 \\
\end{array}\right\}
\left\{ \begin{array}{ccc}
\H & 1 & S^\prime_c \\
\lambda^\prime & S & 1 \\
\end{array}\right\}
\nonumber \\
& & \hspace{-10mm} \times \sum_L (-)^L (2L+1)
\left\{ \begin{array}{ccc}
\lambda & \ell & L \\
\ell^\prime & \lambda^\prime & \kappa \\
\end{array}\right\}
\left\{ \begin{array}{ccc}
J & \ell & S_c \\
\lambda & S & L \\
\end{array}\right\}
\left\{ \begin{array}{ccc}
J & \ell^\prime & S^\prime_c \\
\lambda^\prime & S & L \\
\end{array}\right\}\ .
\label{c16}
\end{eqnarray}
Th spatial function $f^\kappa_{\lambda \lambda^\prime}(R)$
is given by
\begin{eqnarray}
& & f^\kappa_{\lambda \lambda^\prime}(R)
=\int^\infty_0 d\,r~u_{\lambda}(r)
\,u_{\lambda^\prime}(r)\,v_{\kappa}(R, r/2)\ ,\nonumber \\
& & v_{\kappa}(R, r/2)=\frac{1}{2} \int^1_{-1}d\,x
~v\left(\sqrt{R^2+r^2/4-rRx}\right)~P_\kappa(x)\ ,
\nonumber \\
& & v(r)=\frac{R}{2r}\left[{\rm erf}\left(\frac{\sqrt{3}}{2}\frac{r+\rho}{b}
\right)
+{\rm erf}\left(\frac{\sqrt{3}}{2}\frac{r-\rho}{b}\right)\right]\ ,
\label{c17}
\end{eqnarray}
where $u_\lambda(r)$ is the $S$-wave ($\lambda=0)$)
and $D$-wave ($\lambda=2)$ deuteron wave functions
usually denoted by $u(r)$ and $w(r)$, respectively.

%
%

\newcommand{\etal}{{\em et al.}}

\end{document}